\documentclass[aps,prb,twocolumn,superscriptaddress,floatfix]{revtex4-2}

\usepackage{graphicx}
\usepackage{color}
\usepackage[dvipsnames]{xcolor}
\usepackage{amsmath,amssymb,mathrsfs,ulem}
\usepackage{braket}
\usepackage{mathtools}
\usepackage{subfigure}

\usepackage{bbm}

\definecolor{mildblue}{rgb}{.2,0.7,.9}
\definecolor{warmred}{rgb}{.68,0.19,.14}

\definecolor{dante}{rgb}{.2,0.7,.9}

\definecolor{daniel}{rgb}{.0,1,0.5}

\date{\today}

\usepackage{array, multirow, bigdelim, makecell} 

\begin{document}

\title{Realization of a three-dimensional quantum Hall effect in a Zeeman-induced second order topological insulator on a torus}

\author{Zhe Hou}
\affiliation{Department of Physics, University of Basel, Klingelbergstrasse 82, 
CH-4056 Basel, Switzerland}
\author{Clara S. Weber}
\affiliation{Institut f{\"u}r Theorie der Statistischen Physik, RWTH Aachen University, 
52056 Aachen, Germany \\ and JARA - Fundamentals of Future Information Technology, 52056 Aachen, Germany}
\author{Dante M. Kennes}
\affiliation{Institut f{\"u}r Theorie der Statistischen Physik, RWTH Aachen University, 
52056 Aachen, Germany \\ and JARA - Fundamentals of Future Information Technology, 52056 Aachen, Germany}
\affiliation{Max Planck Institute for the Structure and Dynamics of Matter, Center for Free Electron Laser Science, 22761 Hamburg, Germany}
\author{Daniel Loss}
\affiliation{Department of Physics, University of Basel, Klingelbergstrasse 82, 
CH-4056 Basel, Switzerland}
\author{Herbert Schoeller}
\affiliation{Institut f{\"u}r Theorie der Statistischen Physik, RWTH Aachen University, 
52056 Aachen, Germany \\ and JARA - Fundamentals of Future Information Technology, 52056 Aachen, Germany}
\author{Jelena Klinovaja}
\affiliation{Department of Physics, University of Basel, Klingelbergstrasse 82, 
CH-4056 Basel, Switzerland}
\author{Mikhail Pletyukhov}
\email[Email: ]{pletmikh@physik.rwth-aachen.de}
\affiliation{Institut f{\"u}r Theorie der Statistischen Physik, RWTH Aachen University, 
52056 Aachen, Germany \\ and JARA - Fundamentals of Future Information Technology, 52056 Aachen, Germany}

\begin{abstract}
We propose a realization of a quantum Hall effect (QHE) in a second-order topological insulator (SOTI) in three dimensions (3D), which is mediated by hinge states on a torus surface. It results from the nontrivial interplay of the material structure, Zeeman effect, and  the surface curvature. In contrast to the conventional 2D- and 3D-QHE, we show that the 3D-SOTI QHE is {\it not} affected by orbital effects of the applied magnetic field and
exists in the presence of a Zeeman term only, induced e.g. by magnetic doping. To explain the 3D-SOTI QHE, we analyze the boundary charge for a 3D-SOTI and establish its universal dependence on the Aharonov-Bohm flux threading through the torus hole. Exploiting the fundamental relation between the boundary charge and the Hall conductance, 
we demonstrate the universal quantization of the latter, as well as its stability against random disorder potentials and continuous deformations of the torus surface. 
\end{abstract}

\maketitle

{\it Introduction.}---The quantum Hall effect (QHE), observed by Klitzing {\it et al.} \cite{klitzing_etal_prl_80} in 1980, is one of the fundamental phenomena in condensed matter physics. It has since been a driving force for the field of topological insulators (TI) with a plethora of new topological materials discovered in the last decades \cite{kane_mele_1_prl_05,kane_mele_2_prl_05,bernevig_etal_science_06,koenig_etal_science_07,hasan_kane_rmp_10,qi_zhang_rmp_11}. 
As a consequence, many intriguing generalizations such as the anomalous QHE (in the absence of an external magnetic field) in 2D \cite{haldane_prl_88,liu_etal_prl_08,bestwick_etal_prl_15,yu_etal_science_10,chang_etal_science_13,checkelsky_etal_ntp_14,feng_etal_prl_15,kandala_etal_natcom_15,deng_etal_science_20,liu_etal_natmat_20,ge_etal_nsr_20,riberolles_etal_prb_21,yan_etal_prb_21,fukasawa_etal_prb_21,xu_etal_prb_21} and the 3D-QHE \cite{sitte_etal_prl_12,tang_etal_nature_19,xu_etal_ntp_14,schumann_etal_prl_18,zhang_etal_nature_19} have been studied.
Moreover, the concept of TIs has been extended to higher-order TIs, which are governed by a new bulk-boundary correspondence \cite{benalcazar_etal_science_17,benalcazar_etal_prb_17,song_etal_prl_17,langbehn_etal_prl_17,geier_etal_prb_18,imhof_etal_ntp_18,schindler_etal_sca_18,trifunovic_brouwer_prx_19}. For example, a second (third) order TI with dimension $D$ features $D-2$ ($D-3$) dimensional hinge (corner) states. Recently, within 3D second-order TIs (3D-SOTI), an additional class of systems exhibiting the QHE has been proposed \cite{langbehn_etal_prl_17,fu_etal_prr_21,petridis_zilberberg_prr_20,volpez_etal_prl_19,laubscher_etal_prr_19}.

One way to realize a 3D-SOTI is to build on a 3D-TI \cite{fu_etal_prl_07,hasan_kane_rmp_10,qi_zhang_rmp_11,chang_etal_science_13,xu_etal_ntp_12,rienks_etal_nature_19} and introduce anisotropic gaps on different surfaces (boundaries), e.g., with an effective Zeeman term by magnetic doping or using an external magnetic field \cite{khalaf_prb_18,volpez_etal_prl_19,laubscher_etal_prr_19,plekhanov_etal_prr_19,ren_etal_prl_20,laubscher_etal_prb_20,laubscher_etal_prr_20,plekhanov_etal_prr_20,plekhanov_etal_prb_21}.
At special positions where the effective gap closes and reverses sign, topological hinge or corner states emerge, in analogy to 
the Jackiw-Rebbi mode \cite{jackiw_rebbi_prd_76}. 

\begin{figure}[t]
        \centering
        \includegraphics[width=\columnwidth]{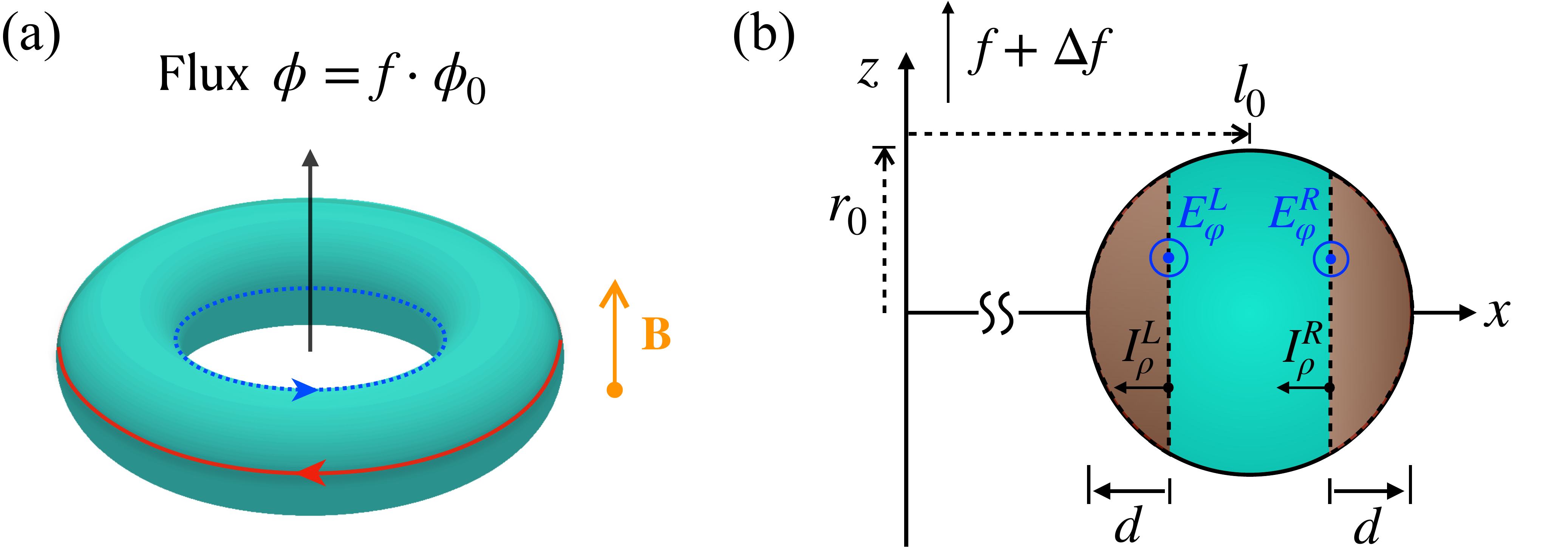}
        \caption{(a) Torus model  and  (b) its 2D cross-section in the $x-z $ plane. The torus hole is pierced by the Aharonov-Bohm flux $\phi = f \cdot \phi_0$.  In addition, there is a constant uniform magnetic field $\bf{B}$ in $z$ direction. The system hosts the clockwise (red) and counter-clockwise (blue) propagating hinge states at the outer and inner equatorial rings of the torus, respectively. The hinge states are localized within the outer (or right) and the inner (or left) boundary regions [shown in brown color in (b)] of the width $d$ (as measured in the radial direction). The electric fields $E^{R,L}_{\varphi}$ and the Hall currents $I^{R,L}_{\rho}$ are generated upon changing $f$ as shown in (b).}
        \label{fig:models}
    \end{figure}

In this Letter, we will consider such a model and discuss a specific realization using a torus geometry (or smooth deformations thereof) as shown in Fig.~\ref{fig:models}, where the emerging hinge states are shown with blue and red arrows. This geometry allows us to insert a magnetic flux $\phi$ through the torus hole, and thus we construct a 
3D analogy of
the Corbino disc which is used 
in the Laughlin argument  
for the conventional 2D-QHE \cite{laughlin_prb_81}. However, in contrast to the Corbino setup, the 3D-SOTI QHE discussed here is {\it not} induced by orbital effects but results purely from the Zeeman effect.

While there are several approaches known to deduce the conventional 2D quantized Hall conductance \cite{laughlin_prb_81,thouless_etal_prl_82,avron_etal_prl_83,kohmoto_aph_85}, 
we will present a generalization of a recently developed boundary charge approach to the case of a 3D-SOTI, which, most importantly, is applicable to {\it both} clean and disordered systems. It is based on recent detailed studies of the boundary charge in one-dimensional (1D) modulated wires \cite{park_etal_prb_16,thakurathi_etal_prb_18,pletyukhov_etal_prb_20,pletyukhov_etal_prbr_20,pletyukhov_etal_prr_20,miles_etal_prb_21,mueller_etal_prb_21,laubscher_etal_prb_21,weber_etal_prl_21,piasotski_etal_prr_21,hayward_etal_pra_21}, together with dimensional extensions to 2D-QHE models \cite{thakurathi_etal_prb_18,pletyukhov_etal_prb_20} revealing the fundamental relationship between the linear dependence of the boundary charge on the phase of the potential in 1D models and the quantized Hall conductance in 2D models, a concept which can even  be generalized to explain the fractional QHE in the presence of interactions \cite{laubscher_etal_prb_21}.

\begin{figure}[t]
        \centering
         \includegraphics[width = 0.8\columnwidth]{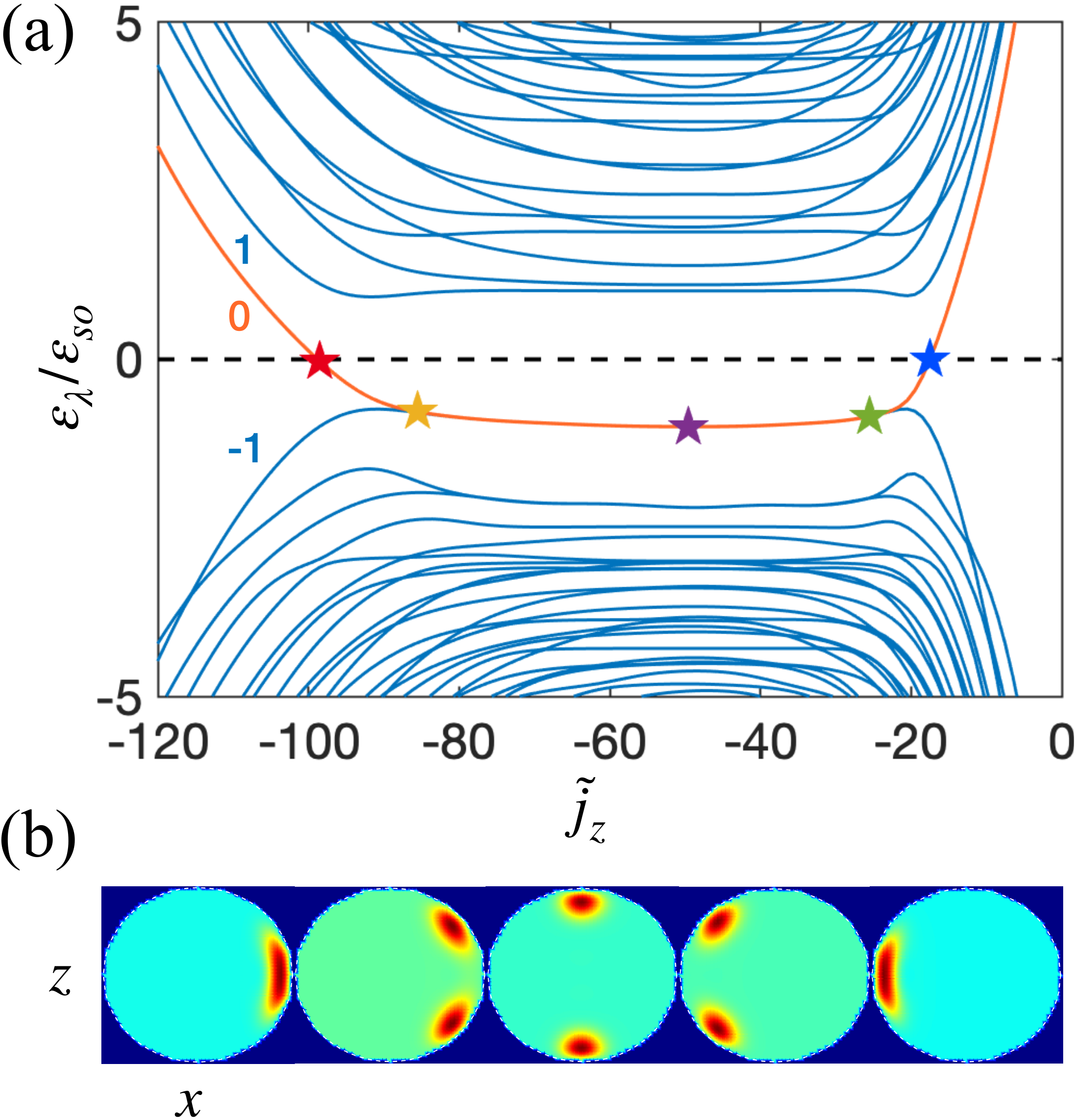}    
        \caption{(a) Low-energy spectrum of torus model in units of spin-orbit energy $\varepsilon_{so} = \frac{m^* \alpha^2}{2}$. Band indices are $\lambda \leq -1$ for valence bands, $\lambda \geq 1$ for conduction bands, and $\lambda =0$ for the band hosting Landau surface states in the plateau region together with the right ($R$) and left ($L$) hinge states in the steep parts of the dispersion. The parameters we use here are: $\delta=3 \, \varepsilon_{so}$, $l_B=l_Z=l_{so}$, $l_0=10 \, l_{so}$, and $r_0=5 \, l_{so}$. The lattice constant was chosen $a=l_{so}/5$ in the tb model calculation. The black dashed line denotes the chemical potential $\mu=0$. Note that the two bands $\lambda=0,-1$ are almost degenerate along the plateau region but separated by an (invisible) small gap.
        (b) Probability density of the five states labeled with star symbols in (a) on the surface band $\lambda=0$ below the chemical potential, shown in the same sequence as in (a). The red and blue stars denote the $R$ and $L$ hinge states, respectively, while the yellow, purple, and green stars denote the Landau surface states with the intermediate angular positions of the density localization. The total angular momentum for the five states are (from left to right): $\tilde{j}_z=$ -98.5, -85.5, -49.5, -25.5, -17.5. }
        \label{fig:torus_spectrum}
\end{figure}

Most importantly, we demonstrate for the 3D-SOTI model on a torus that the boundary charge, driven by the Aharonov-Bohm flux $\phi=f \phi_0$, where $\phi_0$ is the flux quantum, shows a linear relation with respect to $f$, with the slope being universal and quantized to integer numbers $\pm e$, where $e$ is the electron charge. 
Examining all states below the chemical potential, we reach a remarkable conclusion: the whole contribution to the boundary charge comes only from the top-most occupied valence band [shown in orange in Fig.~\ref{fig:torus_spectrum}(a)]. We further include disorder effects by considering a 3D tight-binding (tb) counterpart of our continuum model with broken rotational symmetry and show robustness of the quantized slope. The boundary charge concept, thus, provides a powerful tool in characterizing 3D-SOTIs and universally  explains topological effects in condensed matter systems as well as their stability against random potential disorder and continuous surface deformations.

{\it Model.}---In this Letter we consider a continuum model in a finite three-dimensional region, which is confined to the interior of a torus [see Fig.~\ref{fig:models}(a))], to realize a second order TI with a minimal set of ingredients, given by band inversion, spin-orbit coupling, magnetic flux, and Zeeman term.  The Hamiltonian (with $e=\hbar=1$) is given by 
\begin{align}
    H = \sigma_z \left[ \frac{(\textbf{p}+\textbf{A})^2}{2m^*} - \delta \right] 
    +\alpha \,\sigma_x  (\textbf{p} + \textbf{A}) \cdot \textbf{s} + \varepsilon_Z s_z,
    \label{eq:H_main}
\end{align}
where $\textbf{p}$ and $\textbf{S}=\frac{1}{2} \textbf{s}$ are 3D vectors of the momentum and the physical spin-$\frac{1}{2}$ operators, respectively. The parameters $m^*$ and $\alpha>0$ are the  effective  mass of the electron and the spin-orbit interaction, respectively, while $\varepsilon_Z$ denotes the Zeeman energy. Both the  magnetization direction inducing the Zeeman term   and the uniform magnetic field $\textbf{B}$ generated by the vector potential $\textbf{A}$ are assumed to point along the $z$-axis. We note that $\varepsilon_Z$ is an independent parameter, controllable by magnetic doping. The Pauli matrices $\sigma_i$ span the orbital degrees of freedom. The parameter $\delta >0$ is chosen positive in order to realize the band inversion at finite values of $\alpha$, $\varepsilon_Z$, and $B \equiv |\textbf{B}| $. The vector potential $\textbf{A}= \textbf{A}_{B} + \textbf{A}_{\phi} =  \frac{B \rho }{2} \textbf{e}_{\varphi}+ \frac{\phi}{2 \pi \rho} \textbf{e}_{\varphi}  \equiv   \frac{B \rho }{2} \textbf{e}_{\varphi}+ \frac{f}{\rho} \textbf{e}_{\varphi}$ contains the terms generating both the uniform magnetic field $B $ and the Aharonov-Bohm (AB) magnetic flux $\phi$. Here, we used the cylindrical coordinates $(\rho, \varphi,z)$ and introduced the number of flux quanta $f =\frac{\phi}{\phi_0} \stackrel{!}{=} \frac{\phi}{2 \pi}$.

Besides the spin-orbit length $l_{so}=1/(\alpha m^*)$, we introduce two further typical scales $l_B$ and $l_Z$ associated with the orbital magnetic field and the Zeeman term, respectively, via $B=1/l_B^2$ and $\epsilon_Z=1/(2m^* l_Z^2)$. In the main part of this Letter we will discuss the most critical case of strong orbital fields, where $l_B \sim l_Z$, and will see that the orbital effects do not influence the quantization of the Hall conductance. In the  Supplemental Material (SM) \cite{SM} we will also discuss the case of zero orbital effects with the same conclusion. This shows that our results are expected to be stable for any orbital field. 

In the torus geometry and in the absence of disorder the model \eqref{eq:H_main} possesses the axial symmetry generated by the $z$-component of the total angular momentum operator $J_z = L_z + \frac{1}{2} s_z = - i \frac{\partial}{\partial \varphi} + \frac{1}{2} s_z $. Exploiting this symmetry we reduce the 3D model to an effective 2D model. It is
described by the Hamiltonian (see SM \cite{SM} for details)
\begin{align}
h_{2D} &=  \frac{\sigma_z}{2 m^*} \left[ -  \frac{\partial^2}{\partial x^2}  + \frac{J^2 (x)}{x^2}-  s_z \frac{J (x)}{x^2}  - \frac{\partial^2}{\partial z^2} - 2 m^* \delta \right] \nonumber \\
&+ \alpha \sigma_x \left[ - i s_{x}  \frac{\partial}{\partial x} + s_y \frac{J (x)}{x} - i s_z \frac{\partial}{\partial z} \right] + \varepsilon_Z  s_z  ,
\label{ham2D}
\end{align}
and defined on the disc of radius $r_0$ in the $(x>0,z)$ half-plane, which is displaced away from the $z$-axis by $l_0 >r_0$ [see Fig.~\ref{fig:models}(b)]. We introduce the following continuous parameter $\tilde{j}_z = j_z +f$, where $j_z$ are half-integer eigenvalues of the operator $J_z$, and $f$ is restricted to fractional values, $f \in [0,1)$, interpolating between adjacent values of $j_z$. The parameter $\tilde{j}_z$ appears inside the function $J (x) = \tilde{j}_z + \frac{B x^2}{2}$, where the last term accounts for the orbital $B$-field effect.

{\it Hinge states.}---Approximating \eqref{ham2D} by its tight-binding (tb) lattice counterpart or by solving it directly in an appropriately chosen basis (see SM \cite{SM} for details), we calculate the band structure shown in Fig.~\ref{fig:torus_spectrum}(a). Among the valence and conduction bands $\varepsilon_{\lambda} (\tilde{j}_z)$, labeled by the band indices $\lambda \leq -1$ and $\lambda \geq 1$ and lying below and above the chemical potential $\mu=0$, respectively, there is a distinguished band $\lambda=0$ (orange line) which crosses the chemical potential at the two points $\tilde{j}_{z;R,L}^*$ (with $\varepsilon'_0 (\tilde{j}_{z;R,L}^*)\lessgtr 0 $), see also in Fig.~\ref{fig:torus_charge}(a).

As shown in  Fig.~\ref{fig:torus_spectrum}(b), the states on the negative linear slope near $\tilde{j}_{z,R}^*$ (indicated by the red star) form the right ($R$) hinge mode, since their density is localized near the disc right  side $\theta = \frac{\pi}{2}$ (expressed in the polar coordinates $x= l_0 + r \sin \theta $, $z=r \cos \theta$, with $\theta \in (-\pi,\pi]$ and $r<r_0$, associated with  the disc). They propagate in the clockwise direction along the outer equator of the torus (looking from the top), see the red line in Fig.~\ref{fig:models}(a).  The left ($L$) hinge mode with the positive slope resides near $\tilde{j}^*_{z,L}$ [indicated by the blue star in Fig.~\ref{fig:torus_spectrum}(a)], its states are localized near $\theta \approx -\frac{\pi}{2}$ (inner equator of the torus) and propagate in the counter-clockwise direction, see the blue line in Fig.~\ref{fig:models}(a).

To understand the properties of these hinge modes more deeply we rely on their low-energy description \cite{SM} which is valid in the regime $l_0 \gg r_0 \gg l_{so}$ and $ \frac{r_0}{l_{so}} \gg \frac{l_{so}^2}{l_Z^2} \gg \frac{l_{so}}{r_0}$. It allows us to reveal the spinor structure of the hinge modes as well as to provide quantitative estimates of their properties. 

In particular, we establish that the $R$ and $L$ hinge modes are stabilized by the Zeeman term, staying exponentially localized near the torus surface on the length scale of $l_{so}$. In the angular direction, their localization is gaussian with the variance $\sigma_{\theta} = \frac{l_Z}{\sqrt{r_0 l_{so}/2}}$. The presence of $l_Z$ in the variance is indicative of the Zeeman mechanism of their onset. Moreover, we notice that their angular spread  exceeds the radial localization length, that is $r_0 \sigma_{\theta} \gg l_{so}$. Finally, we approximate the energy dispersion of the $R$ and $L$ hinge modes by $\varepsilon_0^{R,L} (\tilde{j}_z) \approx \mp \frac{\alpha}{l_0} (\tilde{j}_z - \tilde{j}_{z;R,L}^*) $, where the offsets $ \tilde{j}_{z;R,L}^* \approx -\frac{B \cdot \pi (l_0 \pm r_0)^2}{2 \pi}$ account the $B$-field flux quanta through the outer and inner equatorial rings of the torus, i.e. are generated by the orbital effects of the $B$-field.

{\it Plateau region.}---In the broad range  $\tilde{j}_{z,R}^* < \tilde{j}_z < \tilde{j}_{z,L}^*$, there is a nearly degenerate plateau of weakly dispersing Landau surface states [between yellow and green stars in Fig.~\ref{fig:torus_spectrum}(a)]. Looking first at the states in the plateau's middle (that is near the purple star in Fig.~\ref{fig:torus_spectrum}(a) at $\tilde{j}_z \approx - \frac{B \cdot \pi l_0^2}{2 \pi}$), we observe that they are the linear combinations of the so called top and bottom  Landau surface states that are localized near $\theta =0$ and $\theta = \pi$, respectively. Due to the finite torus size, the splitting between $\varepsilon_0$ and $\varepsilon_{-1}$ bands in the  middle of the plateau is estimated by an exponentially small value $\propto \alpha \sqrt{B} e^{-2 B r_0^2}$ \cite{SM}. The density of the state with higher energy, i.e. belonging to the band $\varepsilon_0$, is shown in the central inset of Fig.~\ref{fig:torus_spectrum}(b). This state is drastically different from the above discussed $R$ and $L$ hinge states, since its stabilization occurs due to the orbital $B$-field effect, while the Zeeman term produces only a perturbative contribution $-\varepsilon_Z$ to its energy. The low-energy analysis also predicts that its angular variance  $\sigma_{\theta} =\frac{l_B}{r_0}$ is both independent of $\varepsilon_Z$ and smaller than the variance of the $R$ and $L$ hinges (while the radial localization length $l_{so}$ remains the same).  

The other states belonging to $\varepsilon_0$ connect  the top-bottom Landau surface states in the plateau middle with the $R$ and $L$ hinges on the linear branches of the band $\varepsilon_0$. Their density localization evolves along the disc circumference in the angular direction with the changing $\tilde{j}_z$ value [see Fig.~\ref{fig:torus_spectrum}(b)]. These states retain the same radial localization length $l_{so}$, while their angular variances increase from the top-bottom Landau surface states to the $R$ and $L$ hinges, with the stabilization mechanism accordingly changing from the orbital to the Zeeman effect. 

\begin{figure}[t]
        \centering
        \includegraphics[width = 0.9 \columnwidth]{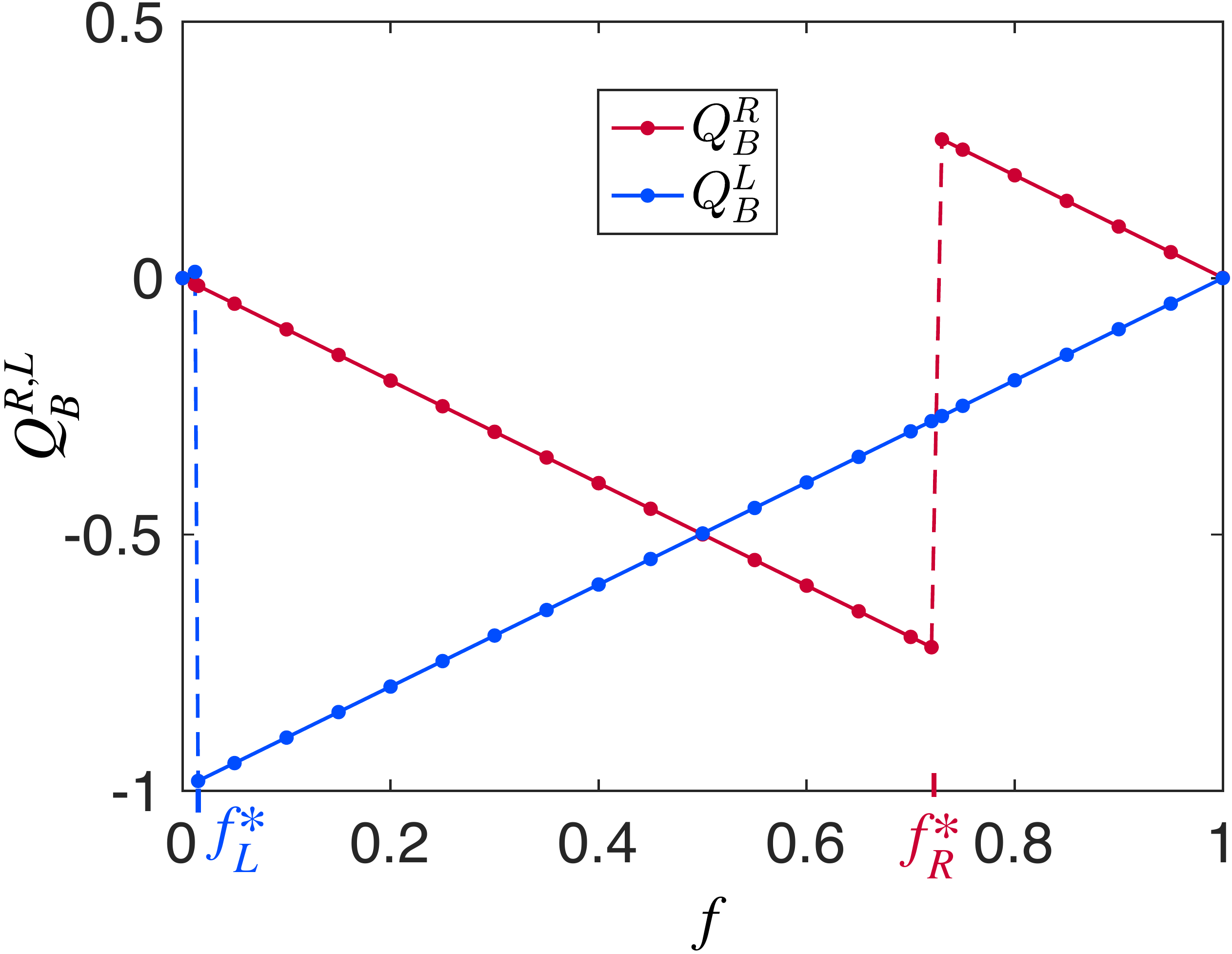}
        \caption{Boundary charges as functions of $f$ feature the unit slopes, which are complemented by the unit jumps at the flux values $f_R^* \approx 0.72$, and $f_L^* \approx 0.01$ (parameters as in Fig.~\ref{fig:torus_spectrum}). At these values an electron enters or leaves the system (see Fig.~\ref{fig:torus_charge} and the caption therein to check that $\tilde{j}_{z,R}^* -f_R^* =-99.5$ and $\tilde{j}_{z,L}^* - f_L^*=-17.5$ are half-integers). The width of the boundary region is set to $d=3 \, l_{so}$. The actual numerical summation runs over the $j_z$ range from -199.5 to 100.5, which is sufficiently wide to ensure convergence. We use a discrete spacing $\Delta f=0.05$ (the slope values and jump positions are calculated from a smaller spacing $\Delta f=0.01$). The slope for the right (left) boundary charge is -0.9993 (0.9952) which is obtained by linear fitting of the longer segment of each curve. These values agree with the corresponding jump sizes.}
        \label{fig:EMF}
    \end{figure}

{\it Boundary charge and 3D-SOTI QHE.}---A convenient way to introduce the Hall conductance is to define it via the charge localized within an appropriately chosen boundary region \cite{park_etal_prb_16,thakurathi_etal_prb_18,pletyukhov_etal_prb_20,pletyukhov_etal_prbr_20}. In our model, the outer ($R$) and the inner ($L$) boundary regions $\mathcal{B}_{R,L}$ [shown in brown in Fig.~\ref{fig:models}(b)] are bounded by the cylindrical shells of the radii $\rho_{R,L} = l_{0} \pm (r_0 - d)$, respectively. Choosing the width $d$ such that $r_0 \gg d \gg l_{so}, \frac{l_Z^2}{l_{so}}$, we ensure that the $R$ and $L$ hinge states are fully confined within $\mathcal{B}_{R,L}$, respectively. It is natural to define the boundary charges $Q_B^{R,L} (f)=\sum_{\lambda} Q_{B,\lambda}^{R,L}(f)$, with
\begin{align}
    Q_{B,\lambda}^{R,L}(f) = \sum_{j_z,\varepsilon_\lambda(\tilde{j}_z)<\mu} \bar{Q}_{B,\lambda}^{R,L} (\tilde{j}_z)  
     - \sum_{j_z,\varepsilon_\lambda(j_z)<\mu} \bar{Q}_{B,\lambda}^{R,L} (j_z),
    \label{QB_f}
\end{align}
where $\bar{Q}_{B,\lambda}^{R,L} (\tilde{j}_z)=\int_{\mathcal{B}_{R,L}} d x d z |\Psi_{\lambda} (x,z; \tilde{j}_z)|^2$ is a partial contribution of the state $(\lambda, j_z)$ evaluated at the AB flux $f \cdot \phi_0$. To make the definition of $Q_{B,\lambda}^{R,L} (f)$ for each band insensitive to a specific choice of $d$, we subtract a large background contribution at zero flux.

The central result of this Letter, supported by the numerical result shown in Fig.~\ref{fig:EMF}, is that the dependence of $Q_B^{R,L}(f)$ on the flux $f$ is piece-wise linear, 
\begin{align}
    Q_{B}^{R,L} (f) =\mp f \pm  \Theta (f-f_{R,L}^*) ,
    \label{lin_dep}
\end{align}
with jumps at $f=f_{R,L}^*$, where $\tilde{j}_{z;R,L}^*-f$ is half-integer. 
The slope value is correlated with the jump size such that the periodicity $ Q_{B}^{R,L} (f)= Q_{B}^{R,L} (f+1)$ is maintained.
Changing the flux adiabatically in time, the immediate consequence of \eqref{lin_dep} is the quantization of the Hall conductance, relating the Hall current to the electric field via $I_{\rho}^{R,L} = \sigma_{\rho \varphi}^{R,L} E_{\varphi}^{R,L}$, see Fig.~\ref{fig:models}(b) and the discussion in Refs.~\cite{thakurathi_etal_prb_18,pletyukhov_etal_prb_20,laubscher_etal_prb_21}. Using Faraday's law and the continuity equation we get $E_\varphi^{R,L}=-\dot{\phi}/(2\pi\rho_{R,L})$ and $I_\rho^{R,L}=\pm \dot{Q}_B^{R,L}/(2\pi\rho_{R,L})$. Together with $\dot{Q}_B^{R,L}\approx \frac{\partial Q_B^{R,L}}{\partial f} \dot{f}$ and $\dot{\phi}=2\pi\dot{f}$, we obtain the quantization of the Hall conductance $\sigma_{\rho \varphi}^{R,L} = \frac{1}{2\pi}\stackrel{!}{=}\frac{e^2}{h}$  away from the jump points $f^*_{R,L}$.

\begin{figure}[t]
         \includegraphics[width = 0.9 \columnwidth]{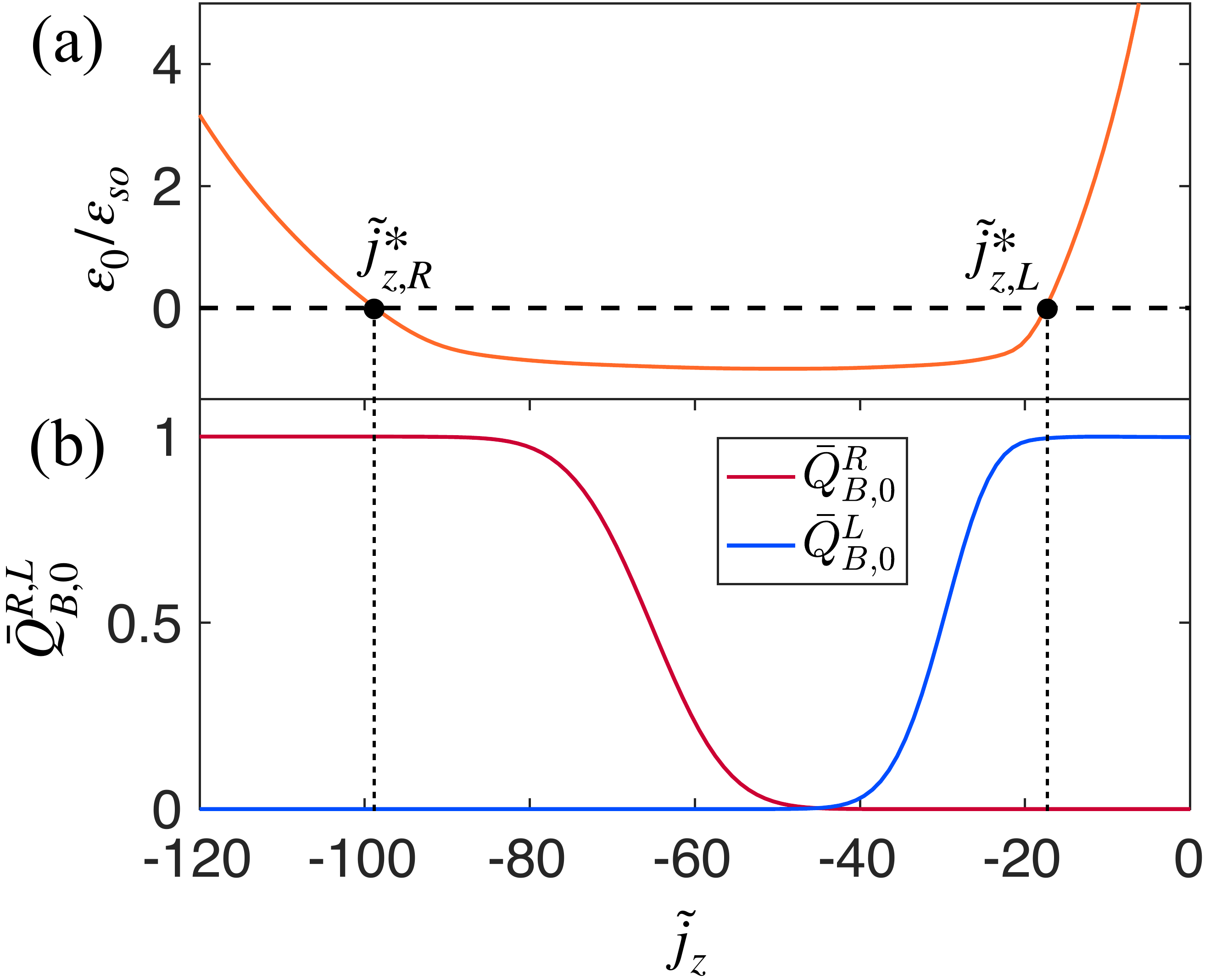}
        \caption{(a) The band $\varepsilon_0(\tilde{j}_z)$  and  (b) its partial charge contributions $\bar{Q}_{B,0}^{R,L}(\tilde{j}_z)$ to the right (red curve) and left (blue curve) boundary regions 
        (parameters as in Fig.~\ref{fig:torus_spectrum}). The roots of the equation $\varepsilon_0(\tilde{j}_z)=\mu=0$ are $\tilde{j}_{z,R}^* \approx -98.78$ and $\tilde{j}_{z,L}^* \approx -17.49$.
        }
        \label{fig:torus_charge}
\end{figure}

The second important result is that the contribution $Q_{B,\lambda}^{R,L}$ from all valence bands with $\lambda \leq -1$ is negligible (see SM \cite{SM} for the numerical verification). This is an intuitively expected result, since the valence band $\varepsilon_{-1}$ is separated from the band $\varepsilon_0$ by the topologically trivial, though exponentially small, gap. The only contribution to the boundary charge is thus provided by the mode $\varepsilon_0$, which is shown in Fig.~\ref{fig:torus_charge}.

Let us provide the analytical arguments for the results presented above. Concerning the discrete jumps of $Q_B^{R,L}(f)$ at $f=f_{R,L}^*$, it is obvious that they appear at those flux values, where a half-integer $j_z$ exists with $j_z+f_{R,L}^*=\tilde{j}_{z;R,L}^*$. These are the points where the $R$/$L$ hinge modes of the $\lambda=0$ band cross the chemical potential, leading to a jump of $Q_{B,\lambda=0}^{R,L}(f)$ by $\pm 1$. In contrast, for all fully occupied bands $\lambda<0$, there is no jump. 

The linear dependence on the flux follows essentially from the smooth behaviour of $\bar{Q}_{B,\lambda}^{R,L}(\tilde{j}_z)$ as function of $\tilde{j}_z$, i.e., it varies typically on the scale $\Delta \tilde{j}_z \sim \tilde{j}_{z,L}^*-\tilde{j}_{z,R}^*$ of the plateau size, see Fig.~\ref{fig:torus_charge}(b) for $\lambda=0$ (the same applies for all $\lambda \ne 0$, except that the asymptotic values are coinciding). The plateau size is roughly related to the number $N_B\sim r_0 l_0/l_B^2$ of flux quanta threaded through the torus surface from the orbital field (for $B=0$ another scale $\sim l_0$ is obtained, see SM \cite{SM}). Therefore, the $n$-th order derivative of $\bar{Q}_{B,\lambda}^{R,L}(\tilde{j}_z)$ will scale with $1/N_B^n$ in the plateau region (outside this region the derivatives are very small), leading with an additional factor $N_B$ from the sum over $j_z$ to the scaling $(\frac{d}{df})^n Q_{B,\lambda}^{R,L}(f)\sim 1/N_B^{n-1}$. Thus, all higher-order derivatives are negligible for $n\ge 2$, provided that $N_B\gg 1$ is fulfilled. This can be achieved to any desired accuracy by increasing the torus radius $l_0$ (this result holds for any $B$, see SM \cite{SM}). Therefore, only the linear term in $f$ contributes significantly to $Q_{B,\lambda}^{R,L}(f)$ for each band. Since the slope value of the linear term of $Q_{B,\lambda}^{R,L}(f)$ is correlated with the discrete jumps due to the periodicity property $Q_{B,\lambda}^{R,L}(f)=Q_{B,\lambda}^{R,L}(f+1)$, we find that the slope is zero for all bands $\lambda<0$ and $\mp 1$ for $\lambda=0$. This shows that the contribution of all valence bands $\lambda<0$ is  negligible and the universal linear slope of the total boundary charge is fully determined by the 
$\lambda=0$ band.

\begin{figure}[t]
        \centering
        \includegraphics[width =  \columnwidth]{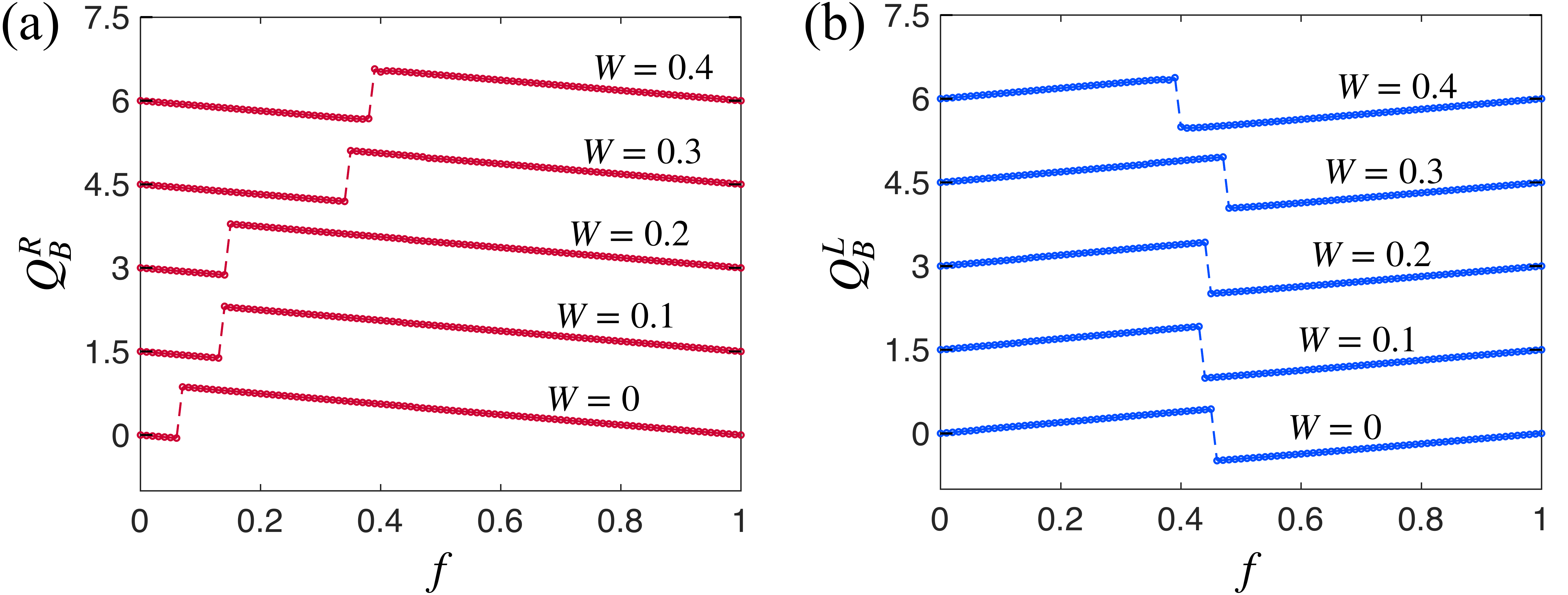}
        \caption{Boundary charges calculated for the 3D tb model in the presence of disorder potentials uniformly distributed on the interval $[-W/2,W/2]$ (we consider one single random disorder configuration; other parameters being as in Fig.~\ref{fig:torus_spectrum}) as function of the flux $f$ (with discrete spacing given by $\Delta f=0.01$). The different curves with increasing disorder strength $W$ (in units of $\varepsilon_{so}$) are shifted upwards with a step value 1.5. The slope value for the right (left) boundary charge is calculated to be -0.915 (0.914). Different disorder strengths $W$ give the same slope value up to the quoted accuracy. Small deviations from the universal  values $\mp 1$ are due to the choice of the larger lattice constant $a_{3D}=l_{so}$ in the 3D calculations (see SM \cite{SM} for further computational details). }
        \label{fig:Disorder}
    \end{figure}

{\it Disorder effects.}---We add random disorder potentials to every site of the 3D tb lattice realization of the Hamiltonian \eqref{eq:H_main}. Disorder breaks the rotational symmetry, the quantum number $j_z$ is no longer conserved, and the band index $\lambda$ is not well defined either. Remarkably, the boundary charge $Q_B^{R,L}$, in which all the states below $\mu$ are included, is still well defined, as well as the relation between the boundary charge and the Hall conductance. In Fig.~\ref{fig:Disorder} we show that the linear slopes of $Q_B^{R,L} (f)$ remain very close to the universal value $\mp 1$, i.e. staying essentially unaffected by disorder. This demonstrates 
the fundamental significance of the boundary charge concept: It is capable of explaining the Hall conductance quantization even in the disordered case, when the simple clean-case explanation presented above is no longer applicable.

{\it Discussion}.--- The proposed 3D-SOTI QHE is robust to disorder which is very promising for its experimental realization using standard 3D-TIs with Zeeman fields induced
by external magnetic fields, magnetic doping, or nearby ferromagnets. It occurs specifically only in 3D systems (for system sizes exceeding the localization length of hinge states) and is insensitive to orbital effects (see SM \cite{SM}), providing various fingerprints for its unique observation. Moreover, it is robust to shape distortions and will even occur in a spherical model with an infinitesimally thin hollow tube along the vertical axis 
(to allow for the Aharonov-Bohm flux insertion).
The obtained results for the boundary charge showcase the same qualitative features as in the torus model (see SM \cite{SM}) and its detailed study opens another avenue of future research.

{\it Acknowledgments}.---This work was supported by the Deutsche Forschungsgemeinschaft via RTG 1995, the Swiss National Science Foundation (SNSF) and by the Deutsche Forschungsgemeinschaft (DFG, German Research Foundation) under Germany's Excellence Strategy - Cluster of Excellence Matter and Light for Quantum Computing (ML4Q) EXC 2004/1 - 390534769. We acknowledge support from the Max Planck-New York City Center for Non-Equilibrium Quantum Phenomena. Simulations were performed with computing resources granted by RWTH Aachen University under projects rwth0752 and rwth0841, and at sciCORE (http://scicore.unibas.ch/) scientific computing center at University of Basel. Funding was received from the European Union's Horizon 2020 research and innovation program (ERC Starting Grant, grant agreement No 757725).

\begin{appendix}
\begin{widetext} 
\section{Effective low-energy theory}
\label{app:eff_ham}

\subsection{Derivation of the effective two-dimensional model by exploiting the rotational symmetry}

In the cylindrical coordinates $(\rho, \varphi, z)$ the Hamiltonian \eqref{eq:H_main} acquires the form
\begin{align}
H &=  \frac{\sigma_z}{2 m^*} \left[ - \frac{1}{\rho} \frac{\partial }{\partial \rho} \rho \frac{\partial}{\partial \rho}  + \frac{1}{\rho^2} \left( -i \frac{\partial}{\partial \varphi} +f + \frac{B \rho^2}{2}\right)^2 - \frac{\partial^2}{\partial z^2} - 2 m^* \delta \right]
\\
&+\alpha \sigma_x \left[ - i s_{\rho}  \frac{\partial}{\partial \rho} + \frac{s_{\varphi}}{\rho}  \left( -i  \frac{\partial}{\partial \varphi} + f + \frac{B \rho^2}{2}  \right) - i s_z \frac{\partial}{\partial z} \right] + \varepsilon_Z s_z  ,
\label{ham3D_cyl}
\end{align}
where
\begin{align}
s_{\rho} &= s_x \cos \varphi + s_y \sin \varphi , \\
s_{\varphi} &= -s_x \sin \varphi + s_y \cos \varphi .
\end{align}

Performing the unitary transformation $U_{\varphi} =e^{-\frac{i}{2} \varphi} e^{\frac{i}{2} s_z \varphi}$, which transforms
\begin{align}
U_{\varphi}  s_{\rho} U_{\varphi}^{\dagger} &= s_x, \\
U_{\varphi} s_{\varphi} U_{\varphi}^{\dagger} &= s_y , \\
U_{\varphi} J_z U_{\varphi }^{\dagger} &=U_{\varphi} \left( - i \frac{\partial}{\partial \varphi} + \frac12 s_z \right) U_{\varphi }^{\dagger} = - i \frac{\partial}{\partial \varphi} + \frac12 \equiv \hat{J}_z,
\end{align}
supporting the periodic boundary conditions in $\varphi$, we obtain
\begin{align}
    h_{3D} &= U_{\varphi} H U_{\varphi}^{\dagger} \\
        &= \frac{\sigma_z}{2 m^*} \left[ - \frac{1}{\rho} \frac{\partial }{\partial \rho} \rho \frac{\partial}{\partial \rho}  + \frac{1}{\rho^2} \left( \hat{J}_z -\frac12 s_z +f + \frac{B \rho^2}{2}\right)^2 - \frac{\partial^2}{\partial z^2} - 2 m^* \delta \right] \\
&+\alpha \sigma_x \left[ - i s_{z}  \frac{\partial}{\partial \rho} + \frac{s_{y}}{\rho}  \left( \hat{J}_z - \frac12 s_z + f + \frac{B \rho^2}{2} \right) - i s_z \frac{\partial}{\partial z} \right] + \varepsilon_Z s_z .
\end{align}
By virtue of the $U_{\varphi}$ transformation we remove the $\varphi$-dependence from the Hamiltonian and thereby explicitly manifest the rotational symmetry about $z$ axis, $[h_{3D}, \hat{J}_z]=U_{\varphi} [H, J_z] U_{\varphi}^{\dagger}=0$. 

The total angular momentum operator $\hat{J}_z$ can be replaced by its half-integer eigenvalue $j_z$. Combining the latter with $f$, we obtain the parameter $\tilde{j}_z = j_z +f$. In addition, we perform the transformation 
\begin{align}
    h_{2D} &= \sqrt{\rho} \, h_{3D} \,\frac{1}{\sqrt{\rho}}
\end{align}
in order to have the euclidean scalar product in the effective two-dimensional model formulated in the half-plane $(x>0,z)$. Finally, renaming $\rho \to x$, we obtain the Hamiltonian \eqref{ham2D}.

\subsection{Coordinate transformation}

The Hamiltonian \eqref{eq:H_main} for the torus model can be alternatively represented in the new  -- toroidal-poloidal -- coordinate system
\begin{align}
    x &= l_0 + r \sin \theta , \\
    z &= r \cos \theta ,
\end{align}
where  $r< r_0 < l_0$ and $\theta \in (-\pi, \pi]$.
Transforming the  differential operators, we obtain
\begin{align}
    \frac{\partial}{\partial x} &= \sin \theta \frac{\partial}{\partial r} +  \frac{\cos \theta}{r} \frac{\partial}{\partial \theta} =\frac12 (\Lambda_+ - \Lambda_-), \\
    \frac{\partial}{\partial z} &= \cos \theta \frac{\partial}{\partial r} -  \frac{\sin \theta}{r} \frac{\partial}{\partial \theta} =\frac{i}{2} (\Lambda_+ + \Lambda_-),
\end{align}
where
\begin{align}
    \Lambda_{\pm} = -i e^{\pm i \theta} \frac{\partial}{\partial r} \pm \frac{e^{\pm i \theta}}{r} \frac{\partial}{\partial \theta}.
\end{align}
In addition, the two-dimensional Laplace operator reads
\begin{align}
    \nabla_{2D}^2 = \frac{\partial^2}{\partial x^2} + \frac{\partial^2}{\partial z^2} = \frac{\partial^2}{\partial r^2} + \frac{1}{r} \frac{\partial}{\partial r} + \frac{1}{r^2} \frac{\partial^2}{\partial \theta^2}.
\end{align}

Then, we rewrite \eqref{ham2D} as
\begin{align}
   h_{2D} &=  \frac{\sigma_z}{2 m^*} \left[  -\nabla_{2D}^2+ \left( \frac{\tilde{j}_z}{x} + \frac{B x}{2}\right)^2 -  s_z \left( \frac{\tilde{j}_z }{x^2} +\frac{B}{2}\right)   - 2 m^* \delta \right] \nonumber \\
&+ \alpha \sigma_x \left[   \frac{s_z - i s_x}{2} \Lambda_+ + \frac{s_z + i s_x}{2} \Lambda_- + s_y  \left( \frac{\tilde{j}_z}{x} + \frac{B x}{2}  \right)  \right] + \varepsilon_Z s_z .
    \label{ham2D_curve}
\end{align}

\subsection{Matrix elements of the 2D model in the basis of Bessel functions}
\label{subsec:Bessel}

To compute matrix elements in the torus model we introduce the eigenbasis of $-\nabla^2_{2D}$ with the open boundary conditions on the disc $r< r_0$
\begin{align}
   u_{n,j_n} (r, \theta) = \langle r, \theta | n,j_n \rangle = \frac{1}{\sqrt{N_{n,j_{n}}}} J_{n} \left( \xi_{n,j_{n}} \frac{r}{r_0}\right) \frac{e^{i n \theta} }{\sqrt{2 \pi}}, \quad n \in \mathbbm{Z}, \quad j_n \in \mathbbm{N}.
   \label{disc_laplace_eigenfunc}
\end{align}
It is expressed in terms of the Bessel function $J_n$ and its zeroes $\xi_{n,j_n}$ enumerated by $j_n$ for each $n$. Being normalized by
\begin{align}
    N_{n,j_{n}} = \frac{r_0^2}{2} [J'_n (\xi_{n,j_{n}})]^2 ,
\end{align}
the eigenfunctions \eqref{disc_laplace_eigenfunc} obey the orthogonality condition
\begin{align}
    \langle n', {j'_{n'}} |   n,j_n \rangle  &= \int_0^{2\pi}  d \theta \int_0^{r_0} dr \, r  \, u^{*}_{n', {j'_{n'}} } (r, \theta )  \, u_{n,j_n} (r, \theta) = \delta_{n'n} \delta_{j'_n , j_n}.
\end{align}

Focusing on the representation \eqref{ham2D_curve}, we analytically evaluate the matrix elements
\begin{align}
     \langle n', {j'_{n'}} | - \nabla_{2D}^2 |  n,j_n \rangle  &= \delta_{n'n} \delta_{j'_n , j_n} \left(\frac{\xi_{n, j_n}}{r_0}\right)^2 , \\
   \langle n', {j'_{n'}} |\Lambda_{\pm}|   n,j_n \rangle  &=   -\frac{\delta_{n', n \pm 1}}{r_0} 
   (-1)^{j_n + j'_{n \pm 1}} \, \text{sgn} \, (n \pm 0^+) \frac{2 i \, \xi_{n, j_n} \, \xi_{n \pm 1, j'_{n \pm 1}}}{\xi_{n, j_n}^2 - \xi_{n\pm 1, j'_{n \pm 1}}^2} ,
\end{align}
and numerically evaluate the matrix elements $\langle n', {j'_{n'}} | x |  n,j_n \rangle$, $\langle n', {j'_{n'}} | x^2 |  n,j_n \rangle$, $\langle n', {j'_{n'}} | \frac{1}{x} |  n,j_n \rangle$, $\langle n', {j'_{n'}} | \frac{1}{x^2} |  n,j_n \rangle$.

Choosing an appropriate high-energy cutoff, we diagonalize the matrix Hamiltonian and compare its spectrum in Fig.~\ref{fig:SM_Comparison} with the tb model spectrum presented in the main text.

\begin{figure}[t]
        \centering
        \includegraphics[width = 0.45 \columnwidth]{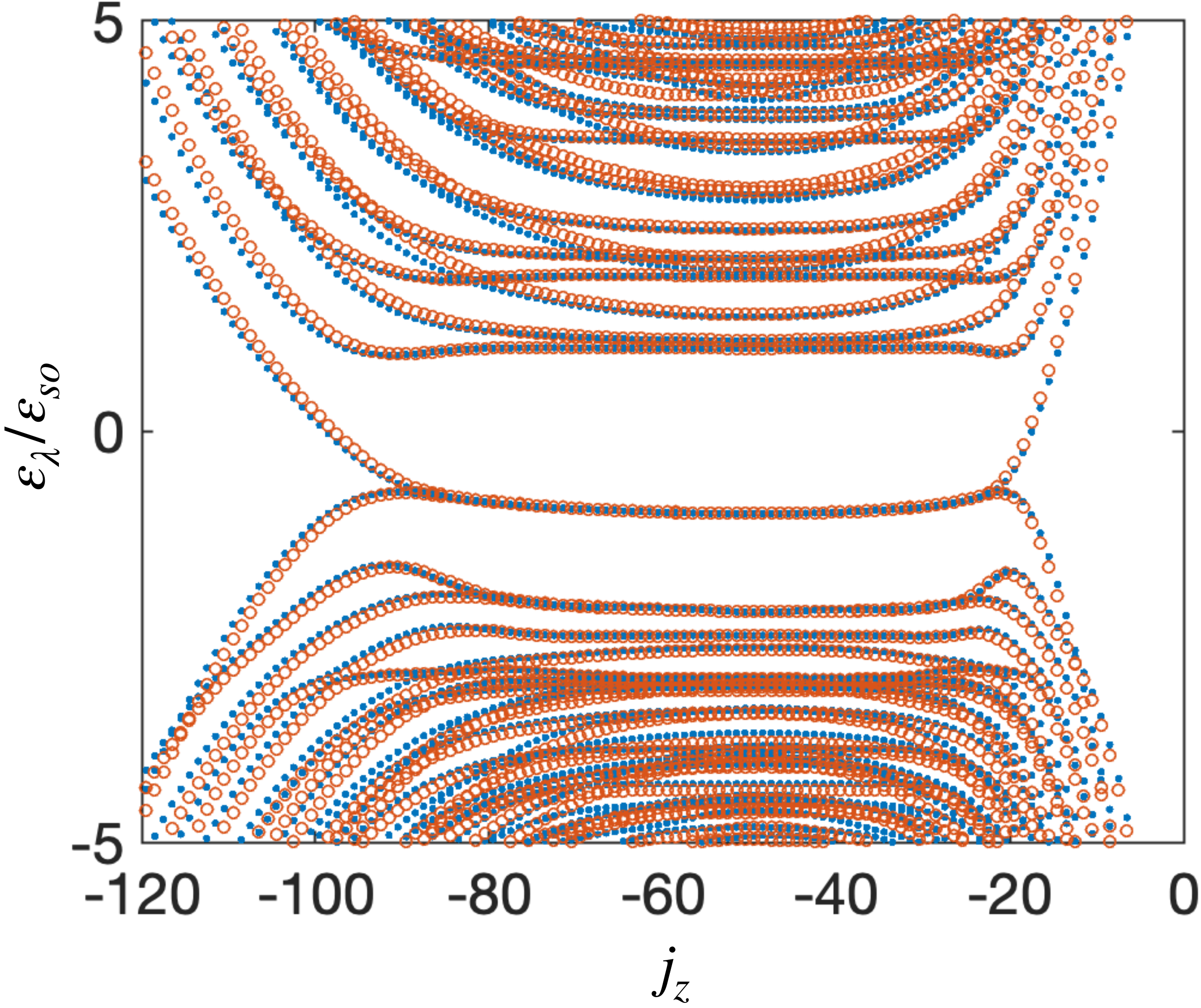}
        \caption{Comparison of the low-energy states calculated within the tb model (blue dots) and the continuum model (orange circles) at $f=0$ (same parameters as in Figs.~\ref{fig:torus_spectrum}, \ref{fig:EMF} and \ref{fig:torus_charge}). Small deviations can be consistently eliminated by reducing the lattice constant value in the tb model and by increasing the high-energy cut-off in the continuum model.}
        \label{fig:SM_Comparison}
    \end{figure}

\subsection{Effective surface model}

To derive an effective surface model, we apply the following unitary transformation to the Hamiltonian \eqref{ham2D_curve} 
\begin{align}
U_{\theta} = e^{-\frac{i}{2}  \theta} e^{\frac{i}{2} s_y \theta},
\label{U_theta}
\end{align}
which maintains the periodic boundary conditions in $\theta$ variable. 

The transformation \eqref{U_theta} acts as follows:
\begin{align}
    & U_{\theta} s_x U_{\theta}^{\dagger} = s_x \cos \theta + s_z \sin \theta, \\
    & U_{\theta} s_z U_{\theta}^{\dagger} = -s_x \sin \theta + s_z \cos \theta, \\
    & U_{\theta} (s_x \sin \theta + s_z \cos \theta) U_{\theta}^{\dagger} 
    = s_z, \\
    & U_{\theta} (s_x \cos \theta - s_z \sin \theta) U_{\theta}^{\dagger} = s_x, 
\end{align}
and
\begin{align}
    U_{\theta} \frac{\partial}{\partial \theta} U_{\theta}^{\dagger}  = \frac{\partial}{\partial \theta} + i \frac{1-s_y}{2} .
\end{align}
The listed transformed operators are to replace in \eqref{ham2D_curve} their initial counterparts.

In addition to \eqref{U_theta}, we transform wavefunctions $\Psi \to \sqrt{r} \Psi$ in order to absorb the functional determinant $r$ into the wavefunction's definition. Accordingly we transform all operators, $O \to  \sqrt{r} O \frac{1}{\sqrt{r}}$. Thus we get
\begin{align}
&\tilde{h}_{2D} = \sqrt{r} U_{\theta} h_{2 D} U_{\theta}^{\dagger} \frac{1}{\sqrt{r}} \nonumber \\
&= \frac{\sigma_z}{2 m^*} \left[ - \frac{\partial^2}{\partial r^2}+ \frac{1}{r^2} \left( -i \frac{\partial}{\partial \theta} + \frac{1}{2}\right)^2 - \frac{s_y}{r^2} \left( -i \frac{\partial}{\partial \theta} + \frac{1}{2}\right)  + \left( \frac{\tilde{j}_z}{x} + \frac{B x}{2}\right)^2 -  (-s_x \sin \theta + s_z \cos \theta) \left( \frac{\tilde{j}_z }{x^2} +\frac{B}{2}\right) - 2 m^* \delta \right] \nonumber \\
&+ \alpha \sigma_x \left[ - i s_z \frac{\partial}{\partial r}  +    \frac{s_x }{r} \left(-i \frac{\partial}{\partial \theta} +\frac{1}{2} \right) + s_y  \left( \frac{\tilde{j}_z}{x} + \frac{B x}{2}  \right)  \right] + \varepsilon_Z  (-s_x \sin \theta + s_z \cos \theta)  .
\label{ham2D_theta}
\end{align}

For a sufficiently large radius $r_0 \gg l_{so}$, we can neglect the curvature effects in deriving the low-energy theory for surface states. Considering first the radial part of \eqref{ham2D_theta}
\begin{align}
\tilde{h}^{(0)}_{2D} = \frac{\sigma_z}{2 m^*} \left[ - \frac{\partial^2}{\partial r^2} - 2 m^* \delta \right] - i \alpha \sigma_x s_z \frac{\partial}{\partial r},
\label{ham_radial}
\end{align}
we establish that surface states are  zero-energy states of this Hamiltonian, and their radial dependence is captured by the ansatz
\begin{align}
    \psi_{surface} (r) \propto e^{- (r_0-r)/l_{so}} \sin [ k_F (r_0-r)],
    \label{ansatz_radial}
\end{align}
with $k_F = \sqrt{2 m^* \delta - (m^* \alpha)^2}$. Inserting it into \eqref{ham_radial}, we obtain an equation for the spinor structure $|\chi_{surface} \rangle$ of the surface states
\begin{align}
   \frac{1 - \sigma_y s_z}{2} |\chi_{surface} \rangle =0 .
\end{align}
Hence we find the projector onto the surface states subspace
\begin{align}
    P_{surface} = \frac{1+\sigma_y s_z}{2}.
\end{align}

Projecting \eqref{ham2D_theta} onto the surface states subspace, averaging the radial degree of freedom in the state \eqref{ansatz_radial}, and additionally approximating $\langle g (r) \rangle  \approx g (r_0)$, we obtain the effective low-energy theory for the surface states
\begin{align}
& h_{surface} = P_{surface} \left[    \sigma_z s_y   \left( \frac{\alpha }{r_0} - \frac{1}{2 m^* r_0^2} \right) \left(-i \frac{\partial}{\partial \theta} +\frac{1}{2} \right) -  \sigma_z s_x \left( \frac{\alpha }{x} - \frac{\sin \theta}{2 m^* x^2} \right) \left( \tilde{j}_z + \frac{B x^2}{2}  \right)   + \varepsilon_Z s_z \cos \theta  \right]  P_{surface} ,
\label{ham_surface1}
\end{align}
with $x = l_0 + r_0 \sin \theta$.

\subsection{Hinge and Landau surface states in the torus model}

1) To describe the right and left hinge states, we consider $l_0  \gg r_0 \gg l_{so}$ in the effective surface Hamiltonian \eqref{ham_surface1} and derive
\begin{align}
& h_{surface} \approx P_{surface} \left[    \sigma_z s_y \frac{\alpha }{r_0}  \left(-i \frac{\partial}{\partial \theta} +\frac{1}{2} \right) -  \sigma_z s_x \frac{\alpha }{l_0}  \left( \tilde{j}_z + \frac{B x^2}{2}  \right)   + \varepsilon_Z  s_z \cos \theta  \right]  P_{surface} .
\label{ham_surface2}
\end{align}

Near $\theta = \pm \frac{\pi}{2}$ we approximate
\begin{align}
h_{surface} \approx P_{surface} \left[  -i   \sigma_z s_y  \frac{\alpha }{r_0}   \frac{\partial}{\partial \theta} + \sigma_z s_y \frac{\alpha}{2 r_0} -  \sigma_z s_x \frac{\alpha }{l_0} \left( \tilde{j}_z + \frac{B (l_0 \pm r_0)^2}{2}  \right)   + 
\varepsilon_Z  s_z \cos \theta  \right]  P_{surface} .
\label{torus_RL}
\end{align}
At $\tilde{j}_z \approx -  \frac{B (l_0 \pm r_0)^2}{2}$ we solve the eigenvalue problem for a hinge state with nearly zero energy
\begin{align}
    \left[  -i   \sigma_z s_y  \frac{\alpha }{r_0}   \frac{\partial}{\partial \theta} + \varepsilon_Z  s_z \cos \theta  \right] e^{\pm \varepsilon_Z\frac{r_0}{\alpha} \sin \theta} | \tilde{\chi}_{R,L} \rangle = 2\varepsilon_Z e^{\pm \varepsilon_Z \frac{r_0}{\alpha} \sin \theta}   s_z \cos \theta  \frac{1 \mp  \sigma_z  s_x}{2}  | \tilde{\chi}_{R,L} \rangle = 0 .
\end{align}

 Thus  $| \tilde{\chi}_{R,L} \rangle$ is a simultaneous eigenstate (to the eigenvalue $+1$) of the projectors $P_{surface} = \frac{1+\sigma_y s_z}{2}$ and $\frac{1 \pm \sigma_z  s_x}{2} $. It reads
 \begin{align}
     | \tilde{\chi}_{R,L} \rangle = \frac12 \left[ \left( \begin{array}{c} 1 \\ i \end{array}  \right)_{\sigma_z} \left( \begin{array}{c} 1 \\ 0 \end{array}  \right)_{s_z} \pm \left( \begin{array}{c} 1 \\ -i \end{array}  \right)_{\sigma_z} \left( \begin{array}{c} 0 \\ 1 \end{array}  \right)_{s_z}  \right] .
 \end{align}
Noticing that $ \langle \tilde{\chi}_{R,L} | \sigma_z s_y | \tilde{\chi}_{R,L} \rangle =0 $, we perturbatively evaluate the energy dispersion in $\tilde{j}_z$ 
 \begin{align}
     \varepsilon_{R,L} (\tilde{j}_z) \approx - \langle  \tilde{\chi}_{R,L} |  \sigma_z s_x  | \tilde{\chi}_{R,L} \rangle \frac{\alpha }{l_0} \left( \tilde{j}_z + \frac{B (l_0 \pm r_0 )^2}{2}  \right)   = \mp \frac{\alpha }{l_0} \left( \tilde{j}_z + \frac{B (l_0 \pm r_0)^2}{2}  \right).
     \label{eq:hinge_disp_LET}
 \end{align}
 
Undoing the $\theta$-transformation, we obtain
\begin{align}
    | \chi_{R,L} \rangle = U_{\theta}^{\dagger} | \tilde{\chi}_{R,L} \rangle = \frac{e^{i \frac{\theta}{2}}}{2} \left[ \left( \begin{array}{c} 1 \\ i \end{array}  \right)_{\sigma_z} \left( \begin{array}{c} \cos \frac{\theta}{2} \\ \sin \frac{\theta}{2} \end{array}  \right)_{s_z} \pm \left( \begin{array}{c} 1 \\ -i \end{array}  \right)_{\sigma_z} \left( \begin{array}{c} -\sin \frac{\theta}{2} \\ \cos \frac{\theta}{2} \end{array}  \right)_{s_z}  \right].
\end{align}
Note that $\sigma_x s_y  | \chi_{R,L} \rangle  = \mp | \chi_{R,L} \rangle$.

2) To describe the top and bottom Landau surface states we consider the values $\tilde{j}_z \approx -  \frac{B l_0^2}{2}$, which correspond to the spatial localization at $\sin \theta \approx 0$, that is $\theta \approx 0$ or $\theta \approx \pi$.

Carefully approximating 
\begin{align}
h_{surface} &\approx 
P_{surface}  \left[  \sigma_x s_x   \frac{\alpha}{r_0} \left( -i  \frac{\partial}{\partial \theta} + \frac12 \right)  + \sigma_x s_y \frac{\alpha}{l_0}  \left( \tilde{j}_z +\frac{B l_0^2}{2}  \right) +  \sigma_x s_y \alpha B  r_0 \sin \theta  + \varepsilon_Z  s_z \cos \theta \right],
\end{align}
we extract the leading order Hamiltonian
\begin{align}
h_{TB}^{(0)} &\approx  - i \sigma_x s_x   P_{surface}  \frac{\alpha}{r_0} \left[      \frac{\partial}{\partial \theta}  -s_z  B  r_0^2 \sin \theta \right].
\end{align}
It possesses the two zero-energy solutions: the top and bottom states $\propto e^ {\pm B  r_0^2 \cos \theta}$ with $s_z=\mp 1$, which are localized near $\theta \approx 0$ and $\theta \approx \pi$, respectively.
Their spinor structure is determined by the relations $\sigma_y s_z = +1$ and $s_z = \mp 1$:
\begin{align}
| \tilde{\chi}_T \rangle = \frac{1}{\sqrt{2}}   \left( \begin{array}{c} 1 \\ -i  \end{array}\right)_{\sigma_z} \left(  \begin{array}{c} 0 \\ 1  \end{array} \right)_{s_z}  , \quad | \tilde{\chi}_B \rangle = \frac{1}{\sqrt{2}}   \left( \begin{array}{c} 1 \\ i \end{array}\right)_{\sigma_z} \left(  \begin{array}{c} 1 \\ 0  \end{array} \right)_{s_z} .
\end{align}
Undoing the $\theta$-transformation, we obtain
\begin{align}
| \chi_T \rangle &= U_{\theta}^{\dagger}| \tilde{\chi}_T \rangle   = \frac{e^{\frac{i}{2}  \theta} }{\sqrt{2}} \left( \begin{array}{c} 1 \\ -i  \end{array}\right)_{\sigma_z}   \left(  \begin{array}{c} -\sin \frac{\theta}{2} \\ \cos \frac{\theta}{2}  \end{array} \right)_{s_z} , \\
| \chi_B \rangle &= U_{\theta}^{\dagger}| \tilde{\chi}_B \rangle  = \frac{ e^{\frac{i}{2} \theta} }{\sqrt{2}}
\left( \begin{array}{c} 1 \\ i \end{array}\right)_{\sigma_z}  \left(  \begin{array}{c} \cos \frac{\theta}{2} \\ \sin \frac{\theta}{2}  \end{array} \right)_{s_z}.
\end{align}

Energy of these states is obtained from the perturbative $\tilde{j}_z$-independent  correction
\begin{align}
\varepsilon_{T,B}  \approx -  \varepsilon_Z  .
\end{align}
The degeneracy is lifted by the exponentially small overlap of the two states, which leads to the symmetric and antisymmetric combinations of the top and bottom states. In particular, projecting the initially neglected small terms
\begin{align}
    \Delta h_{surface} &\approx 
P_{surface}  \left[  \sigma_x s_x   \frac{\alpha}{2r_0}   + \sigma_x s_y \frac{\alpha}{l_0}  \left( \tilde{j}_z +\frac{B l_0^2}{2}  \right)  + \varepsilon_Z  s_z \cos \theta \right]
\end{align}
onto the subspace spanned by the top and bottom states found above in the leading approximation, we obtain the following effective $2 \times 2$ Hamiltonian
\begin{align}
    \Delta h_{surface}^{eff} = - \varepsilon_Z  + \frac{\alpha}{2 r_0 \mathcal{N}} \left( \begin{array}{cc} 0 & i +  \frac{2 r_0}{l_0} (\tilde{j}_z +\frac{B l_0^2}{2} )  \\ -i + \frac{2 r_0}{l_0} (\tilde{j}_z +\frac{B l_0^2}{2} ) & 0 \end{array}\right) ,
    \label{h_TB_eff}
\end{align}
where
\begin{align}
    \mathcal{N} = \frac{1}{2 \pi} \int_{-\pi}^{\pi} d \theta e^{\pm 2 B r_0^2 \cos \theta} \approx \frac{e^{2B r_0^2}}{2 \sqrt{\pi B r_0^2}}, \quad B r_0^2 \gg 1.
\end{align}
Diagonalizing \eqref{h_TB_eff}, we obtain the energy splitting
\begin{align}
    \Delta \varepsilon (\tilde{j}_z) \approx \frac{\alpha}{r_0 \mathcal{N}} \sqrt{1+ \frac{4 r_0^2}{l_0^2} \left( \tilde{j}_z +\frac{B l_0^2}{2} \right)^2 }.
\end{align}
It is minimal in the plateau's middle,
\begin{align}
     \Delta \varepsilon (\tilde{j}_z \approx - \frac{B l_0^2}{2}) \approx \frac{\alpha}{r_0 \mathcal{N}} \approx 2 \alpha \sqrt{\pi B} e^{-B r_0^2} ,
\end{align}
giving an estimate of the splitting between $\varepsilon_{-1}$ and $\varepsilon_0$ bands.

\begin{figure}[t]
        \centering
        \includegraphics[width = 0.99 \columnwidth]{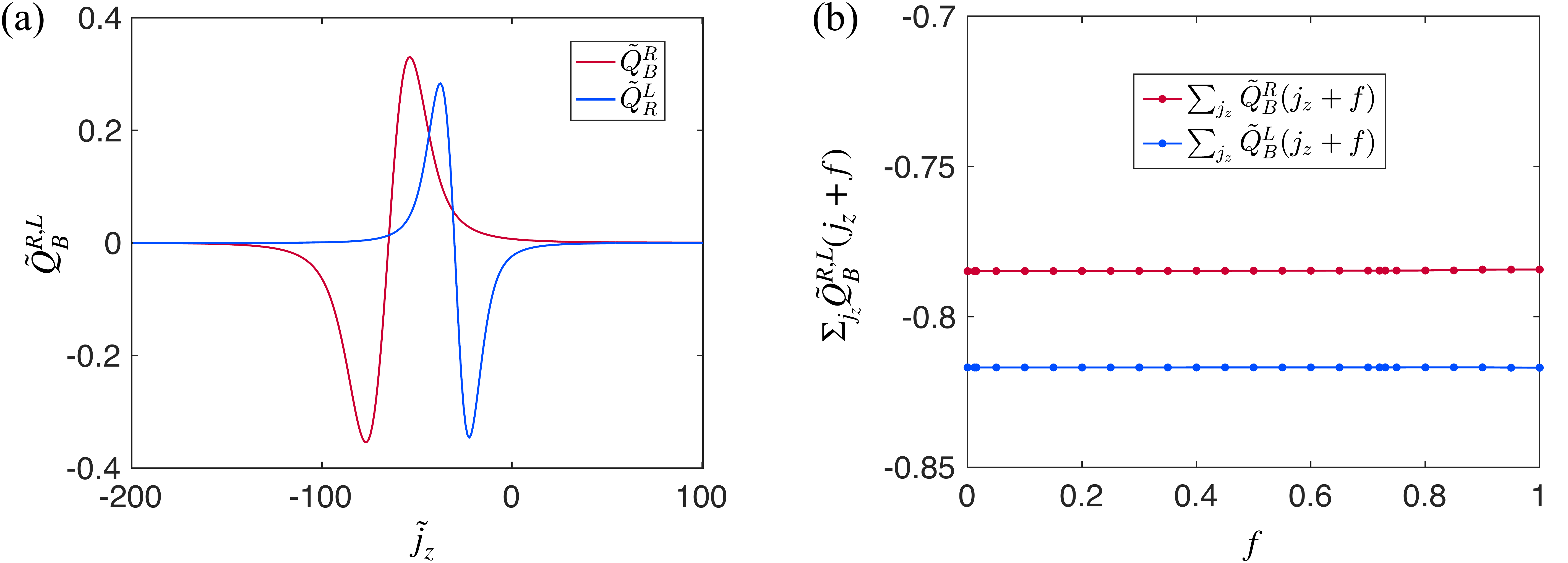}
        \caption{(a) $\tilde{j}_z$-resolved boundary charges as defined in Eq. \eqref{QRL_sum_lambda}. (b)  The summed values $\sum_{j_z} \tilde{Q}_{B}^{R,L} (j_z+f) $ (same parameters as in Figs.~\ref{fig:torus_spectrum}, \ref{fig:EMF} and \ref{fig:torus_charge}). In the numerical summation $j_z$ ranges from -199.5 to 100.5.}
        \label{fig:SM_EMF}
    \end{figure}

\section{Vanishing contribution to the boundary charge from the valence bands $\varepsilon_{\lambda \leq -1}$}

Studying numerically the quantity 
\begin{align}
    \tilde{Q}_B^{R,L} (\tilde{j}_z)= \sum_{\lambda \leq -1}  [ \bar{Q}^{R,L}_{B,\lambda} (\tilde{j}_z) - \bar{Q}^{R,L}_{B,\lambda} (|\tilde{j}_z | \to \infty) ],
    \label{QRL_sum_lambda}
\end{align}
we obtain the results shown in  Fig.~\ref{fig:SM_EMF}(a). Evaluating the sum $\sum_{j_z} \tilde{Q}_B^{R,L} (j_z + f)$ we obtain a small number of $O (1)$ which is constant for all values of $f$ (see  Fig.~\ref{fig:SM_EMF}(b)). Then
\begin{align}
   \sum_{\lambda \leq -1} \sum_{j_z}  [ \bar{Q}^{R,L}_{B,\lambda} (j_z +f) - \bar{Q}^{R,L}_{B,\lambda} (j_z ) ] = \sum_{j_z} [ \tilde{Q}_B^{R,L} (j_z + f) - \tilde{Q}_B^{R,L} (j_z)] =0,
\end{align}
which proves the vanishing contribution of the bands $\varepsilon_{\lambda \leq -1}$ to the boundary charge.

\section{Details of the tight-binding modelling and calculation of the boundary charge}

\subsection{2D tight-binding model as a discrete version of the continuum model}

To numerically solve the effective 2D continuum model shown in Eq. \eqref{ham2D} of the main text, we first replace the continuous coordinates by discrete sites, and substitute the derivatives with finite differences \cite{datta1997electronic}. The site labelled by $i$ has the position ${\bf r}_{i} = ma \hat{{\bf e}}_x + n a \hat{{\bf e}}_z$, where $m, n$ are integer numbers, $a$ is a lattice constant chosen to be small as compared to other scales in the model, and $\hat{{\bf e}}_{\zeta}$ with $\zeta=x,z$ is the unit vector pointing towards the $\zeta$-direction. In the lattice representation, the differential operators are replaced as: $\frac{\partial^2}{\partial \zeta^2} \psi(x,z) |_{{\bf r}_{i}} = \frac{1}{a^2}  \left[ \psi({\bf r}_i + a \hat{{\bf e}}_\zeta)  - 2\psi({\bf r}_i)  + \psi({\bf r}_i - a\hat{{\bf e}}_\zeta) \right]$, and $\frac{\partial }{\partial \zeta} \psi(x,z)|_{{\bf r}_i} = \frac{1}{a} \left[ \psi({\bf r}_i + a \hat{{\bf e}}_\zeta) - \psi({\bf r}_i ) \right]$, where $\psi(x,z)$ is the 2D wavefunction.  With this approximation we get the matrix representation of the Hamiltonian \eqref{ham2D} which is in the tb form and only contains the nearest coupling terms between adjacent sites. To get the geometry of a disc as shown in Fig. 1(b) of the main text, we consider a rectangular grid with length and width slightly larger than $2r_0$, and choose the center of the grid to coincide with that of the disc. By applying an infinitely large on-site potential outside the disc boundary, we achieve a realization of the open boundary conditions on the  disc boundary. By exact diagonalization of the Hamiltonian matrix, we obtained the band structure, the density of states, and the boundary charge. In Fig. \ref{fig:SM_Comparison} we compare the low-energy states obtained from the 2D tb model by setting $a=l_{so}/5$, and the one from directly solving the continuum model \eqref{ham2D} in the basis introduced in the subsection \ref{subsec:Bessel}. The good agreement validates the accuracy of the 2D tb approximation.

\subsection{Tight-binding Hamiltonian of the 3D TI under magnetic fields}

To calculate the boundary charge in the presence of disorder potential, we use a 3D tb Hamiltonian of a TI \cite{shen2017topological,hosur2011majorana}. The Hamiltonian defined on a cubic lattice is written as
\begin{align}
    H_{3D}^{tb}= \sum_i t |{\bf r}_i \rangle \mathcal{M}_i \langle {\bf r}_i | + \sum_i \sum_{\zeta=x,y,z} t \left( |{\bf r}_i + a_{3D}\hat{{\bf e}}_{\zeta} \rangle \mathcal{T}_{\zeta}
   e^{i\phi_{{\bf r}_i, {\bf r}_i + a_{3D}\hat{{\bf e}}_{\zeta}}} \langle {\bf r}_i | +  h.c. \right),
   \label{H_3Dtb}
\end{align}
 where $|{\bf r}_i \rangle$ is the Wannier basis denoting the lattice site with real-space position ${\bf r}_i$, and has both orbital and spin degrees of freedom encoded in the components $|{\bf r}_i \rangle = (|{\bf r}_i^{1\uparrow} \rangle, |{\bf r}_i^{1\downarrow} \rangle, |{\bf r}_i^{2\uparrow}\rangle, |{\bf r}_i^{2\downarrow} \rangle )^T$  with $\sigma=1,2$ and $s=\uparrow, \downarrow$. Here $t$ and $a_{3D}$ represent the hopping amplitude and the lattice constant of the 3D tb model, respectively. The onsite matrix elements are given by $\mathcal{M}_i = m_0 \sigma_z +\Delta_z s_z +U_i \equiv \mathcal{M}_0 +U_i$, where $\Delta_z$ is the dimensionless Zeeman energy in units of $t$, and $U_i$ is the on-site disorder potential which is uniformly distributed within $[-W/2, W/2]$ with $W$ being the characteristic disorder strength. The hopping matrix elements are given by $\mathcal{T}_{\zeta} = -\frac{\alpha_{tb}}{2i} \sigma_x s_{\zeta} - \frac{m_1}{2} \sigma_z$.
 
 The orbital effect of the magnetic field is included by adding a phase factor $\phi_{{\bf r}_i, {\bf r}_i+a_{3D} \hat{{\bf e}}_{\zeta}} = -2\pi \int_{{\bf r}_i} ^{{\bf r}_i +a_{3D} \hat{{\bf e}}_{\zeta}} {\bf A}\cdot d {\bf r} /\phi_0$ to the hopping matrices $\mathcal{T}_{\zeta}$, with  the vector potential ${\bf A}$ containing contributions which generate  both the uniform magnetic field ${\bf B}$ and the AB flux. In the real calculations to get the geometry of a torus, we start with a cuboid of the size $N_x \times N_y \times N_z \cdot a^3_{3D}$, and restrict the actual volume to the interior of the torus  $\left( \sqrt{x^2+y^2} -l_0 \right)^2 + z^2 < r^2_0$ by applying an infinite onsite potential outside it. To simplify the numerical calculation we use the Landau gauge $(-By,0,0)$ for ${\bf A}_{B}$, and the discontinuous gauge $2\pi f \delta(\varphi) \nabla \varphi$ for ${\bf A}_{\phi}$.

Let us now relate the parameters of the above introduced 3D tb model to the parameters of the Hamiltonian \eqref{eq:H_main} in the main text. To this end, we first omit in the 3D tb Hamiltonian the disorder term $U_i$ and the orbital effects of the magnetic fields, and relax the boundary conditions. This restores the translational invariance and allows us to introduce the Bloch Hamiltonian 
\begin{align}
    H_{3D}^{tb} =& \sum_{\bf k} | {\bf k} \rangle \left[ t \mathcal{M}_0 + \sum_{\zeta=x,y,z} \left( t \mathcal{T}_{\zeta} e^{-i {\bf k}\cdot \hat{{\bf e}}_{\zeta} \, a_{3D}} + t \mathcal{T}_{\zeta}^\dagger e^{i {\bf k}\cdot \hat{{\bf e}}_{\zeta} \, a_{3D}} \right) \right] \langle {\bf k}| 
    \equiv \sum_{\bf k} | {\bf k}\rangle h_{3D} ({\bf k}) \langle {\bf k} |,
\end{align}
with
\begin{align}
h_{3D} ({\bf k}) =& t \alpha_{tb} \sigma_x (s_x \sin{k_x a_{3D}} + s_y \sin{k_y a_{3D}} + s_z \sin{k_z a_{3D}} ) + t \Delta_z s_z \nonumber \\
& + t \sigma_z \left[ m_0 -m_1 (\cos{k_x a_{3D}} + \cos{k_y a_{3D}} + \cos{k_z a_{3D}}) \right] .
\end{align}
Approximating this expression around the ${\bf \Gamma} = (0,0,0)$ point in the Brillouin zone by the virtue of relations $\sin{k_{\zeta} a_{3D}} \approx k_{\zeta} a_{3D}$ and $\cos{k_{\zeta} a_{3D}} \approx 1-\frac{1}{2} k^2_{\zeta} a^2_{3D}$, we get
\begin{align}
    h_{3D} ({\bf k}) \approx
    t \sigma_z s_0 \left[ \frac{m_1 a^2_{3D}}{2} k^2 - ( 3m_1 - m_0) \right] 
   + t \alpha_{tb} \cdot a_{3D} \sigma_x ({\bf k}\cdot {\bf s}) 
    +t\Delta_z s_z .
    \label{kpmodel}
\end{align}
To achieve the equivalence between \eqref{kpmodel} and \eqref{eq:H_main}, we identify $\frac{t m_1 a^2_{3D}}{2} = \frac{1}{2m^*}$, $t (3m_1 -m_0) =\delta$, $t \alpha_{tb} \cdot a_{3D} =\alpha$, and $t \Delta_z =\varepsilon_Z$. 
By setting the hopping energy $t=\varepsilon_{so}$ and the lattice constant $a_{3D}=l_{so}$, we get the following values of the dimensionless parameters in the 3D tb Hamiltonian: $m_1=2, m_0=3, \alpha_{tb}=2,$ and $\Delta_z=1$, in accordance with the parameter values of the Hamiltonian \eqref{eq:H_main}
specified in the caption of Fig.~\ref{fig:torus_spectrum}. The orbital effect of the magnetic field is incorporated by substituting $\bf k$ with ${\bf k}+{\bf A}$. The magnetic flux through each unit cell projected onto the $x-y$ plane generated by the uniform magnetic field $\bf B$ is calculated to be $\phi_{xy} = B a^2_{3D} =1$. This observation helps us to calculate numerically the phase factor $\phi_{{\bf r}_i, {\bf r}_i+a_{3D}\hat{{\bf e}}_{\zeta}} $ appearing in  \eqref{H_3Dtb}. 

Having established the correspondence in the low-energy description between the 3D tb Hamiltonian and its continuum model counterpart, we further use the tb model to evaluate the boundary charge in the presence of on-site disorder breaking the rotational symmetry. The corresponding results are shown in Fig. \ref{fig:Disorder} of the main text.

\subsection{Green's function method in calculating the boundary charge}

Since we have used a 3D cuboid to simulate the torus model, the starting lattice has $N_x \times N_y \times N_z$ sites. Accounting the spin and orbital degrees of freedom, we obtain the Hilbert space dimension of the tb Hamiltonian to be $N_H=4\times N_x \times N_y \times N_z$. For the torus with $l_0=10 \,  l_{so}$ and $r_0=5 \, l_{so}$, we need at least $N_x=N_y=31$, and $N_z=11$. Exactly diagonalizing such a huge matrix of the dimension $N_H=42284$ requires a large amount of virtual computer memory rendering the calculation very time-consuming. Instead of this direct approach, we adhere to the recursive Green's function method \cite{do_anp_14} to calculate the boundary charge. This method allows us to separate the task into hundreds of jobs which can be computed in parallel on a high-performance computing cluster. 

The total charge within the region $\mathcal{B}$ below the chemical potential of the 3D lattice system is defined as:
$Q^{tb}_{\mathcal{B}}= \int_{- \infty} ^{\infty} d\varepsilon \sum_{i \in {\mathcal{B}}} \Theta(\mu-\varepsilon) \rho_i (\varepsilon) = \sum_{i \in \mathcal{B}} Q^{tb}(i)$, where $Q^{tb}(i)\equiv \int_{-\infty}^\mu d \varepsilon \rho_i (\varepsilon) = \sum_{\varepsilon_\lambda \leq \mu} \rho_i (\varepsilon_\lambda)$ is the total charge at site $i$. Here $\varepsilon_\lambda$ is the discrete eigenvalues of the isolated system with eigenstate $|\lambda \rangle $, $\rho_i (\varepsilon_{\lambda}) = |\langle {\bf r}_i | \lambda \rangle |^2$, and $\rho_i (\varepsilon) = \sum_{\varepsilon_{\lambda}} \rho_i (\varepsilon_{\lambda}) \delta (\varepsilon - \varepsilon_{\lambda}) $ is the local density of states at site $i$. Further, we represent
\begin{align}
Q^{tb}(i) = \frac{1}{2\pi i}\int_{\mathcal{C}} dz G_i (z)
\end{align}
in terms of the site-diagonal components $G_i (z) = \langle {\bf r}_i | G (z) | {\bf r}_i \rangle = \sum_{\varepsilon_{\lambda}} \frac{\rho_i (\varepsilon_{\lambda})}{z- \varepsilon_{\lambda}}$ of the Green's function $G (z)= (z - H_{3D}^{tb})^{-1}$. In this expression,
a counterclockwise integration path $\mathcal{C}$ in the complex-$z$ plane encompasses the poles of $G_i (z)$ at $z=\varepsilon_{\lambda}$ lying below the chemical potential $\mu$. 

To facilitate the integration, we use a rectangular loop $\mathcal{C}$ with the four corners in the complex plane: $z_{1(2)}=\mu +0^+ \mp b_{M} i$, and $z_{3(4)}=-a_{M}\pm b_{M}i$, where $b_{M}$ and $a_{M}$ are both large and positive real numbers denoting the absolute maximum values of  the imaginary and real part of $z$, respectively. This allows us to express
\begin{align}
Q^{tb}(i) = \frac{1}{2\pi} \int_{-b_{M}}^{b_{M}} db \, {\rm Re} \left[G_i (\mu + 0^+ +bi) \right] 
+ \frac{1}{\pi} \int_{-a_{M}}^{\mu + 0^+} da \, {\rm Im} \left[ G_i (a-b_{M}i) \right] - \frac{1}{2\pi} \int_{-b_{M}}^{b_{M}} db \, {\rm Re} \left[ G_i (-a_{M} + bi) \right].
\end{align}
The last two integrals are insensitive to the parameter $f$ (which can be ever numerically tested), because they account contributions from the states $|\varepsilon_{\lambda}| \sim a_M , b_M$ lying far below the Fermi surface. This observation leads us to the operational formula
\begin{align}
Q^{tb}(i) = \frac{1}{2\pi} \int^{b_M}_{-b_M} db \, {\rm Re} \left[ G_i (\mu +0^+ +bi)
\right] + {\rm const}.
\end{align}
To calculate $G_i (z)$, we employ the recursive Green's function method \cite{do_anp_14} which avoids a direct calculation of the inverse of the huge Hamiltonian matrix.

\begin{figure}[t]
        \centering
        \includegraphics[width = 0.99 \columnwidth]{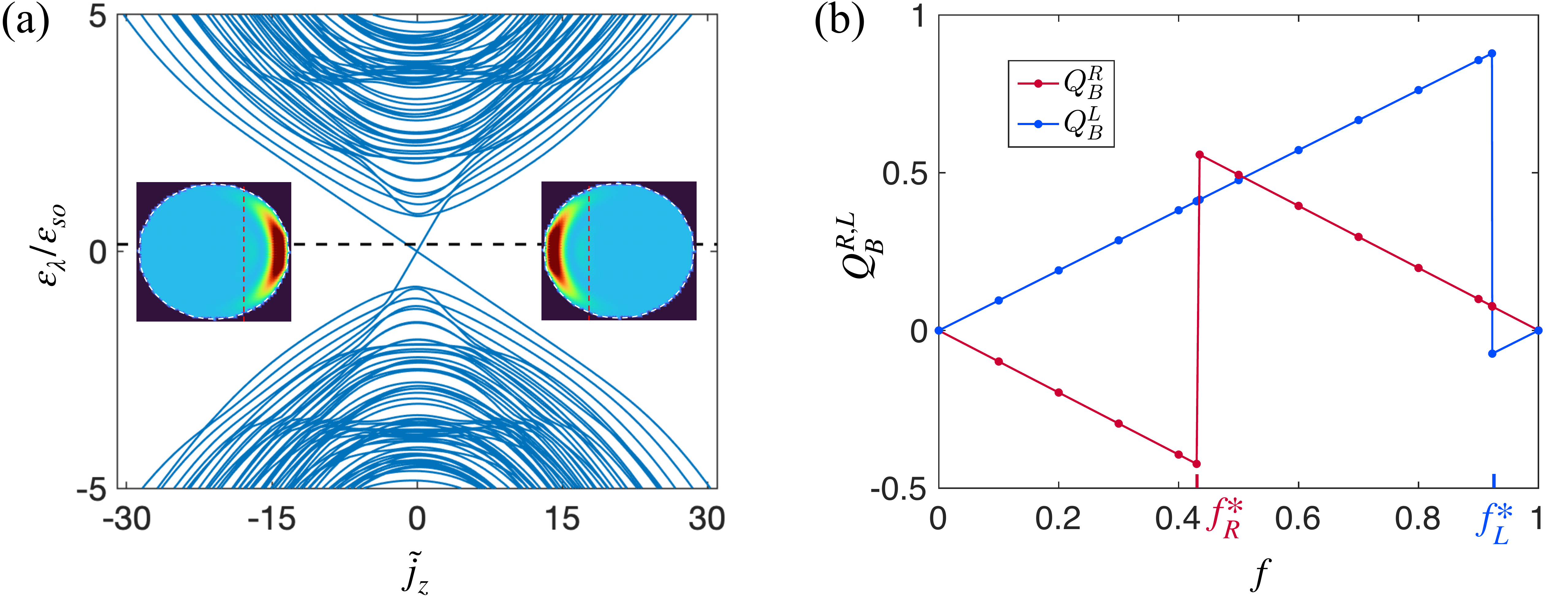}
        \caption{(a) Low-energy spectrum of the torus model (in units of spin-orbit energy $\varepsilon_{so}$) without magnetic orbital effect ($B=0$). The parameters we use here are: $\delta=3 \, \varepsilon_{so}$, $l_0=10 \, l_{so}$, $r_0=5 \, l_{so}$, and $\varepsilon_Z=\varepsilon_{so}$. The chemical potential is set to $\mu=0.15$. (b) Right and left boundary charges as functions of $f$ feature the linear slopes with the values -0.9859 and 0.9523, respectively, which are complemented by the jumps with the corresponding amplitudes at the flux values $f_R^* \approx 0.43$, and $f_L^* \approx 0.92$.  The actual numerical summation runs over the $j_z$ range from -149.5 to 150.5, which is sufficient to ensure convergence. The width of the boundary regions is $d=3 \, l_{so}$.  It accommodates approximately 98.53 (95.23) percents of the weight of the right (left) hinge state at the energy $\approx 0.1508 \, \varepsilon_{so}$ ($\approx 0.1490 \, \varepsilon_{so}$), see insets in (a), and this explains the deviation of the jump (and the slope) size from the universal unit value. }
        \label{fig:SM_no_orbital}
    \end{figure}

\section{3D QHE in the absence of the orbital $B$-field effect}

In Fig. \ref{fig:SM_no_orbital}(a) we show the band structure of our model in the absence of the orbital $B$-field effect. It features the two mid-gap modes of the $R$ and $L$ hinge states traversing the surface gap and connecting the hinge modes with bulk surface states. In Fig. \ref{fig:SM_no_orbital}(b) we show the right and left boundary charges (defined within the same boundary regions as in Fig. \ref{fig:models}(b)) which receive contributions from all occupied states below the chemical potential $\mu$. They feature the linear dependencies on $f$ with the slope values -0.9859 and 0.9523, which are complemented by the jumps of the corresponding sizes.

In the absence of the orbital $B$-field effect there is no plateau region of surface states connecting the $R$ and $L$ hinges. Therefore for the analytical explanation it is easier to consider the total boundary charge and write it as a sum over all total angular momenta 
\begin{align}
    \label{eq:total_QB}
    Q_{B}^{R,L}(f) &= \sum_{j_z} \left[\bar{Q}_{B}^{R,L}(j_z+f) - \bar{Q}_{B}^{R,L}(j_z)\right],
\end{align}
where $\bar{Q}_B^{R,L}(j_z+f)$ includes the sum over all bands and the condition that the energy must be smaller than $\mu$
\begin{align}
    \label{eq:tilde_QB_jz}
    \bar{Q}_B^{R,L}(\tilde{j_z}) = 
    \sum_\lambda \Theta(\mu - \varepsilon_\lambda(\tilde{j}_z)) \bar{Q}_{B,\lambda}^{R,L}(\tilde{j}_z) .
\end{align}
To estimate the typical scale $\Delta j_z$ on which $\bar{Q}_B^{R,L}(j_z + f)$ is expected to vary on $j_z$ (up to jumps occurring when the energy of the highest valence band crosses the chemical potential), we have to compare the scale of the momentum parallel to the surface $\sim j_z/l_0$ with various other inverse length scales appearing in the Hamiltonian (\ref{ham2D_curve}) (like $\varepsilon_Z/\alpha\sim l_{so}/l_Z^2$, $m^* \alpha = 1/l_{so}$ , or $\sqrt{m^*\delta}$). 
Since all these inverse length scales are independent of $l_0$, the scale $\Delta j_z$ will always increase with the torus radius $l_0$ and we find, analog to the arguments presented in the main text, that the boundary charge $Q_B^{R,L}(f)$ is dominated by its linear term to any desired degree of accuracy by increasing the torus radius. The linear behaviour together with the jump properties and the periodicity of $Q_B^{R,L}(f)$ then proves the universal slopes of $Q_B^{R,L}(f)$ in the same manner as discussed in the main text. 

In the presence of strong orbital fields $l_B \sim l_Z$ we note that the typical scale $\Delta j_z$ arises from the combination $j_z + f + B x^2/2$ in the Hamiltonian (\ref{ham2D_curve}). With $x=l_0 + r \sin\theta$ and $B=1/l_B^2$ we get
\begin{align}
    \label{eq:B_x2}
    \frac{B x^2}{2} = \frac{l_0^2}{2 l_B^2} + \frac{r l_0}{l_B^2} \sin\theta + \frac{r^2}{2 l_B^2}\sin^2\theta .
\end{align}
The first term leads to an unimportant shift of $j_z$. With $\sin\theta\sim O(1)$ and $r\sim r_0 < l_0$, the last two terms determine the scale $\Delta j_z \sim l_0 r_0 /l_B^2$ which is of the order of the number of fluxes threaded through the torus surface from the orbital field, cf. the main text.

As a result, for any value of the orbital field $B$, we obtain that the scale $\Delta j_z$ increases linearly with the torus radius $l_0$. Therefore, the universal form of the boundary charge is unaffected by the orbital effects. We note that our estimate of the higher-order derivatives 
\begin{align}
    \label{eq:higher_order_derivative}
    \frac{d^n}{df^n} Q_B^{R,L}(f) \sim \frac{1}{(\Delta j_z)^{n-1}}
\end{align}
is quite conservative. The reason is that the sum over $j_z$ for the higher-order derivatives 
$(\frac{d}{d \tilde{j}_z})^n \bar{Q}_B^{R,L}(\tilde{j_z})$, with $n\ge 2$, of (\ref{eq:total_QB}) contains many terms with different signs which partially cancel each other. This can be seen from replacing sums by integrals and noting that the asymptotic values of $(\frac{d}{d \tilde{j}_z})^{n-1} \bar{Q}_B^{R,L}(\tilde{j_z})$, with $n\ge 2$, go to zero for $\tilde{j}_z\rightarrow\pm\infty$. The same holds close to the point where the highest valence band leads to a jump of $\bar{Q}_B^{R,L}(\tilde{j}_z)$ when some hinge modes crosses the chemical potential. The variation in $j_z$ becomes very weak close to the jumping point if the boundary region is chosen much larger than the localization length of all hinge modes. The vanishing integral of all higher-order derivatives with $n\ge 2$ then indicates that also the sum is expected to be very small providing another reason why the linear behaviour is so robust with an error expected to be even much smaller than $1/\Delta j_z$. This is indeed the case in Fig.~\ref{fig:SM_no_orbital}, where $\Delta j_z \sim l_0/l_{so}=10$, but the deviation from the linear behaviour is much smaller than $10\%$.

\begin{figure}[t]
        \centering
        \includegraphics[width = 0.99 \columnwidth]{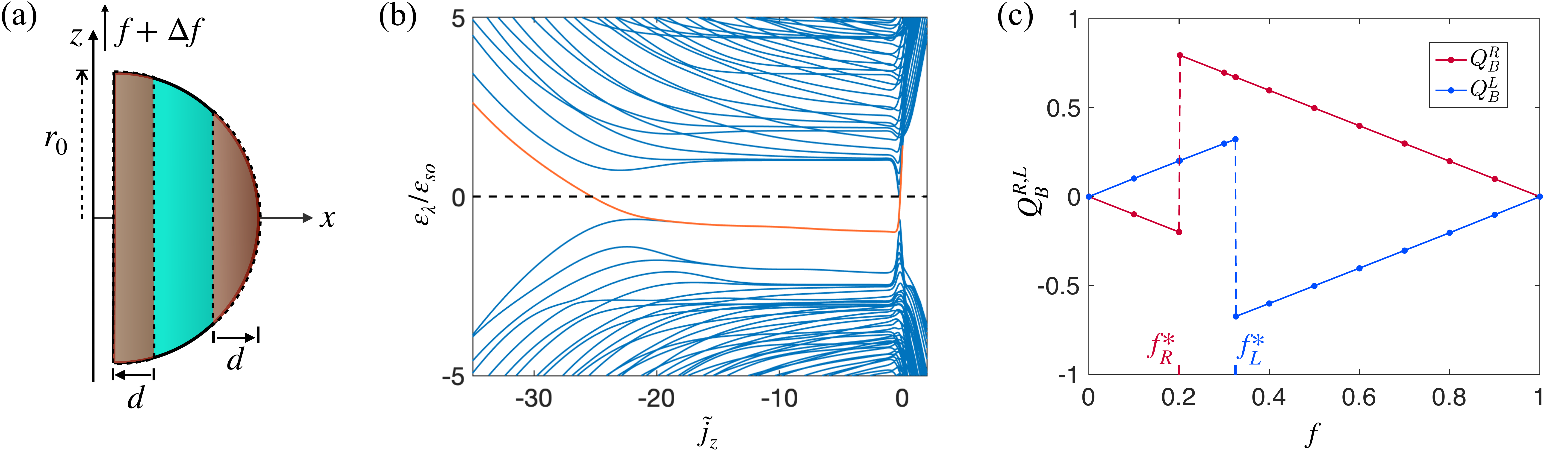}
        \caption{(a) Sketch of the half-disc. (b) Low-energy spectrum of the half-disc model at $B \neq 0$ in units of spin-orbit energy $\varepsilon_{so}$. The parameters we use here are: $\delta=3 \, \varepsilon_{so}$, $l_B=l_{so}$, $l_0=0$, $r_0=8 \, l_{so}$, and $\varepsilon_Z=\varepsilon_{so}$. The lattice constant is chosen to be $a = l_{so}/3$. The chemical potential is set to $\mu=0$. (c) Right and left boundary charges as functions of $f$ feature the linear slopes with the values -0.9962 and 0.9993, respectively, which are complemented by the unit jumps at the flux values $f_R^* \approx 0.2$, and $f_L^* \approx 0.32$. The width of the boundary regions is $d=2.5 \, l_{so}$. The actual numerical summation runs over the $j_z$ range from -59.5 to 30.5, which is sufficient to ensure convergence. }
        \label{fig:SM_sphere}
    \end{figure}
    
\section{3D QHE in the spherical geometry}

In Fig. \ref{fig:SM_sphere} we show a two-dimensional projection of the sphere in the right half-plane $(x>0 , z)$ appearing in the form of the half-disc. The model (with the magnetic orbital effects) is equipped with the open boundary conditions along its whole circumference including the vertical section. Vanishing of the wavefunction on the vertical section is equivalent to drilling an infinitesimally thin hole in the sphere, which allows for an insertion of the AB flux.

The band structure shown in Fig. \ref{fig:SM_sphere}(b) features a distinguished band (marked in orange) of states residing along the circumference of the half-disc. The states on the negative gentle slope are the hinge states localized near the spherical equator, while the states on the positive steep slope are spatially localized along the vertical axis.

In Fig. \ref{fig:SM_sphere}(c) the right and left boundary charges in the corresponding brown regions of the panel (a) are shown as functions of $f$. They receive the whole contribution from the orange band of the panel (b). The calculated slopes are very close to the universal unit values, which is analytically explained by the same arguments as in the torus case.

\end{widetext}

\end{appendix}

\bibliography{citations}

\begin{thebibliography}{70}%
\makeatletter
\providecommand \@ifxundefined [1]{%
 \@ifx{#1\undefined}
}%
\providecommand \@ifnum [1]{%
 \ifnum #1\expandafter \@firstoftwo
 \else \expandafter \@secondoftwo
 \fi
}%
\providecommand \@ifx [1]{%
 \ifx #1\expandafter \@firstoftwo
 \else \expandafter \@secondoftwo
 \fi
}%
\providecommand \natexlab [1]{#1}%
\providecommand \enquote  [1]{``#1''}%
\providecommand \bibnamefont  [1]{#1}%
\providecommand \bibfnamefont [1]{#1}%
\providecommand \citenamefont [1]{#1}%
\providecommand \href@noop [0]{\@secondoftwo}%
\providecommand \href [0]{\begingroup \@sanitize@url \@href}%
\providecommand \@href[1]{\@@startlink{#1}\@@href}%
\providecommand \@@href[1]{\endgroup#1\@@endlink}%
\providecommand \@sanitize@url [0]{\catcode `\\12\catcode `\$12\catcode
  `\&12\catcode `\#12\catcode `\^12\catcode `\_12\catcode `\%12\relax}%
\providecommand \@@startlink[1]{}%
\providecommand \@@endlink[0]{}%
\providecommand \url  [0]{\begingroup\@sanitize@url \@url }%
\providecommand \@url [1]{\endgroup\@href {#1}{\urlprefix }}%
\providecommand \urlprefix  [0]{URL }%
\providecommand \Eprint [0]{\href }%
\providecommand \doibase [0]{https://doi.org/}%
\providecommand \selectlanguage [0]{\@gobble}%
\providecommand \bibinfo  [0]{\@secondoftwo}%
\providecommand \bibfield  [0]{\@secondoftwo}%
\providecommand \translation [1]{[#1]}%
\providecommand \BibitemOpen [0]{}%
\providecommand \bibitemStop [0]{}%
\providecommand \bibitemNoStop [0]{.\EOS\space}%
\providecommand \EOS [0]{\spacefactor3000\relax}%
\providecommand \BibitemShut  [1]{\csname bibitem#1\endcsname}%
\let\auto@bib@innerbib\@empty
\bibitem [{\citenamefont {Klitzing}\ \emph {et~al.}(1980)\citenamefont
  {Klitzing}, \citenamefont {Dorda},\ and\ \citenamefont
  {Pepper}}]{klitzing_etal_prl_80}%
  \BibitemOpen
  \bibfield  {author} {\bibinfo {author} {\bibfnamefont {K.~v.}\ \bibnamefont
  {Klitzing}}, \bibinfo {author} {\bibfnamefont {G.}~\bibnamefont {Dorda}},\
  and\ \bibinfo {author} {\bibfnamefont {M.}~\bibnamefont {Pepper}},\
  }\bibfield  {title} {\bibinfo {title} {New method for high-accuracy
  determination of the fine-structure constant based on quantized hall
  resistance},\ }\href {https://doi.org/10.1103/PhysRevLett.45.494} {\bibfield
  {journal} {\bibinfo  {journal} {Phys. Rev. Lett.}\ }\textbf {\bibinfo
  {volume} {45}},\ \bibinfo {pages} {494} (\bibinfo {year} {1980})}\BibitemShut
  {NoStop}%
\bibitem [{\citenamefont {Kane}\ and\ \citenamefont
  {Mele}(2005{\natexlab{a}})}]{kane_mele_1_prl_05}%
  \BibitemOpen
  \bibfield  {author} {\bibinfo {author} {\bibfnamefont {C.~L.}\ \bibnamefont
  {Kane}}\ and\ \bibinfo {author} {\bibfnamefont {E.~J.}\ \bibnamefont
  {Mele}},\ }\bibfield  {title} {\bibinfo {title} {Quantum spin hall effect in
  graphene},\ }\href@noop {} {\bibfield  {journal} {\bibinfo  {journal} {Phys.
  Rev. Lett.}\ }\textbf {\bibinfo {volume} {95}},\ \bibinfo {pages} {226801}
  (\bibinfo {year} {2005}{\natexlab{a}})}\BibitemShut {NoStop}%
\bibitem [{\citenamefont {Kane}\ and\ \citenamefont
  {Mele}(2005{\natexlab{b}})}]{kane_mele_2_prl_05}%
  \BibitemOpen
  \bibfield  {author} {\bibinfo {author} {\bibfnamefont {C.~L.}\ \bibnamefont
  {Kane}}\ and\ \bibinfo {author} {\bibfnamefont {E.~J.}\ \bibnamefont
  {Mele}},\ }\bibfield  {title} {\bibinfo {title} {${Z}_{2}$ topological order
  and the quantum spin hall effect},\ }\href@noop {} {\bibfield  {journal}
  {\bibinfo  {journal} {Phys. Rev. Lett.}\ }\textbf {\bibinfo {volume} {95}},\
  \bibinfo {pages} {146802} (\bibinfo {year} {2005}{\natexlab{b}})}\BibitemShut
  {NoStop}%
\bibitem [{\citenamefont {Bernevig}\ \emph {et~al.}(2006)\citenamefont
  {Bernevig}, \citenamefont {Hughes},\ and\ \citenamefont
  {Zhang}}]{bernevig_etal_science_06}%
  \BibitemOpen
  \bibfield  {author} {\bibinfo {author} {\bibfnamefont {B.~A.}\ \bibnamefont
  {Bernevig}}, \bibinfo {author} {\bibfnamefont {T.~L.}\ \bibnamefont
  {Hughes}},\ and\ \bibinfo {author} {\bibfnamefont {S.-C.}\ \bibnamefont
  {Zhang}},\ }\bibfield  {title} {\bibinfo {title} {Quantum spin hall effect
  and topological phase transition in hgte quantum wells},\ }\href
  {https://doi.org/10.1126/science.1133734} {\bibfield  {journal} {\bibinfo
  {journal} {Science}\ }\textbf {\bibinfo {volume} {314}},\ \bibinfo {pages}
  {1757–1761} (\bibinfo {year} {2006})}\BibitemShut {NoStop}%
\bibitem [{\citenamefont {K{\"o}nig}\ \emph {et~al.}(2007)\citenamefont
  {K{\"o}nig}, \citenamefont {Wiedmann}, \citenamefont {Br{\"u}ne},
  \citenamefont {Roth}, \citenamefont {Buhmann}, \citenamefont {Molenkamp},
  \citenamefont {Qi},\ and\ \citenamefont {Zhang}}]{koenig_etal_science_07}%
  \BibitemOpen
  \bibfield  {author} {\bibinfo {author} {\bibfnamefont {M.}~\bibnamefont
  {K{\"o}nig}}, \bibinfo {author} {\bibfnamefont {S.}~\bibnamefont {Wiedmann}},
  \bibinfo {author} {\bibfnamefont {C.}~\bibnamefont {Br{\"u}ne}}, \bibinfo
  {author} {\bibfnamefont {A.}~\bibnamefont {Roth}}, \bibinfo {author}
  {\bibfnamefont {H.}~\bibnamefont {Buhmann}}, \bibinfo {author} {\bibfnamefont
  {L.~W.}\ \bibnamefont {Molenkamp}}, \bibinfo {author} {\bibfnamefont {X.-L.}\
  \bibnamefont {Qi}},\ and\ \bibinfo {author} {\bibfnamefont {S.-C.}\
  \bibnamefont {Zhang}},\ }\bibfield  {title} {\bibinfo {title} {Quantum spin
  hall insulator state in hgte quantum wells},\ }\href
  {https://doi.org/10.1126/science.1148047} {\bibfield  {journal} {\bibinfo
  {journal} {Science}\ }\textbf {\bibinfo {volume} {318}},\ \bibinfo {pages}
  {766} (\bibinfo {year} {2007})}\BibitemShut {NoStop}%
\bibitem [{\citenamefont {Hasan}\ and\ \citenamefont
  {Kane}(2010)}]{hasan_kane_rmp_10}%
  \BibitemOpen
  \bibfield  {author} {\bibinfo {author} {\bibfnamefont {M.~Z.}\ \bibnamefont
  {Hasan}}\ and\ \bibinfo {author} {\bibfnamefont {C.~L.}\ \bibnamefont
  {Kane}},\ }\bibfield  {title} {\bibinfo {title} {Colloquium: Topological
  insulators},\ }\href {https://doi.org/10.1103/RevModPhys.82.3045} {\bibfield
  {journal} {\bibinfo  {journal} {Rev. Mod. Phys.}\ }\textbf {\bibinfo {volume}
  {82}},\ \bibinfo {pages} {3045} (\bibinfo {year} {2010})}\BibitemShut
  {NoStop}%
\bibitem [{\citenamefont {Qi}\ and\ \citenamefont
  {Zhang}(2011)}]{qi_zhang_rmp_11}%
  \BibitemOpen
  \bibfield  {author} {\bibinfo {author} {\bibfnamefont {X.-L.}\ \bibnamefont
  {Qi}}\ and\ \bibinfo {author} {\bibfnamefont {S.-C.}\ \bibnamefont {Zhang}},\
  }\bibfield  {title} {\bibinfo {title} {Topological insulators and
  superconductors},\ }\href {https://doi.org/10.1103/RevModPhys.83.1057}
  {\bibfield  {journal} {\bibinfo  {journal} {Rev. Mod. Phys.}\ }\textbf
  {\bibinfo {volume} {83}},\ \bibinfo {pages} {1057} (\bibinfo {year}
  {2011})}\BibitemShut {NoStop}%
\bibitem [{\citenamefont {Haldane}(1988)}]{haldane_prl_88}%
  \BibitemOpen
  \bibfield  {author} {\bibinfo {author} {\bibfnamefont {F.~D.~M.}\
  \bibnamefont {Haldane}},\ }\bibfield  {title} {\bibinfo {title} {Model for a
  quantum hall effect without landau levels: Condensed-matter realization of
  the "parity anomaly"},\ }\href {https://doi.org/10.1103/PhysRevLett.61.2015}
  {\bibfield  {journal} {\bibinfo  {journal} {Phys. Rev. Lett.}\ }\textbf
  {\bibinfo {volume} {61}},\ \bibinfo {pages} {2015} (\bibinfo {year}
  {1988})}\BibitemShut {NoStop}%
\bibitem [{\citenamefont {Liu}\ \emph {et~al.}(2008)\citenamefont {Liu},
  \citenamefont {Qi}, \citenamefont {Dai}, \citenamefont {Fang},\ and\
  \citenamefont {Zhang}}]{liu_etal_prl_08}%
  \BibitemOpen
  \bibfield  {author} {\bibinfo {author} {\bibfnamefont {C.-X.}\ \bibnamefont
  {Liu}}, \bibinfo {author} {\bibfnamefont {X.-L.}\ \bibnamefont {Qi}},
  \bibinfo {author} {\bibfnamefont {X.}~\bibnamefont {Dai}}, \bibinfo {author}
  {\bibfnamefont {Z.}~\bibnamefont {Fang}},\ and\ \bibinfo {author}
  {\bibfnamefont {S.-C.}\ \bibnamefont {Zhang}},\ }\bibfield  {title} {\bibinfo
  {title} {Quantum anomalous hall effect in
  ${\mathrm{hg}}_{1\ensuremath{-}y}{\mathrm{mn}}_{y}\mathrm{Te}$ quantum
  wells},\ }\href {https://doi.org/10.1103/PhysRevLett.101.146802} {\bibfield
  {journal} {\bibinfo  {journal} {Phys. Rev. Lett.}\ }\textbf {\bibinfo
  {volume} {101}},\ \bibinfo {pages} {146802} (\bibinfo {year}
  {2008})}\BibitemShut {NoStop}%
\bibitem [{\citenamefont {Bestwick}\ \emph {et~al.}(2015)\citenamefont
  {Bestwick}, \citenamefont {Fox}, \citenamefont {Kou}, \citenamefont {Pan},
  \citenamefont {Wang},\ and\ \citenamefont
  {Goldhaber-Gordon}}]{bestwick_etal_prl_15}%
  \BibitemOpen
  \bibfield  {author} {\bibinfo {author} {\bibfnamefont {A.~J.}\ \bibnamefont
  {Bestwick}}, \bibinfo {author} {\bibfnamefont {E.~J.}\ \bibnamefont {Fox}},
  \bibinfo {author} {\bibfnamefont {X.}~\bibnamefont {Kou}}, \bibinfo {author}
  {\bibfnamefont {L.}~\bibnamefont {Pan}}, \bibinfo {author} {\bibfnamefont
  {K.~L.}\ \bibnamefont {Wang}},\ and\ \bibinfo {author} {\bibfnamefont
  {D.}~\bibnamefont {Goldhaber-Gordon}},\ }\bibfield  {title} {\bibinfo {title}
  {Precise quantization of the anomalous hall effect near zero magnetic
  field},\ }\href {https://doi.org/10.1103/PhysRevLett.114.187201} {\bibfield
  {journal} {\bibinfo  {journal} {Phys. Rev. Lett.}\ }\textbf {\bibinfo
  {volume} {114}},\ \bibinfo {pages} {187201} (\bibinfo {year}
  {2015})}\BibitemShut {NoStop}%
\bibitem [{\citenamefont {Yu}\ \emph {et~al.}(2010)\citenamefont {Yu},
  \citenamefont {Zhang}, \citenamefont {Zhang}, \citenamefont {Zhang},
  \citenamefont {Dai},\ and\ \citenamefont {Fang}}]{yu_etal_science_10}%
  \BibitemOpen
  \bibfield  {author} {\bibinfo {author} {\bibfnamefont {R.}~\bibnamefont
  {Yu}}, \bibinfo {author} {\bibfnamefont {W.}~\bibnamefont {Zhang}}, \bibinfo
  {author} {\bibfnamefont {H.-J.}\ \bibnamefont {Zhang}}, \bibinfo {author}
  {\bibfnamefont {S.-C.}\ \bibnamefont {Zhang}}, \bibinfo {author}
  {\bibfnamefont {X.}~\bibnamefont {Dai}},\ and\ \bibinfo {author}
  {\bibfnamefont {Z.}~\bibnamefont {Fang}},\ }\bibfield  {title} {\bibinfo
  {title} {Quantized anomalous hall effect in magnetic topological
  insulators},\ }\href {https://doi.org/10.1126/science.1187485} {\bibfield
  {journal} {\bibinfo  {journal} {Science}\ }\textbf {\bibinfo {volume}
  {329}},\ \bibinfo {pages} {61} (\bibinfo {year} {2010})}\BibitemShut
  {NoStop}%
\bibitem [{\citenamefont {Chang}\ \emph {et~al.}(2013)\citenamefont {Chang},
  \citenamefont {Zhang}, \citenamefont {Feng}, \citenamefont {Shen},
  \citenamefont {Zhang}, \citenamefont {Guo}, \citenamefont {Li}, \citenamefont
  {Ou}, \citenamefont {Wei}, \citenamefont {Wang}, \citenamefont {Ji},
  \citenamefont {Feng}, \citenamefont {Ji}, \citenamefont {Chen}, \citenamefont
  {Jia}, \citenamefont {Dai}, \citenamefont {Fang}, \citenamefont {Zhang},
  \citenamefont {He}, \citenamefont {Wang}, \citenamefont {Lu}, \citenamefont
  {Ma},\ and\ \citenamefont {Xue}}]{chang_etal_science_13}%
  \BibitemOpen
  \bibfield  {author} {\bibinfo {author} {\bibfnamefont {C.-Z.}\ \bibnamefont
  {Chang}}, \bibinfo {author} {\bibfnamefont {J.}~\bibnamefont {Zhang}},
  \bibinfo {author} {\bibfnamefont {X.}~\bibnamefont {Feng}}, \bibinfo {author}
  {\bibfnamefont {J.}~\bibnamefont {Shen}}, \bibinfo {author} {\bibfnamefont
  {Z.}~\bibnamefont {Zhang}}, \bibinfo {author} {\bibfnamefont
  {M.}~\bibnamefont {Guo}}, \bibinfo {author} {\bibfnamefont {K.}~\bibnamefont
  {Li}}, \bibinfo {author} {\bibfnamefont {Y.}~\bibnamefont {Ou}}, \bibinfo
  {author} {\bibfnamefont {P.}~\bibnamefont {Wei}}, \bibinfo {author}
  {\bibfnamefont {L.-L.}\ \bibnamefont {Wang}}, \bibinfo {author}
  {\bibfnamefont {Z.-Q.}\ \bibnamefont {Ji}}, \bibinfo {author} {\bibfnamefont
  {Y.}~\bibnamefont {Feng}}, \bibinfo {author} {\bibfnamefont {S.}~\bibnamefont
  {Ji}}, \bibinfo {author} {\bibfnamefont {X.}~\bibnamefont {Chen}}, \bibinfo
  {author} {\bibfnamefont {J.}~\bibnamefont {Jia}}, \bibinfo {author}
  {\bibfnamefont {X.}~\bibnamefont {Dai}}, \bibinfo {author} {\bibfnamefont
  {Z.}~\bibnamefont {Fang}}, \bibinfo {author} {\bibfnamefont {S.-C.}\
  \bibnamefont {Zhang}}, \bibinfo {author} {\bibfnamefont {K.}~\bibnamefont
  {He}}, \bibinfo {author} {\bibfnamefont {Y.}~\bibnamefont {Wang}}, \bibinfo
  {author} {\bibfnamefont {L.}~\bibnamefont {Lu}}, \bibinfo {author}
  {\bibfnamefont {X.-C.}\ \bibnamefont {Ma}},\ and\ \bibinfo {author}
  {\bibfnamefont {Q.-K.}\ \bibnamefont {Xue}},\ }\bibfield  {title} {\bibinfo
  {title} {Experimental observation of the quantum anomalous hall effect in a
  magnetic topological insulator},\ }\href@noop {} {\bibfield  {journal}
  {\bibinfo  {journal} {Science}\ }\textbf {\bibinfo {volume} {340}},\ \bibinfo
  {pages} {167} (\bibinfo {year} {2013})}\BibitemShut {NoStop}%
\bibitem [{\citenamefont {Checkelsky}\ \emph {et~al.}(2014)\citenamefont
  {Checkelsky}, \citenamefont {Yoshimi}, \citenamefont {Tsukazaki},
  \citenamefont {Takahashi}, \citenamefont {Kozuka}, \citenamefont {Falson},
  \citenamefont {Kawasaki},\ and\ \citenamefont
  {Tokura}}]{checkelsky_etal_ntp_14}%
  \BibitemOpen
  \bibfield  {author} {\bibinfo {author} {\bibfnamefont {J.~G.}\ \bibnamefont
  {Checkelsky}}, \bibinfo {author} {\bibfnamefont {R.}~\bibnamefont {Yoshimi}},
  \bibinfo {author} {\bibfnamefont {A.}~\bibnamefont {Tsukazaki}}, \bibinfo
  {author} {\bibfnamefont {K.~S.}\ \bibnamefont {Takahashi}}, \bibinfo {author}
  {\bibfnamefont {Y.}~\bibnamefont {Kozuka}}, \bibinfo {author} {\bibfnamefont
  {J.}~\bibnamefont {Falson}}, \bibinfo {author} {\bibfnamefont
  {M.}~\bibnamefont {Kawasaki}},\ and\ \bibinfo {author} {\bibfnamefont
  {Y.}~\bibnamefont {Tokura}},\ }\bibfield  {title} {\bibinfo {title}
  {Trajectory of the anomalous hall effect towards the quantized state in a
  ferromagnetic topological insulator},\ }\href
  {https://doi.org/10.1038/nphys3053} {\bibfield  {journal} {\bibinfo
  {journal} {Nature Physics}\ }\textbf {\bibinfo {volume} {10}},\ \bibinfo
  {pages} {731} (\bibinfo {year} {2014})}\BibitemShut {NoStop}%
\bibitem [{\citenamefont {Feng}\ \emph {et~al.}(2015)\citenamefont {Feng},
  \citenamefont {Feng}, \citenamefont {Ou}, \citenamefont {Wang}, \citenamefont
  {Liu}, \citenamefont {Zhang}, \citenamefont {Zhao}, \citenamefont {Jiang},
  \citenamefont {Zhang}, \citenamefont {He}, \citenamefont {Ma}, \citenamefont
  {Xue},\ and\ \citenamefont {Wang}}]{feng_etal_prl_15}%
  \BibitemOpen
  \bibfield  {author} {\bibinfo {author} {\bibfnamefont {Y.}~\bibnamefont
  {Feng}}, \bibinfo {author} {\bibfnamefont {X.}~\bibnamefont {Feng}}, \bibinfo
  {author} {\bibfnamefont {Y.}~\bibnamefont {Ou}}, \bibinfo {author}
  {\bibfnamefont {J.}~\bibnamefont {Wang}}, \bibinfo {author} {\bibfnamefont
  {C.}~\bibnamefont {Liu}}, \bibinfo {author} {\bibfnamefont {L.}~\bibnamefont
  {Zhang}}, \bibinfo {author} {\bibfnamefont {D.}~\bibnamefont {Zhao}},
  \bibinfo {author} {\bibfnamefont {G.}~\bibnamefont {Jiang}}, \bibinfo
  {author} {\bibfnamefont {S.-C.}\ \bibnamefont {Zhang}}, \bibinfo {author}
  {\bibfnamefont {K.}~\bibnamefont {He}}, \bibinfo {author} {\bibfnamefont
  {X.}~\bibnamefont {Ma}}, \bibinfo {author} {\bibfnamefont {Q.-K.}\
  \bibnamefont {Xue}},\ and\ \bibinfo {author} {\bibfnamefont {Y.}~\bibnamefont
  {Wang}},\ }\bibfield  {title} {\bibinfo {title} {Observation of the zero hall
  plateau in a quantum anomalous hall insulator},\ }\href
  {https://doi.org/10.1103/PhysRevLett.115.126801} {\bibfield  {journal}
  {\bibinfo  {journal} {Phys. Rev. Lett.}\ }\textbf {\bibinfo {volume} {115}},\
  \bibinfo {pages} {126801} (\bibinfo {year} {2015})}\BibitemShut {NoStop}%
\bibitem [{\citenamefont {Kandala}\ \emph {et~al.}(2015)\citenamefont
  {Kandala}, \citenamefont {Richardella}, \citenamefont {Kempinger},
  \citenamefont {Liu},\ and\ \citenamefont {Samarth}}]{kandala_etal_natcom_15}%
  \BibitemOpen
  \bibfield  {author} {\bibinfo {author} {\bibfnamefont {A.}~\bibnamefont
  {Kandala}}, \bibinfo {author} {\bibfnamefont {A.}~\bibnamefont
  {Richardella}}, \bibinfo {author} {\bibfnamefont {S.}~\bibnamefont
  {Kempinger}}, \bibinfo {author} {\bibfnamefont {C.-X.}\ \bibnamefont {Liu}},\
  and\ \bibinfo {author} {\bibfnamefont {N.}~\bibnamefont {Samarth}},\
  }\bibfield  {title} {\bibinfo {title} {Giant anisotropic magnetoresistance in
  a quantum anomalous hall insulator},\ }\href
  {https://doi.org/10.1038/ncomms8434} {\bibfield  {journal} {\bibinfo
  {journal} {Nature Communications}\ }\textbf {\bibinfo {volume} {6}},\
  \bibinfo {pages} {7434} (\bibinfo {year} {2015})}\BibitemShut {NoStop}%
\bibitem [{\citenamefont {Deng}\ \emph {et~al.}(2020)\citenamefont {Deng},
  \citenamefont {Yu}, \citenamefont {Shi}, \citenamefont {Guo}, \citenamefont
  {Xu}, \citenamefont {Wang}, \citenamefont {Chen},\ and\ \citenamefont
  {Zhang}}]{deng_etal_science_20}%
  \BibitemOpen
  \bibfield  {author} {\bibinfo {author} {\bibfnamefont {Y.}~\bibnamefont
  {Deng}}, \bibinfo {author} {\bibfnamefont {Y.}~\bibnamefont {Yu}}, \bibinfo
  {author} {\bibfnamefont {M.~Z.}\ \bibnamefont {Shi}}, \bibinfo {author}
  {\bibfnamefont {Z.}~\bibnamefont {Guo}}, \bibinfo {author} {\bibfnamefont
  {Z.}~\bibnamefont {Xu}}, \bibinfo {author} {\bibfnamefont {J.}~\bibnamefont
  {Wang}}, \bibinfo {author} {\bibfnamefont {X.~H.}\ \bibnamefont {Chen}},\
  and\ \bibinfo {author} {\bibfnamefont {Y.}~\bibnamefont {Zhang}},\ }\bibfield
   {title} {\bibinfo {title} {Quantum anomalous hall effect in intrinsic
  magnetic topological insulator mnbi<sub>2</sub>te<sub>4</sub>},\ }\href
  {https://doi.org/10.1126/science.aax8156} {\bibfield  {journal} {\bibinfo
  {journal} {Science}\ }\textbf {\bibinfo {volume} {367}},\ \bibinfo {pages}
  {895} (\bibinfo {year} {2020})}\BibitemShut {NoStop}%
\bibitem [{\citenamefont {Liu}\ \emph {et~al.}(2020)\citenamefont {Liu},
  \citenamefont {Wang}, \citenamefont {Li}, \citenamefont {Wu}, \citenamefont
  {Li}, \citenamefont {Li}, \citenamefont {He}, \citenamefont {Xu},
  \citenamefont {Zhang},\ and\ \citenamefont {Wang}}]{liu_etal_natmat_20}%
  \BibitemOpen
  \bibfield  {author} {\bibinfo {author} {\bibfnamefont {C.}~\bibnamefont
  {Liu}}, \bibinfo {author} {\bibfnamefont {Y.}~\bibnamefont {Wang}}, \bibinfo
  {author} {\bibfnamefont {H.}~\bibnamefont {Li}}, \bibinfo {author}
  {\bibfnamefont {Y.}~\bibnamefont {Wu}}, \bibinfo {author} {\bibfnamefont
  {Y.}~\bibnamefont {Li}}, \bibinfo {author} {\bibfnamefont {J.}~\bibnamefont
  {Li}}, \bibinfo {author} {\bibfnamefont {K.}~\bibnamefont {He}}, \bibinfo
  {author} {\bibfnamefont {Y.}~\bibnamefont {Xu}}, \bibinfo {author}
  {\bibfnamefont {J.}~\bibnamefont {Zhang}},\ and\ \bibinfo {author}
  {\bibfnamefont {Y.}~\bibnamefont {Wang}},\ }\bibfield  {title} {\bibinfo
  {title} {Robust axion insulator and chern insulator phases in a
  two-dimensional antiferromagnetic topological insulator},\ }\href
  {https://doi.org/10.1038/s41563-019-0573-3} {\bibfield  {journal} {\bibinfo
  {journal} {Nature Materials}\ }\textbf {\bibinfo {volume} {19}},\ \bibinfo
  {pages} {522} (\bibinfo {year} {2020})}\BibitemShut {NoStop}%
\bibitem [{\citenamefont {Ge}\ \emph {et~al.}(2020)\citenamefont {Ge},
  \citenamefont {Liu}, \citenamefont {Li}, \citenamefont {Li}, \citenamefont
  {Luo}, \citenamefont {Wu}, \citenamefont {Xu},\ and\ \citenamefont
  {Wang}}]{ge_etal_nsr_20}%
  \BibitemOpen
  \bibfield  {author} {\bibinfo {author} {\bibfnamefont {J.}~\bibnamefont
  {Ge}}, \bibinfo {author} {\bibfnamefont {Y.}~\bibnamefont {Liu}}, \bibinfo
  {author} {\bibfnamefont {J.}~\bibnamefont {Li}}, \bibinfo {author}
  {\bibfnamefont {H.}~\bibnamefont {Li}}, \bibinfo {author} {\bibfnamefont
  {T.}~\bibnamefont {Luo}}, \bibinfo {author} {\bibfnamefont {Y.}~\bibnamefont
  {Wu}}, \bibinfo {author} {\bibfnamefont {Y.}~\bibnamefont {Xu}},\ and\
  \bibinfo {author} {\bibfnamefont {J.}~\bibnamefont {Wang}},\ }\bibfield
  {title} {\bibinfo {title} {{High-Chern-number and high-temperature quantum
  Hall effect without Landau levels}},\ }\href
  {https://doi.org/10.1093/nsr/nwaa089} {\bibfield  {journal} {\bibinfo
  {journal} {National Science Review}\ }\textbf {\bibinfo {volume} {7}},\
  \bibinfo {pages} {1280} (\bibinfo {year} {2020})}\BibitemShut {NoStop}%
\bibitem [{\citenamefont {Riberolles}\ \emph {et~al.}(2021)\citenamefont
  {Riberolles}, \citenamefont {Zhang}, \citenamefont {Gordon}, \citenamefont
  {Butch}, \citenamefont {Ke}, \citenamefont {Yan},\ and\ \citenamefont
  {McQueeney}}]{riberolles_etal_prb_21}%
  \BibitemOpen
  \bibfield  {author} {\bibinfo {author} {\bibfnamefont {S.~X.~M.}\
  \bibnamefont {Riberolles}}, \bibinfo {author} {\bibfnamefont
  {Q.}~\bibnamefont {Zhang}}, \bibinfo {author} {\bibfnamefont
  {E.}~\bibnamefont {Gordon}}, \bibinfo {author} {\bibfnamefont {N.~P.}\
  \bibnamefont {Butch}}, \bibinfo {author} {\bibfnamefont {L.}~\bibnamefont
  {Ke}}, \bibinfo {author} {\bibfnamefont {J.-Q.}\ \bibnamefont {Yan}},\ and\
  \bibinfo {author} {\bibfnamefont {R.~J.}\ \bibnamefont {McQueeney}},\
  }\bibfield  {title} {\bibinfo {title} {Evolution of magnetic interactions in
  sb-substituted ${\mathrm{mnbi}}_{2}{\mathrm{te}}_{4}$},\ }\href
  {https://doi.org/10.1103/PhysRevB.104.064401} {\bibfield  {journal} {\bibinfo
   {journal} {Phys. Rev. B}\ }\textbf {\bibinfo {volume} {104}},\ \bibinfo
  {pages} {064401} (\bibinfo {year} {2021})}\BibitemShut {NoStop}%
\bibitem [{\citenamefont {Yan}\ \emph {et~al.}(2021)\citenamefont {Yan},
  \citenamefont {Fernandez-Mulligan}, \citenamefont {Mei}, \citenamefont {Lee},
  \citenamefont {Protic}, \citenamefont {Fukumori}, \citenamefont {Yan},
  \citenamefont {Liu}, \citenamefont {Mao},\ and\ \citenamefont
  {Yang}}]{yan_etal_prb_21}%
  \BibitemOpen
  \bibfield  {author} {\bibinfo {author} {\bibfnamefont {C.}~\bibnamefont
  {Yan}}, \bibinfo {author} {\bibfnamefont {S.}~\bibnamefont
  {Fernandez-Mulligan}}, \bibinfo {author} {\bibfnamefont {R.}~\bibnamefont
  {Mei}}, \bibinfo {author} {\bibfnamefont {S.~H.}\ \bibnamefont {Lee}},
  \bibinfo {author} {\bibfnamefont {N.}~\bibnamefont {Protic}}, \bibinfo
  {author} {\bibfnamefont {R.}~\bibnamefont {Fukumori}}, \bibinfo {author}
  {\bibfnamefont {B.}~\bibnamefont {Yan}}, \bibinfo {author} {\bibfnamefont
  {C.}~\bibnamefont {Liu}}, \bibinfo {author} {\bibfnamefont {Z.}~\bibnamefont
  {Mao}},\ and\ \bibinfo {author} {\bibfnamefont {S.}~\bibnamefont {Yang}},\
  }\bibfield  {title} {\bibinfo {title} {Origins of electronic bands in the
  antiferromagnetic topological insulator
  ${\mathrm{mnbi}}_{2}{\mathrm{te}}_{4}$},\ }\href
  {https://doi.org/10.1103/PhysRevB.104.L041102} {\bibfield  {journal}
  {\bibinfo  {journal} {Phys. Rev. B}\ }\textbf {\bibinfo {volume} {104}},\
  \bibinfo {pages} {L041102} (\bibinfo {year} {2021})}\BibitemShut {NoStop}%
\bibitem [{\citenamefont {Fukasawa}\ \emph {et~al.}(2021)\citenamefont
  {Fukasawa}, \citenamefont {Kusaka}, \citenamefont {Sumida}, \citenamefont
  {Hashizume}, \citenamefont {Ichinokura}, \citenamefont {Takeda},
  \citenamefont {Ideta}, \citenamefont {Tanaka}, \citenamefont {Shimizu},
  \citenamefont {Hitosugi},\ and\ \citenamefont
  {Hirahara}}]{fukasawa_etal_prb_21}%
  \BibitemOpen
  \bibfield  {author} {\bibinfo {author} {\bibfnamefont {T.}~\bibnamefont
  {Fukasawa}}, \bibinfo {author} {\bibfnamefont {S.}~\bibnamefont {Kusaka}},
  \bibinfo {author} {\bibfnamefont {K.}~\bibnamefont {Sumida}}, \bibinfo
  {author} {\bibfnamefont {M.}~\bibnamefont {Hashizume}}, \bibinfo {author}
  {\bibfnamefont {S.}~\bibnamefont {Ichinokura}}, \bibinfo {author}
  {\bibfnamefont {Y.}~\bibnamefont {Takeda}}, \bibinfo {author} {\bibfnamefont
  {S.}~\bibnamefont {Ideta}}, \bibinfo {author} {\bibfnamefont
  {K.}~\bibnamefont {Tanaka}}, \bibinfo {author} {\bibfnamefont
  {R.}~\bibnamefont {Shimizu}}, \bibinfo {author} {\bibfnamefont
  {T.}~\bibnamefont {Hitosugi}},\ and\ \bibinfo {author} {\bibfnamefont
  {T.}~\bibnamefont {Hirahara}},\ }\bibfield  {title} {\bibinfo {title}
  {Absence of ferromagnetism in
  ${\mathrm{mnbi}}_{2}{\mathrm{te}}_{4}/{\mathrm{bi}}_{2}{\mathrm{te}}_{3}$
  down to 6 k},\ }\href {https://doi.org/10.1103/PhysRevB.103.205405}
  {\bibfield  {journal} {\bibinfo  {journal} {Phys. Rev. B}\ }\textbf {\bibinfo
  {volume} {103}},\ \bibinfo {pages} {205405} (\bibinfo {year}
  {2021})}\BibitemShut {NoStop}%
\bibitem [{\citenamefont {Xu}\ \emph {et~al.}(2021)\citenamefont {Xu},
  \citenamefont {Zhang}, \citenamefont {Alizade}, \citenamefont {Jahangirli},
  \citenamefont {Lyzwa}, \citenamefont {Sheveleva}, \citenamefont {Marsik},
  \citenamefont {Li}, \citenamefont {Yao}, \citenamefont {Wang}, \citenamefont
  {Shen}, \citenamefont {Dai}, \citenamefont {Kataev}, \citenamefont {Otrokov},
  \citenamefont {Chulkov}, \citenamefont {Mamedov},\ and\ \citenamefont
  {Bernhard}}]{xu_etal_prb_21}%
  \BibitemOpen
  \bibfield  {author} {\bibinfo {author} {\bibfnamefont {B.}~\bibnamefont
  {Xu}}, \bibinfo {author} {\bibfnamefont {Y.}~\bibnamefont {Zhang}}, \bibinfo
  {author} {\bibfnamefont {E.~H.}\ \bibnamefont {Alizade}}, \bibinfo {author}
  {\bibfnamefont {Z.~A.}\ \bibnamefont {Jahangirli}}, \bibinfo {author}
  {\bibfnamefont {F.}~\bibnamefont {Lyzwa}}, \bibinfo {author} {\bibfnamefont
  {E.}~\bibnamefont {Sheveleva}}, \bibinfo {author} {\bibfnamefont
  {P.}~\bibnamefont {Marsik}}, \bibinfo {author} {\bibfnamefont {Y.~K.}\
  \bibnamefont {Li}}, \bibinfo {author} {\bibfnamefont {Y.~G.}\ \bibnamefont
  {Yao}}, \bibinfo {author} {\bibfnamefont {Z.~W.}\ \bibnamefont {Wang}},
  \bibinfo {author} {\bibfnamefont {B.}~\bibnamefont {Shen}}, \bibinfo {author}
  {\bibfnamefont {Y.~M.}\ \bibnamefont {Dai}}, \bibinfo {author} {\bibfnamefont
  {V.}~\bibnamefont {Kataev}}, \bibinfo {author} {\bibfnamefont {M.~M.}\
  \bibnamefont {Otrokov}}, \bibinfo {author} {\bibfnamefont {E.~V.}\
  \bibnamefont {Chulkov}}, \bibinfo {author} {\bibfnamefont {N.~T.}\
  \bibnamefont {Mamedov}},\ and\ \bibinfo {author} {\bibfnamefont
  {C.}~\bibnamefont {Bernhard}},\ }\bibfield  {title} {\bibinfo {title}
  {Infrared study of the multiband low-energy excitations of the topological
  antiferromagnet ${\mathrm{mnbi}}_{2}{\mathrm{te}}_{4}$},\ }\href
  {https://doi.org/10.1103/PhysRevB.103.L121103} {\bibfield  {journal}
  {\bibinfo  {journal} {Phys. Rev. B}\ }\textbf {\bibinfo {volume} {103}},\
  \bibinfo {pages} {L121103} (\bibinfo {year} {2021})}\BibitemShut {NoStop}%
\bibitem [{\citenamefont {Sitte}\ \emph {et~al.}(2012)\citenamefont {Sitte},
  \citenamefont {Rosch}, \citenamefont {Altman},\ and\ \citenamefont
  {Fritz}}]{sitte_etal_prl_12}%
  \BibitemOpen
  \bibfield  {author} {\bibinfo {author} {\bibfnamefont {M.}~\bibnamefont
  {Sitte}}, \bibinfo {author} {\bibfnamefont {A.}~\bibnamefont {Rosch}},
  \bibinfo {author} {\bibfnamefont {E.}~\bibnamefont {Altman}},\ and\ \bibinfo
  {author} {\bibfnamefont {L.}~\bibnamefont {Fritz}},\ }\bibfield  {title}
  {\bibinfo {title} {Topological insulators in magnetic fields: Quantum hall
  effect and edge channels with a nonquantized $\ensuremath{\theta}$ term},\
  }\href {https://doi.org/10.1103/PhysRevLett.108.126807} {\bibfield  {journal}
  {\bibinfo  {journal} {Phys. Rev. Lett.}\ }\textbf {\bibinfo {volume} {108}},\
  \bibinfo {pages} {126807} (\bibinfo {year} {2012})}\BibitemShut {NoStop}%
\bibitem [{\citenamefont {Tang}\ \emph {et~al.}(2019)\citenamefont {Tang},
  \citenamefont {Ren}, \citenamefont {Wang}, \citenamefont {Zhong},
  \citenamefont {Schneeloch}, \citenamefont {Yang}, \citenamefont {Yang},
  \citenamefont {Lee}, \citenamefont {Gu}, \citenamefont {Qiao},\ and\
  \citenamefont {Zhang}}]{tang_etal_nature_19}%
  \BibitemOpen
  \bibfield  {author} {\bibinfo {author} {\bibfnamefont {F.}~\bibnamefont
  {Tang}}, \bibinfo {author} {\bibfnamefont {Y.}~\bibnamefont {Ren}}, \bibinfo
  {author} {\bibfnamefont {P.}~\bibnamefont {Wang}}, \bibinfo {author}
  {\bibfnamefont {R.}~\bibnamefont {Zhong}}, \bibinfo {author} {\bibfnamefont
  {J.}~\bibnamefont {Schneeloch}}, \bibinfo {author} {\bibfnamefont {S.~A.}\
  \bibnamefont {Yang}}, \bibinfo {author} {\bibfnamefont {K.}~\bibnamefont
  {Yang}}, \bibinfo {author} {\bibfnamefont {P.~A.}\ \bibnamefont {Lee}},
  \bibinfo {author} {\bibfnamefont {G.}~\bibnamefont {Gu}}, \bibinfo {author}
  {\bibfnamefont {Z.}~\bibnamefont {Qiao}},\ and\ \bibinfo {author}
  {\bibfnamefont {L.}~\bibnamefont {Zhang}},\ }\bibfield  {title} {\bibinfo
  {title} {Three-dimensional quantum hall effect and metal--insulator
  transition in zrte5},\ }\href {https://doi.org/10.1038/s41586-019-1180-9}
  {\bibfield  {journal} {\bibinfo  {journal} {Nature}\ }\textbf {\bibinfo
  {volume} {569}},\ \bibinfo {pages} {537} (\bibinfo {year}
  {2019})}\BibitemShut {NoStop}%
\bibitem [{\citenamefont {Xu}\ \emph {et~al.}(2014)\citenamefont {Xu},
  \citenamefont {Miotkowski}, \citenamefont {Liu}, \citenamefont {Tian},
  \citenamefont {Nam}, \citenamefont {Alidoust}, \citenamefont {Hu},
  \citenamefont {Shih}, \citenamefont {Hasan},\ and\ \citenamefont
  {Chen}}]{xu_etal_ntp_14}%
  \BibitemOpen
  \bibfield  {author} {\bibinfo {author} {\bibfnamefont {Y.}~\bibnamefont
  {Xu}}, \bibinfo {author} {\bibfnamefont {I.}~\bibnamefont {Miotkowski}},
  \bibinfo {author} {\bibfnamefont {C.}~\bibnamefont {Liu}}, \bibinfo {author}
  {\bibfnamefont {J.}~\bibnamefont {Tian}}, \bibinfo {author} {\bibfnamefont
  {H.}~\bibnamefont {Nam}}, \bibinfo {author} {\bibfnamefont {N.}~\bibnamefont
  {Alidoust}}, \bibinfo {author} {\bibfnamefont {J.}~\bibnamefont {Hu}},
  \bibinfo {author} {\bibfnamefont {C.-K.}\ \bibnamefont {Shih}}, \bibinfo
  {author} {\bibfnamefont {M.~Z.}\ \bibnamefont {Hasan}},\ and\ \bibinfo
  {author} {\bibfnamefont {Y.~P.}\ \bibnamefont {Chen}},\ }\bibfield  {title}
  {\bibinfo {title} {Observation of topological surface state quantum hall
  effect in an intrinsic three-dimensional topological insulator},\ }\href
  {https://doi.org/10.1038/nphys3140} {\bibfield  {journal} {\bibinfo
  {journal} {Nature Physics}\ }\textbf {\bibinfo {volume} {10}},\ \bibinfo
  {pages} {956} (\bibinfo {year} {2014})}\BibitemShut {NoStop}%
\bibitem [{\citenamefont {Schumann}\ \emph {et~al.}(2018)\citenamefont
  {Schumann}, \citenamefont {Galletti}, \citenamefont {Kealhofer},
  \citenamefont {Kim}, \citenamefont {Goyal},\ and\ \citenamefont
  {Stemmer}}]{schumann_etal_prl_18}%
  \BibitemOpen
  \bibfield  {author} {\bibinfo {author} {\bibfnamefont {T.}~\bibnamefont
  {Schumann}}, \bibinfo {author} {\bibfnamefont {L.}~\bibnamefont {Galletti}},
  \bibinfo {author} {\bibfnamefont {D.~A.}\ \bibnamefont {Kealhofer}}, \bibinfo
  {author} {\bibfnamefont {H.}~\bibnamefont {Kim}}, \bibinfo {author}
  {\bibfnamefont {M.}~\bibnamefont {Goyal}},\ and\ \bibinfo {author}
  {\bibfnamefont {S.}~\bibnamefont {Stemmer}},\ }\bibfield  {title} {\bibinfo
  {title} {Observation of the quantum hall effect in confined films of the
  three-dimensional dirac semimetal ${\mathrm{cd}}_{3}{\mathrm{as}}_{2}$},\
  }\href {https://doi.org/10.1103/PhysRevLett.120.016801} {\bibfield  {journal}
  {\bibinfo  {journal} {Phys. Rev. Lett.}\ }\textbf {\bibinfo {volume} {120}},\
  \bibinfo {pages} {016801} (\bibinfo {year} {2018})}\BibitemShut {NoStop}%
\bibitem [{\citenamefont {Zhang}\ \emph {et~al.}(2019)\citenamefont {Zhang},
  \citenamefont {Zhang}, \citenamefont {Yuan}, \citenamefont {Lu},
  \citenamefont {Zhang}, \citenamefont {Narayan}, \citenamefont {Liu},
  \citenamefont {Zhang}, \citenamefont {Ni}, \citenamefont {Liu}, \citenamefont
  {Choi}, \citenamefont {Suslov}, \citenamefont {Sanvito}, \citenamefont {Pi},
  \citenamefont {Lu}, \citenamefont {Potter},\ and\ \citenamefont
  {Xiu}}]{zhang_etal_nature_19}%
  \BibitemOpen
  \bibfield  {author} {\bibinfo {author} {\bibfnamefont {C.}~\bibnamefont
  {Zhang}}, \bibinfo {author} {\bibfnamefont {Y.}~\bibnamefont {Zhang}},
  \bibinfo {author} {\bibfnamefont {X.}~\bibnamefont {Yuan}}, \bibinfo {author}
  {\bibfnamefont {S.}~\bibnamefont {Lu}}, \bibinfo {author} {\bibfnamefont
  {J.}~\bibnamefont {Zhang}}, \bibinfo {author} {\bibfnamefont
  {A.}~\bibnamefont {Narayan}}, \bibinfo {author} {\bibfnamefont
  {Y.}~\bibnamefont {Liu}}, \bibinfo {author} {\bibfnamefont {H.}~\bibnamefont
  {Zhang}}, \bibinfo {author} {\bibfnamefont {Z.}~\bibnamefont {Ni}}, \bibinfo
  {author} {\bibfnamefont {R.}~\bibnamefont {Liu}}, \bibinfo {author}
  {\bibfnamefont {E.~S.}\ \bibnamefont {Choi}}, \bibinfo {author}
  {\bibfnamefont {A.}~\bibnamefont {Suslov}}, \bibinfo {author} {\bibfnamefont
  {S.}~\bibnamefont {Sanvito}}, \bibinfo {author} {\bibfnamefont
  {L.}~\bibnamefont {Pi}}, \bibinfo {author} {\bibfnamefont {H.-Z.}\
  \bibnamefont {Lu}}, \bibinfo {author} {\bibfnamefont {A.~C.}\ \bibnamefont
  {Potter}},\ and\ \bibinfo {author} {\bibfnamefont {F.}~\bibnamefont {Xiu}},\
  }\bibfield  {title} {\bibinfo {title} {Quantum hall effect based on weyl
  orbits in cd3as2},\ }\href {https://doi.org/10.1038/s41586-018-0798-3}
  {\bibfield  {journal} {\bibinfo  {journal} {Nature}\ }\textbf {\bibinfo
  {volume} {565}},\ \bibinfo {pages} {331} (\bibinfo {year}
  {2019})}\BibitemShut {NoStop}%
\bibitem [{\citenamefont {Benalcazar}\ \emph
  {et~al.}(2017{\natexlab{a}})\citenamefont {Benalcazar}, \citenamefont
  {Bernevig},\ and\ \citenamefont {Hughes}}]{benalcazar_etal_science_17}%
  \BibitemOpen
  \bibfield  {author} {\bibinfo {author} {\bibfnamefont {W.~A.}\ \bibnamefont
  {Benalcazar}}, \bibinfo {author} {\bibfnamefont {B.~A.}\ \bibnamefont
  {Bernevig}},\ and\ \bibinfo {author} {\bibfnamefont {T.~L.}\ \bibnamefont
  {Hughes}},\ }\bibfield  {title} {\bibinfo {title} {Quantized electric
  multipole insulators},\ }\href@noop {} {\bibfield  {journal} {\bibinfo
  {journal} {Science}\ }\textbf {\bibinfo {volume} {357}},\ \bibinfo {pages}
  {61} (\bibinfo {year} {2017}{\natexlab{a}})}\BibitemShut {NoStop}%
\bibitem [{\citenamefont {Benalcazar}\ \emph
  {et~al.}(2017{\natexlab{b}})\citenamefont {Benalcazar}, \citenamefont
  {Bernevig},\ and\ \citenamefont {Hughes}}]{benalcazar_etal_prb_17}%
  \BibitemOpen
  \bibfield  {author} {\bibinfo {author} {\bibfnamefont {W.~A.}\ \bibnamefont
  {Benalcazar}}, \bibinfo {author} {\bibfnamefont {B.~A.}\ \bibnamefont
  {Bernevig}},\ and\ \bibinfo {author} {\bibfnamefont {T.~L.}\ \bibnamefont
  {Hughes}},\ }\bibfield  {title} {\bibinfo {title} {Electric multipole
  moments, topological multipole moment pumping, and chiral hinge states in
  crystalline insulators},\ }\href@noop {} {\bibfield  {journal} {\bibinfo
  {journal} {Phys. Rev. B}\ }\textbf {\bibinfo {volume} {96}},\ \bibinfo
  {pages} {245115} (\bibinfo {year} {2017}{\natexlab{b}})}\BibitemShut
  {NoStop}%
\bibitem [{\citenamefont {Song}\ \emph {et~al.}(2017)\citenamefont {Song},
  \citenamefont {Fang},\ and\ \citenamefont {Fang}}]{song_etal_prl_17}%
  \BibitemOpen
  \bibfield  {author} {\bibinfo {author} {\bibfnamefont {Z.}~\bibnamefont
  {Song}}, \bibinfo {author} {\bibfnamefont {Z.}~\bibnamefont {Fang}},\ and\
  \bibinfo {author} {\bibfnamefont {C.}~\bibnamefont {Fang}},\ }\bibfield
  {title} {\bibinfo {title} {$(d\ensuremath{-}2)$-dimensional edge states of
  rotation symmetry protected topological states},\ }\href@noop {} {\bibfield
  {journal} {\bibinfo  {journal} {Phys. Rev. Lett.}\ }\textbf {\bibinfo
  {volume} {119}},\ \bibinfo {pages} {246402} (\bibinfo {year}
  {2017})}\BibitemShut {NoStop}%
\bibitem [{\citenamefont {Langbehn}\ \emph {et~al.}(2017)\citenamefont
  {Langbehn}, \citenamefont {Peng}, \citenamefont {Trifunovic}, \citenamefont
  {von Oppen},\ and\ \citenamefont {Brouwer}}]{langbehn_etal_prl_17}%
  \BibitemOpen
  \bibfield  {author} {\bibinfo {author} {\bibfnamefont {J.}~\bibnamefont
  {Langbehn}}, \bibinfo {author} {\bibfnamefont {Y.}~\bibnamefont {Peng}},
  \bibinfo {author} {\bibfnamefont {L.}~\bibnamefont {Trifunovic}}, \bibinfo
  {author} {\bibfnamefont {F.}~\bibnamefont {von Oppen}},\ and\ \bibinfo
  {author} {\bibfnamefont {P.~W.}\ \bibnamefont {Brouwer}},\ }\bibfield
  {title} {\bibinfo {title} {Reflection-symmetric second-order topological
  insulators and superconductors},\ }\href@noop {} {\bibfield  {journal}
  {\bibinfo  {journal} {Phys. Rev. Lett.}\ }\textbf {\bibinfo {volume} {119}},\
  \bibinfo {pages} {246401} (\bibinfo {year} {2017})}\BibitemShut {NoStop}%
\bibitem [{\citenamefont {Geier}\ \emph {et~al.}(2018)\citenamefont {Geier},
  \citenamefont {Trifunovic}, \citenamefont {Hoskam},\ and\ \citenamefont
  {Brouwer}}]{geier_etal_prb_18}%
  \BibitemOpen
  \bibfield  {author} {\bibinfo {author} {\bibfnamefont {M.}~\bibnamefont
  {Geier}}, \bibinfo {author} {\bibfnamefont {L.}~\bibnamefont {Trifunovic}},
  \bibinfo {author} {\bibfnamefont {M.}~\bibnamefont {Hoskam}},\ and\ \bibinfo
  {author} {\bibfnamefont {P.~W.}\ \bibnamefont {Brouwer}},\ }\bibfield
  {title} {\bibinfo {title} {Second-order topological insulators and
  superconductors with an order-two crystalline symmetry},\ }\href@noop {}
  {\bibfield  {journal} {\bibinfo  {journal} {Phys. Rev. B}\ }\textbf {\bibinfo
  {volume} {97}},\ \bibinfo {pages} {205135} (\bibinfo {year}
  {2018})}\BibitemShut {NoStop}%
\bibitem [{\citenamefont {Imhof}\ \emph {et~al.}(2018)\citenamefont {Imhof},
  \citenamefont {Berger}, \citenamefont {Bayer}, \citenamefont {Brehm},
  \citenamefont {Molenkamp}, \citenamefont {Kiessling}, \citenamefont
  {Schindler}, \citenamefont {Lee}, \citenamefont {Greiter}, \citenamefont
  {Neupert},\ and\ \citenamefont {Thomale}}]{imhof_etal_ntp_18}%
  \BibitemOpen
  \bibfield  {author} {\bibinfo {author} {\bibfnamefont {S.}~\bibnamefont
  {Imhof}}, \bibinfo {author} {\bibfnamefont {C.}~\bibnamefont {Berger}},
  \bibinfo {author} {\bibfnamefont {F.}~\bibnamefont {Bayer}}, \bibinfo
  {author} {\bibfnamefont {J.}~\bibnamefont {Brehm}}, \bibinfo {author}
  {\bibfnamefont {L.~W.}\ \bibnamefont {Molenkamp}}, \bibinfo {author}
  {\bibfnamefont {T.}~\bibnamefont {Kiessling}}, \bibinfo {author}
  {\bibfnamefont {F.}~\bibnamefont {Schindler}}, \bibinfo {author}
  {\bibfnamefont {C.~H.}\ \bibnamefont {Lee}}, \bibinfo {author} {\bibfnamefont
  {M.}~\bibnamefont {Greiter}}, \bibinfo {author} {\bibfnamefont
  {T.}~\bibnamefont {Neupert}},\ and\ \bibinfo {author} {\bibfnamefont
  {R.}~\bibnamefont {Thomale}},\ }\bibfield  {title} {\bibinfo {title}
  {Topolectrical-circuit realization of topological corner modes},\ }\href@noop
  {} {\bibfield  {journal} {\bibinfo  {journal} {Nature Physics}\ }\textbf
  {\bibinfo {volume} {14}},\ \bibinfo {pages} {925} (\bibinfo {year}
  {2018})}\BibitemShut {NoStop}%
\bibitem [{\citenamefont {Schindler}\ \emph {et~al.}(2018)\citenamefont
  {Schindler}, \citenamefont {Cook}, \citenamefont {Vergniory}, \citenamefont
  {Wang}, \citenamefont {Parkin}, \citenamefont {Bernevig},\ and\ \citenamefont
  {Neupert}}]{schindler_etal_sca_18}%
  \BibitemOpen
  \bibfield  {author} {\bibinfo {author} {\bibfnamefont {F.}~\bibnamefont
  {Schindler}}, \bibinfo {author} {\bibfnamefont {A.~M.}\ \bibnamefont {Cook}},
  \bibinfo {author} {\bibfnamefont {M.~G.}\ \bibnamefont {Vergniory}}, \bibinfo
  {author} {\bibfnamefont {Z.}~\bibnamefont {Wang}}, \bibinfo {author}
  {\bibfnamefont {S.~S.~P.}\ \bibnamefont {Parkin}}, \bibinfo {author}
  {\bibfnamefont {B.~A.}\ \bibnamefont {Bernevig}},\ and\ \bibinfo {author}
  {\bibfnamefont {T.}~\bibnamefont {Neupert}},\ }\bibfield  {title} {\bibinfo
  {title} {Higher-order topological insulators},\ }\href@noop {} {\bibfield
  {journal} {\bibinfo  {journal} {Science Advances}\ }\textbf {\bibinfo
  {volume} {4}},\ \bibinfo {pages} {eaat0346} (\bibinfo {year}
  {2018})}\BibitemShut {NoStop}%
\bibitem [{\citenamefont {Trifunovic}\ and\ \citenamefont
  {Brouwer}(2019)}]{trifunovic_brouwer_prx_19}%
  \BibitemOpen
  \bibfield  {author} {\bibinfo {author} {\bibfnamefont {L.}~\bibnamefont
  {Trifunovic}}\ and\ \bibinfo {author} {\bibfnamefont {P.~W.}\ \bibnamefont
  {Brouwer}},\ }\bibfield  {title} {\bibinfo {title} {Higher-order
  bulk-boundary correspondence for topological crystalline phases},\
  }\href@noop {} {\bibfield  {journal} {\bibinfo  {journal} {Phys. Rev. X}\
  }\textbf {\bibinfo {volume} {9}},\ \bibinfo {pages} {011012} (\bibinfo {year}
  {2019})}\BibitemShut {NoStop}%
\bibitem [{\citenamefont {Fu}\ \emph {et~al.}(2021)\citenamefont {Fu},
  \citenamefont {Hu},\ and\ \citenamefont {Shen}}]{fu_etal_prr_21}%
  \BibitemOpen
  \bibfield  {author} {\bibinfo {author} {\bibfnamefont {B.}~\bibnamefont
  {Fu}}, \bibinfo {author} {\bibfnamefont {Z.-A.}\ \bibnamefont {Hu}},\ and\
  \bibinfo {author} {\bibfnamefont {S.-Q.}\ \bibnamefont {Shen}},\ }\bibfield
  {title} {\bibinfo {title} {Bulk-hinge correspondence and three-dimensional
  quantum anomalous hall effect in second-order topological insulators},\
  }\href {https://doi.org/10.1103/PhysRevResearch.3.033177} {\bibfield
  {journal} {\bibinfo  {journal} {Phys. Rev. Research}\ }\textbf {\bibinfo
  {volume} {3}},\ \bibinfo {pages} {033177} (\bibinfo {year}
  {2021})}\BibitemShut {NoStop}%
\bibitem [{\citenamefont {Petrides}\ and\ \citenamefont
  {Zilberberg}(2020)}]{petridis_zilberberg_prr_20}%
  \BibitemOpen
  \bibfield  {author} {\bibinfo {author} {\bibfnamefont {I.}~\bibnamefont
  {Petrides}}\ and\ \bibinfo {author} {\bibfnamefont {O.}~\bibnamefont
  {Zilberberg}},\ }\bibfield  {title} {\bibinfo {title} {Higher-order
  topological insulators, topological pumps and the quantum hall effect in high
  dimensions},\ }\href {https://doi.org/10.1103/PhysRevResearch.2.022049}
  {\bibfield  {journal} {\bibinfo  {journal} {Phys. Rev. Research}\ }\textbf
  {\bibinfo {volume} {2}},\ \bibinfo {pages} {022049} (\bibinfo {year}
  {2020})}\BibitemShut {NoStop}%
\bibitem [{\citenamefont {Volpez}\ \emph {et~al.}(2019)\citenamefont {Volpez},
  \citenamefont {Loss},\ and\ \citenamefont {Klinovaja}}]{volpez_etal_prl_19}%
  \BibitemOpen
  \bibfield  {author} {\bibinfo {author} {\bibfnamefont {Y.}~\bibnamefont
  {Volpez}}, \bibinfo {author} {\bibfnamefont {D.}~\bibnamefont {Loss}},\ and\
  \bibinfo {author} {\bibfnamefont {J.}~\bibnamefont {Klinovaja}},\ }\bibfield
  {title} {\bibinfo {title} {Second-order topological superconductivity in
  $\ensuremath{\pi}$-junction rashba layers},\ }\href@noop {} {\bibfield
  {journal} {\bibinfo  {journal} {Phys. Rev. Lett.}\ }\textbf {\bibinfo
  {volume} {122}},\ \bibinfo {pages} {126402} (\bibinfo {year}
  {2019})}\BibitemShut {NoStop}%
\bibitem [{\citenamefont {Laubscher}\ \emph {et~al.}(2019)\citenamefont
  {Laubscher}, \citenamefont {Loss},\ and\ \citenamefont
  {Klinovaja}}]{laubscher_etal_prr_19}%
  \BibitemOpen
  \bibfield  {author} {\bibinfo {author} {\bibfnamefont {K.}~\bibnamefont
  {Laubscher}}, \bibinfo {author} {\bibfnamefont {D.}~\bibnamefont {Loss}},\
  and\ \bibinfo {author} {\bibfnamefont {J.}~\bibnamefont {Klinovaja}},\
  }\bibfield  {title} {\bibinfo {title} {Fractional topological
  superconductivity and parafermion corner states},\ }\href@noop {} {\bibfield
  {journal} {\bibinfo  {journal} {Phys. Rev. Research}\ }\textbf {\bibinfo
  {volume} {1}},\ \bibinfo {pages} {032017} (\bibinfo {year}
  {2019})}\BibitemShut {NoStop}%
\bibitem [{\citenamefont {Fu}\ \emph {et~al.}(2007)\citenamefont {Fu},
  \citenamefont {Kane},\ and\ \citenamefont {Mele}}]{fu_etal_prl_07}%
  \BibitemOpen
  \bibfield  {author} {\bibinfo {author} {\bibfnamefont {L.}~\bibnamefont
  {Fu}}, \bibinfo {author} {\bibfnamefont {C.~L.}\ \bibnamefont {Kane}},\ and\
  \bibinfo {author} {\bibfnamefont {E.~J.}\ \bibnamefont {Mele}},\ }\bibfield
  {title} {\bibinfo {title} {Topological insulators in three dimensions},\
  }\href {https://doi.org/10.1103/PhysRevLett.98.106803} {\bibfield  {journal}
  {\bibinfo  {journal} {Phys. Rev. Lett.}\ }\textbf {\bibinfo {volume} {98}},\
  \bibinfo {pages} {106803} (\bibinfo {year} {2007})}\BibitemShut {NoStop}%
\bibitem [{\citenamefont {Xu}\ \emph {et~al.}(2012)\citenamefont {Xu},
  \citenamefont {Neupane}, \citenamefont {Liu}, \citenamefont {Zhang},
  \citenamefont {Richardella}, \citenamefont {Andrew~Wray}, \citenamefont
  {Alidoust}, \citenamefont {Leandersson}, \citenamefont {Balasubramanian},
  \citenamefont {S{\'a}nchez-Barriga}, \citenamefont {Rader}, \citenamefont
  {Landolt}, \citenamefont {Slomski}, \citenamefont {Hugo~Dil}, \citenamefont
  {Osterwalder}, \citenamefont {Chang}, \citenamefont {Jeng}, \citenamefont
  {Lin}, \citenamefont {Bansil}, \citenamefont {Samarth},\ and\ \citenamefont
  {Zahid~Hasan}}]{xu_etal_ntp_12}%
  \BibitemOpen
  \bibfield  {author} {\bibinfo {author} {\bibfnamefont {S.-Y.}\ \bibnamefont
  {Xu}}, \bibinfo {author} {\bibfnamefont {M.}~\bibnamefont {Neupane}},
  \bibinfo {author} {\bibfnamefont {C.}~\bibnamefont {Liu}}, \bibinfo {author}
  {\bibfnamefont {D.}~\bibnamefont {Zhang}}, \bibinfo {author} {\bibfnamefont
  {A.}~\bibnamefont {Richardella}}, \bibinfo {author} {\bibfnamefont
  {L.}~\bibnamefont {Andrew~Wray}}, \bibinfo {author} {\bibfnamefont
  {N.}~\bibnamefont {Alidoust}}, \bibinfo {author} {\bibfnamefont
  {M.}~\bibnamefont {Leandersson}}, \bibinfo {author} {\bibfnamefont
  {T.}~\bibnamefont {Balasubramanian}}, \bibinfo {author} {\bibfnamefont
  {J.}~\bibnamefont {S{\'a}nchez-Barriga}}, \bibinfo {author} {\bibfnamefont
  {O.}~\bibnamefont {Rader}}, \bibinfo {author} {\bibfnamefont
  {G.}~\bibnamefont {Landolt}}, \bibinfo {author} {\bibfnamefont
  {B.}~\bibnamefont {Slomski}}, \bibinfo {author} {\bibfnamefont
  {J.}~\bibnamefont {Hugo~Dil}}, \bibinfo {author} {\bibfnamefont
  {J.}~\bibnamefont {Osterwalder}}, \bibinfo {author} {\bibfnamefont {T.-R.}\
  \bibnamefont {Chang}}, \bibinfo {author} {\bibfnamefont {H.-T.}\ \bibnamefont
  {Jeng}}, \bibinfo {author} {\bibfnamefont {H.}~\bibnamefont {Lin}}, \bibinfo
  {author} {\bibfnamefont {A.}~\bibnamefont {Bansil}}, \bibinfo {author}
  {\bibfnamefont {N.}~\bibnamefont {Samarth}},\ and\ \bibinfo {author}
  {\bibfnamefont {M.}~\bibnamefont {Zahid~Hasan}},\ }\bibfield  {title}
  {\bibinfo {title} {Hedgehog spin texture and berry's phase tuning in a
  magnetic topological insulator},\ }\href {https://doi.org/10.1038/nphys2351}
  {\bibfield  {journal} {\bibinfo  {journal} {Nature Physics}\ }\textbf
  {\bibinfo {volume} {8}},\ \bibinfo {pages} {616} (\bibinfo {year}
  {2012})}\BibitemShut {NoStop}%
\bibitem [{\citenamefont {Rienks}\ \emph {et~al.}(2019)\citenamefont {Rienks},
  \citenamefont {Wimmer}, \citenamefont {S{\'a}nchez-Barriga}, \citenamefont
  {Caha}, \citenamefont {Mandal}, \citenamefont {R{\r{u}}{\v{z}}i{\v{c}}ka},
  \citenamefont {Ney}, \citenamefont {Steiner}, \citenamefont {Volobuev},
  \citenamefont {Groiss}, \citenamefont {Albu}, \citenamefont {Kothleitner},
  \citenamefont {Michali{\v{c}}ka}, \citenamefont {Khan}, \citenamefont
  {Min{\'a}r}, \citenamefont {Ebert}, \citenamefont {Bauer}, \citenamefont
  {Freyse}, \citenamefont {Varykhalov}, \citenamefont {Rader},\ and\
  \citenamefont {Springholz}}]{rienks_etal_nature_19}%
  \BibitemOpen
  \bibfield  {author} {\bibinfo {author} {\bibfnamefont {E.~D.~L.}\
  \bibnamefont {Rienks}}, \bibinfo {author} {\bibfnamefont {S.}~\bibnamefont
  {Wimmer}}, \bibinfo {author} {\bibfnamefont {J.}~\bibnamefont
  {S{\'a}nchez-Barriga}}, \bibinfo {author} {\bibfnamefont {O.}~\bibnamefont
  {Caha}}, \bibinfo {author} {\bibfnamefont {P.~S.}\ \bibnamefont {Mandal}},
  \bibinfo {author} {\bibfnamefont {J.}~\bibnamefont
  {R{\r{u}}{\v{z}}i{\v{c}}ka}}, \bibinfo {author} {\bibfnamefont
  {A.}~\bibnamefont {Ney}}, \bibinfo {author} {\bibfnamefont {H.}~\bibnamefont
  {Steiner}}, \bibinfo {author} {\bibfnamefont {V.~V.}\ \bibnamefont
  {Volobuev}}, \bibinfo {author} {\bibfnamefont {H.}~\bibnamefont {Groiss}},
  \bibinfo {author} {\bibfnamefont {M.}~\bibnamefont {Albu}}, \bibinfo {author}
  {\bibfnamefont {G.}~\bibnamefont {Kothleitner}}, \bibinfo {author}
  {\bibfnamefont {J.}~\bibnamefont {Michali{\v{c}}ka}}, \bibinfo {author}
  {\bibfnamefont {S.~A.}\ \bibnamefont {Khan}}, \bibinfo {author}
  {\bibfnamefont {J.}~\bibnamefont {Min{\'a}r}}, \bibinfo {author}
  {\bibfnamefont {H.}~\bibnamefont {Ebert}}, \bibinfo {author} {\bibfnamefont
  {G.}~\bibnamefont {Bauer}}, \bibinfo {author} {\bibfnamefont
  {F.}~\bibnamefont {Freyse}}, \bibinfo {author} {\bibfnamefont
  {A.}~\bibnamefont {Varykhalov}}, \bibinfo {author} {\bibfnamefont
  {O.}~\bibnamefont {Rader}},\ and\ \bibinfo {author} {\bibfnamefont
  {G.}~\bibnamefont {Springholz}},\ }\bibfield  {title} {\bibinfo {title}
  {Large magnetic gap at the dirac point in bi2te3/mnbi2te4 heterostructures},\
  }\href@noop {} {\bibfield  {journal} {\bibinfo  {journal} {Nature}\ }\textbf
  {\bibinfo {volume} {576}},\ \bibinfo {pages} {423} (\bibinfo {year}
  {2019})}\BibitemShut {NoStop}%
\bibitem [{\citenamefont {Khalaf}(2018)}]{khalaf_prb_18}%
  \BibitemOpen
  \bibfield  {author} {\bibinfo {author} {\bibfnamefont {E.}~\bibnamefont
  {Khalaf}},\ }\bibfield  {title} {\bibinfo {title} {Higher-order topological
  insulators and superconductors protected by inversion symmetry},\ }\href@noop
  {} {\bibfield  {journal} {\bibinfo  {journal} {Phys. Rev. B}\ }\textbf
  {\bibinfo {volume} {97}},\ \bibinfo {pages} {205136} (\bibinfo {year}
  {2018})}\BibitemShut {NoStop}%
\bibitem [{\citenamefont {Plekhanov}\ \emph {et~al.}(2019)\citenamefont
  {Plekhanov}, \citenamefont {Thakurathi}, \citenamefont {Loss},\ and\
  \citenamefont {Klinovaja}}]{plekhanov_etal_prr_19}%
  \BibitemOpen
  \bibfield  {author} {\bibinfo {author} {\bibfnamefont {K.}~\bibnamefont
  {Plekhanov}}, \bibinfo {author} {\bibfnamefont {M.}~\bibnamefont
  {Thakurathi}}, \bibinfo {author} {\bibfnamefont {D.}~\bibnamefont {Loss}},\
  and\ \bibinfo {author} {\bibfnamefont {J.}~\bibnamefont {Klinovaja}},\
  }\bibfield  {title} {\bibinfo {title} {Floquet second-order topological
  superconductor driven via ferromagnetic resonance},\ }\href@noop {}
  {\bibfield  {journal} {\bibinfo  {journal} {Phys. Rev. Research}\ }\textbf
  {\bibinfo {volume} {1}},\ \bibinfo {pages} {032013} (\bibinfo {year}
  {2019})}\BibitemShut {NoStop}%
\bibitem [{\citenamefont {Ren}\ \emph {et~al.}(2020)\citenamefont {Ren},
  \citenamefont {Qiao},\ and\ \citenamefont {Niu}}]{ren_etal_prl_20}%
  \BibitemOpen
  \bibfield  {author} {\bibinfo {author} {\bibfnamefont {Y.}~\bibnamefont
  {Ren}}, \bibinfo {author} {\bibfnamefont {Z.}~\bibnamefont {Qiao}},\ and\
  \bibinfo {author} {\bibfnamefont {Q.}~\bibnamefont {Niu}},\ }\bibfield
  {title} {\bibinfo {title} {Engineering corner states from two-dimensional
  topological insulators},\ }\href@noop {} {\bibfield  {journal} {\bibinfo
  {journal} {Phys. Rev. Lett.}\ }\textbf {\bibinfo {volume} {124}},\ \bibinfo
  {pages} {166804} (\bibinfo {year} {2020})}\BibitemShut {NoStop}%
\bibitem [{\citenamefont {Laubscher}\ \emph
  {et~al.}(2020{\natexlab{a}})\citenamefont {Laubscher}, \citenamefont
  {Chughtai}, \citenamefont {Loss},\ and\ \citenamefont
  {Klinovaja}}]{laubscher_etal_prb_20}%
  \BibitemOpen
  \bibfield  {author} {\bibinfo {author} {\bibfnamefont {K.}~\bibnamefont
  {Laubscher}}, \bibinfo {author} {\bibfnamefont {D.}~\bibnamefont {Chughtai}},
  \bibinfo {author} {\bibfnamefont {D.}~\bibnamefont {Loss}},\ and\ \bibinfo
  {author} {\bibfnamefont {J.}~\bibnamefont {Klinovaja}},\ }\bibfield  {title}
  {\bibinfo {title} {Kramers pairs of majorana corner states in a topological
  insulator bilayer},\ }\href@noop {} {\bibfield  {journal} {\bibinfo
  {journal} {Phys. Rev. B}\ }\textbf {\bibinfo {volume} {102}},\ \bibinfo
  {pages} {195401} (\bibinfo {year} {2020}{\natexlab{a}})}\BibitemShut
  {NoStop}%
\bibitem [{\citenamefont {Laubscher}\ \emph
  {et~al.}(2020{\natexlab{b}})\citenamefont {Laubscher}, \citenamefont {Loss},\
  and\ \citenamefont {Klinovaja}}]{laubscher_etal_prr_20}%
  \BibitemOpen
  \bibfield  {author} {\bibinfo {author} {\bibfnamefont {K.}~\bibnamefont
  {Laubscher}}, \bibinfo {author} {\bibfnamefont {D.}~\bibnamefont {Loss}},\
  and\ \bibinfo {author} {\bibfnamefont {J.}~\bibnamefont {Klinovaja}},\
  }\bibfield  {title} {\bibinfo {title} {Majorana and parafermion corner states
  from two coupled sheets of bilayer graphene},\ }\href@noop {} {\bibfield
  {journal} {\bibinfo  {journal} {Phys. Rev. Research}\ }\textbf {\bibinfo
  {volume} {2}},\ \bibinfo {pages} {013330} (\bibinfo {year}
  {2020}{\natexlab{b}})}\BibitemShut {NoStop}%
\bibitem [{\citenamefont {Plekhanov}\ \emph {et~al.}(2020)\citenamefont
  {Plekhanov}, \citenamefont {Ronetti}, \citenamefont {Loss},\ and\
  \citenamefont {Klinovaja}}]{plekhanov_etal_prr_20}%
  \BibitemOpen
  \bibfield  {author} {\bibinfo {author} {\bibfnamefont {K.}~\bibnamefont
  {Plekhanov}}, \bibinfo {author} {\bibfnamefont {F.}~\bibnamefont {Ronetti}},
  \bibinfo {author} {\bibfnamefont {D.}~\bibnamefont {Loss}},\ and\ \bibinfo
  {author} {\bibfnamefont {J.}~\bibnamefont {Klinovaja}},\ }\bibfield  {title}
  {\bibinfo {title} {Hinge states in a system of coupled rashba layers},\
  }\href@noop {} {\bibfield  {journal} {\bibinfo  {journal} {Phys. Rev.
  Research}\ }\textbf {\bibinfo {volume} {2}},\ \bibinfo {pages} {013083}
  (\bibinfo {year} {2020})}\BibitemShut {NoStop}%
\bibitem [{\citenamefont {Plekhanov}\ \emph {et~al.}(2021)\citenamefont
  {Plekhanov}, \citenamefont {M\"uller}, \citenamefont {Volpez}, \citenamefont
  {Kennes}, \citenamefont {Schoeller}, \citenamefont {Loss},\ and\
  \citenamefont {Klinovaja}}]{plekhanov_etal_prb_21}%
  \BibitemOpen
  \bibfield  {author} {\bibinfo {author} {\bibfnamefont {K.}~\bibnamefont
  {Plekhanov}}, \bibinfo {author} {\bibfnamefont {N.}~\bibnamefont {M\"uller}},
  \bibinfo {author} {\bibfnamefont {Y.}~\bibnamefont {Volpez}}, \bibinfo
  {author} {\bibfnamefont {D.~M.}\ \bibnamefont {Kennes}}, \bibinfo {author}
  {\bibfnamefont {H.}~\bibnamefont {Schoeller}}, \bibinfo {author}
  {\bibfnamefont {D.}~\bibnamefont {Loss}},\ and\ \bibinfo {author}
  {\bibfnamefont {J.}~\bibnamefont {Klinovaja}},\ }\bibfield  {title} {\bibinfo
  {title} {Quadrupole spin polarization as signature of second-order
  topological superconductors},\ }\href@noop {} {\bibfield  {journal} {\bibinfo
   {journal} {Phys. Rev. B}\ }\textbf {\bibinfo {volume} {103}},\ \bibinfo
  {pages} {L041401} (\bibinfo {year} {2021})}\BibitemShut {NoStop}%
\bibitem [{\citenamefont {Jackiw}\ and\ \citenamefont
  {Rebbi}(1976)}]{jackiw_rebbi_prd_76}%
  \BibitemOpen
  \bibfield  {author} {\bibinfo {author} {\bibfnamefont {R.}~\bibnamefont
  {Jackiw}}\ and\ \bibinfo {author} {\bibfnamefont {C.}~\bibnamefont {Rebbi}},\
  }\bibfield  {title} {\bibinfo {title} {Solitons with fermion number
  \textonehalf{}},\ }\href {https://doi.org/10.1103/PhysRevD.13.3398}
  {\bibfield  {journal} {\bibinfo  {journal} {Phys. Rev. D}\ }\textbf {\bibinfo
  {volume} {13}},\ \bibinfo {pages} {3398} (\bibinfo {year}
  {1976})}\BibitemShut {NoStop}%
\bibitem [{\citenamefont {Laughlin}(1981)}]{laughlin_prb_81}%
  \BibitemOpen
  \bibfield  {author} {\bibinfo {author} {\bibfnamefont {R.~B.}\ \bibnamefont
  {Laughlin}},\ }\bibfield  {title} {\bibinfo {title} {Quantized hall
  conductivity in two dimensions},\ }\href
  {https://doi.org/10.1103/PhysRevB.23.5632} {\bibfield  {journal} {\bibinfo
  {journal} {Phys. Rev. B}\ }\textbf {\bibinfo {volume} {23}},\ \bibinfo
  {pages} {5632} (\bibinfo {year} {1981})}\BibitemShut {NoStop}%
\bibitem [{\citenamefont {Thouless}\ \emph {et~al.}(1982)\citenamefont
  {Thouless}, \citenamefont {Kohmoto}, \citenamefont {Nightingale},\ and\
  \citenamefont {den Nijs}}]{thouless_etal_prl_82}%
  \BibitemOpen
  \bibfield  {author} {\bibinfo {author} {\bibfnamefont {D.~J.}\ \bibnamefont
  {Thouless}}, \bibinfo {author} {\bibfnamefont {M.}~\bibnamefont {Kohmoto}},
  \bibinfo {author} {\bibfnamefont {M.~P.}\ \bibnamefont {Nightingale}},\ and\
  \bibinfo {author} {\bibfnamefont {M.}~\bibnamefont {den Nijs}},\ }\bibfield
  {title} {\bibinfo {title} {Quantized hall conductance in a two-dimensional
  periodic potential},\ }\href {https://doi.org/10.1103/PhysRevLett.49.405}
  {\bibfield  {journal} {\bibinfo  {journal} {Phys. Rev. Lett.}\ }\textbf
  {\bibinfo {volume} {49}},\ \bibinfo {pages} {405} (\bibinfo {year}
  {1982})}\BibitemShut {NoStop}%
\bibitem [{\citenamefont {Avron}\ \emph {et~al.}(1983)\citenamefont {Avron},
  \citenamefont {Seiler},\ and\ \citenamefont {Simon}}]{avron_etal_prl_83}%
  \BibitemOpen
  \bibfield  {author} {\bibinfo {author} {\bibfnamefont {J.~E.}\ \bibnamefont
  {Avron}}, \bibinfo {author} {\bibfnamefont {R.}~\bibnamefont {Seiler}},\ and\
  \bibinfo {author} {\bibfnamefont {B.}~\bibnamefont {Simon}},\ }\bibfield
  {title} {\bibinfo {title} {Homotopy and quantization in condensed matter
  physics},\ }\href {https://doi.org/10.1103/PhysRevLett.51.51} {\bibfield
  {journal} {\bibinfo  {journal} {Phys. Rev. Lett.}\ }\textbf {\bibinfo
  {volume} {51}},\ \bibinfo {pages} {51} (\bibinfo {year} {1983})}\BibitemShut
  {NoStop}%
\bibitem [{\citenamefont {Kohmoto}(1985)}]{kohmoto_aph_85}%
  \BibitemOpen
  \bibfield  {author} {\bibinfo {author} {\bibfnamefont {M.}~\bibnamefont
  {Kohmoto}},\ }\bibfield  {title} {\bibinfo {title} {Topological invariant and
  the quantization of the hall conductance},\ }\href
  {https://doi.org/https://doi.org/10.1016/0003-4916(85)90148-4} {\bibfield
  {journal} {\bibinfo  {journal} {Annals of Physics}\ }\textbf {\bibinfo
  {volume} {160}},\ \bibinfo {pages} {343} (\bibinfo {year}
  {1985})}\BibitemShut {NoStop}%
\bibitem [{\citenamefont {Park}\ \emph {et~al.}(2016)\citenamefont {Park},
  \citenamefont {Yang}, \citenamefont {Klinovaja}, \citenamefont {Stano},\ and\
  \citenamefont {Loss}}]{park_etal_prb_16}%
  \BibitemOpen
  \bibfield  {author} {\bibinfo {author} {\bibfnamefont {J.-H.}\ \bibnamefont
  {Park}}, \bibinfo {author} {\bibfnamefont {G.}~\bibnamefont {Yang}}, \bibinfo
  {author} {\bibfnamefont {J.}~\bibnamefont {Klinovaja}}, \bibinfo {author}
  {\bibfnamefont {P.}~\bibnamefont {Stano}},\ and\ \bibinfo {author}
  {\bibfnamefont {D.}~\bibnamefont {Loss}},\ }\bibfield  {title} {\bibinfo
  {title} {Fractional boundary charges in quantum dot arrays with density
  modulation},\ }\href {https://doi.org/10.1103/PhysRevB.94.075416} {\bibfield
  {journal} {\bibinfo  {journal} {Phys. Rev. B}\ }\textbf {\bibinfo {volume}
  {94}},\ \bibinfo {pages} {075416} (\bibinfo {year} {2016})}\BibitemShut
  {NoStop}%
\bibitem [{\citenamefont {Thakurathi}\ \emph {et~al.}(2018)\citenamefont
  {Thakurathi}, \citenamefont {Klinovaja},\ and\ \citenamefont
  {Loss}}]{thakurathi_etal_prb_18}%
  \BibitemOpen
  \bibfield  {author} {\bibinfo {author} {\bibfnamefont {M.}~\bibnamefont
  {Thakurathi}}, \bibinfo {author} {\bibfnamefont {J.}~\bibnamefont
  {Klinovaja}},\ and\ \bibinfo {author} {\bibfnamefont {D.}~\bibnamefont
  {Loss}},\ }\bibfield  {title} {\bibinfo {title} {From fractional boundary
  charges to quantized hall conductance},\ }\href
  {https://doi.org/10.1103/PhysRevB.98.245404} {\bibfield  {journal} {\bibinfo
  {journal} {Phys. Rev. B}\ }\textbf {\bibinfo {volume} {98}},\ \bibinfo
  {pages} {245404} (\bibinfo {year} {2018})}\BibitemShut {NoStop}%
\bibitem [{\citenamefont {Pletyukhov}\ \emph
  {et~al.}(2020{\natexlab{a}})\citenamefont {Pletyukhov}, \citenamefont
  {Kennes}, \citenamefont {Klinovaja}, \citenamefont {Loss},\ and\
  \citenamefont {Schoeller}}]{pletyukhov_etal_prb_20}%
  \BibitemOpen
  \bibfield  {author} {\bibinfo {author} {\bibfnamefont {M.}~\bibnamefont
  {Pletyukhov}}, \bibinfo {author} {\bibfnamefont {D.~M.}\ \bibnamefont
  {Kennes}}, \bibinfo {author} {\bibfnamefont {J.}~\bibnamefont {Klinovaja}},
  \bibinfo {author} {\bibfnamefont {D.}~\bibnamefont {Loss}},\ and\ \bibinfo
  {author} {\bibfnamefont {H.}~\bibnamefont {Schoeller}},\ }\bibfield  {title}
  {\bibinfo {title} {Surface charge theorem and topological constraints for
  edge states: Analytical study of one-dimensional nearest-neighbor
  tight-binding models},\ }\href {https://doi.org/10.1103/PhysRevB.101.165304}
  {\bibfield  {journal} {\bibinfo  {journal} {Phys. Rev. B}\ }\textbf {\bibinfo
  {volume} {101}},\ \bibinfo {pages} {165304} (\bibinfo {year}
  {2020}{\natexlab{a}})}\BibitemShut {NoStop}%
\bibitem [{\citenamefont {Pletyukhov}\ \emph
  {et~al.}(2020{\natexlab{b}})\citenamefont {Pletyukhov}, \citenamefont
  {Kennes}, \citenamefont {Klinovaja}, \citenamefont {Loss},\ and\
  \citenamefont {Schoeller}}]{pletyukhov_etal_prbr_20}%
  \BibitemOpen
  \bibfield  {author} {\bibinfo {author} {\bibfnamefont {M.}~\bibnamefont
  {Pletyukhov}}, \bibinfo {author} {\bibfnamefont {D.~M.}\ \bibnamefont
  {Kennes}}, \bibinfo {author} {\bibfnamefont {J.}~\bibnamefont {Klinovaja}},
  \bibinfo {author} {\bibfnamefont {D.}~\bibnamefont {Loss}},\ and\ \bibinfo
  {author} {\bibfnamefont {H.}~\bibnamefont {Schoeller}},\ }\bibfield  {title}
  {\bibinfo {title} {Topological invariants to characterize universality of
  boundary charge in one-dimensional insulators beyond symmetry constraints},\
  }\href {https://doi.org/10.1103/PhysRevB.101.161106} {\bibfield  {journal}
  {\bibinfo  {journal} {Phys. Rev. B}\ }\textbf {\bibinfo {volume} {101}},\
  \bibinfo {pages} {161106(R)} (\bibinfo {year}
  {2020}{\natexlab{b}})}\BibitemShut {NoStop}%
\bibitem [{\citenamefont {Pletyukhov}\ \emph
  {et~al.}(2020{\natexlab{c}})\citenamefont {Pletyukhov}, \citenamefont
  {Kennes}, \citenamefont {Piasotski}, \citenamefont {Klinovaja}, \citenamefont
  {Loss},\ and\ \citenamefont {Schoeller}}]{pletyukhov_etal_prr_20}%
  \BibitemOpen
  \bibfield  {author} {\bibinfo {author} {\bibfnamefont {M.}~\bibnamefont
  {Pletyukhov}}, \bibinfo {author} {\bibfnamefont {D.~M.}\ \bibnamefont
  {Kennes}}, \bibinfo {author} {\bibfnamefont {K.}~\bibnamefont {Piasotski}},
  \bibinfo {author} {\bibfnamefont {J.}~\bibnamefont {Klinovaja}}, \bibinfo
  {author} {\bibfnamefont {D.}~\bibnamefont {Loss}},\ and\ \bibinfo {author}
  {\bibfnamefont {H.}~\bibnamefont {Schoeller}},\ }\bibfield  {title} {\bibinfo
  {title} {Rational boundary charge in one-dimensional systems with interaction
  and disorder},\ }\href@noop {} {\bibfield  {journal} {\bibinfo  {journal}
  {Phys. Rev. Research}\ }\textbf {\bibinfo {volume} {2}},\ \bibinfo {pages}
  {033345} (\bibinfo {year} {2020}{\natexlab{c}})}\BibitemShut {NoStop}%
\bibitem [{\citenamefont {Miles}\ \emph {et~al.}(2021)\citenamefont {Miles},
  \citenamefont {Kennes}, \citenamefont {Schoeller},\ and\ \citenamefont
  {Pletyukhov}}]{miles_etal_prb_21}%
  \BibitemOpen
  \bibfield  {author} {\bibinfo {author} {\bibfnamefont {S.}~\bibnamefont
  {Miles}}, \bibinfo {author} {\bibfnamefont {D.~M.}\ \bibnamefont {Kennes}},
  \bibinfo {author} {\bibfnamefont {H.}~\bibnamefont {Schoeller}},\ and\
  \bibinfo {author} {\bibfnamefont {M.}~\bibnamefont {Pletyukhov}},\ }\bibfield
   {title} {\bibinfo {title} {Universal properties of boundary and interface
  charges in continuum models of one-dimensional insulators},\ }\href
  {https://doi.org/10.1103/PhysRevB.104.155409} {\bibfield  {journal} {\bibinfo
   {journal} {Phys. Rev. B}\ }\textbf {\bibinfo {volume} {104}},\ \bibinfo
  {pages} {155409} (\bibinfo {year} {2021})}\BibitemShut {NoStop}%
\bibitem [{\citenamefont {M\"uller}\ \emph {et~al.}(2021)\citenamefont
  {M\"uller}, \citenamefont {Piasotski}, \citenamefont {Kennes}, \citenamefont
  {Schoeller},\ and\ \citenamefont {Pletyukhov}}]{mueller_etal_prb_21}%
  \BibitemOpen
  \bibfield  {author} {\bibinfo {author} {\bibfnamefont {N.}~\bibnamefont
  {M\"uller}}, \bibinfo {author} {\bibfnamefont {K.}~\bibnamefont {Piasotski}},
  \bibinfo {author} {\bibfnamefont {D.~M.}\ \bibnamefont {Kennes}}, \bibinfo
  {author} {\bibfnamefont {H.}~\bibnamefont {Schoeller}},\ and\ \bibinfo
  {author} {\bibfnamefont {M.}~\bibnamefont {Pletyukhov}},\ }\bibfield  {title}
  {\bibinfo {title} {Universal properties of boundary and interface charges in
  multichannel one-dimensional models without symmetry constraints},\ }\href
  {https://doi.org/10.1103/PhysRevB.104.125447} {\bibfield  {journal} {\bibinfo
   {journal} {Phys. Rev. B}\ }\textbf {\bibinfo {volume} {104}},\ \bibinfo
  {pages} {125447} (\bibinfo {year} {2021})}\BibitemShut {NoStop}%
\bibitem [{\citenamefont {Laubscher}\ \emph {et~al.}(2021)\citenamefont
  {Laubscher}, \citenamefont {Weber}, \citenamefont {Kennes}, \citenamefont
  {Pletyukhov}, \citenamefont {Schoeller}, \citenamefont {Loss},\ and\
  \citenamefont {Klinovaja}}]{laubscher_etal_prb_21}%
  \BibitemOpen
  \bibfield  {author} {\bibinfo {author} {\bibfnamefont {K.}~\bibnamefont
  {Laubscher}}, \bibinfo {author} {\bibfnamefont {C.~S.}\ \bibnamefont
  {Weber}}, \bibinfo {author} {\bibfnamefont {D.~M.}\ \bibnamefont {Kennes}},
  \bibinfo {author} {\bibfnamefont {M.}~\bibnamefont {Pletyukhov}}, \bibinfo
  {author} {\bibfnamefont {H.}~\bibnamefont {Schoeller}}, \bibinfo {author}
  {\bibfnamefont {D.}~\bibnamefont {Loss}},\ and\ \bibinfo {author}
  {\bibfnamefont {J.}~\bibnamefont {Klinovaja}},\ }\bibfield  {title} {\bibinfo
  {title} {Fractional boundary charges with quantized slopes in interacting
  one- and two-dimensional systems},\ }\href
  {https://doi.org/10.1103/PhysRevB.104.035432} {\bibfield  {journal} {\bibinfo
   {journal} {Phys. Rev. B}\ }\textbf {\bibinfo {volume} {104}},\ \bibinfo
  {pages} {035432} (\bibinfo {year} {2021})}\BibitemShut {NoStop}%
\bibitem [{\citenamefont {Weber}\ \emph {et~al.}(2021)\citenamefont {Weber},
  \citenamefont {Piasotski}, \citenamefont {Pletyukhov}, \citenamefont
  {Klinovaja}, \citenamefont {Loss}, \citenamefont {Schoeller},\ and\
  \citenamefont {Kennes}}]{weber_etal_prl_21}%
  \BibitemOpen
  \bibfield  {author} {\bibinfo {author} {\bibfnamefont {C.~S.}\ \bibnamefont
  {Weber}}, \bibinfo {author} {\bibfnamefont {K.}~\bibnamefont {Piasotski}},
  \bibinfo {author} {\bibfnamefont {M.}~\bibnamefont {Pletyukhov}}, \bibinfo
  {author} {\bibfnamefont {J.}~\bibnamefont {Klinovaja}}, \bibinfo {author}
  {\bibfnamefont {D.}~\bibnamefont {Loss}}, \bibinfo {author} {\bibfnamefont
  {H.}~\bibnamefont {Schoeller}},\ and\ \bibinfo {author} {\bibfnamefont
  {D.~M.}\ \bibnamefont {Kennes}},\ }\bibfield  {title} {\bibinfo {title}
  {Universality of boundary charge fluctuations},\ }\href
  {https://doi.org/10.1103/PhysRevLett.126.016803} {\bibfield  {journal}
  {\bibinfo  {journal} {Phys. Rev. Lett.}\ }\textbf {\bibinfo {volume} {126}},\
  \bibinfo {pages} {016803} (\bibinfo {year} {2021})}\BibitemShut {NoStop}%
\bibitem [{\citenamefont {Piasotski}\ \emph {et~al.}(2021)\citenamefont
  {Piasotski}, \citenamefont {Pletyukhov}, \citenamefont {Weber}, \citenamefont
  {Klinovaja}, \citenamefont {Kennes},\ and\ \citenamefont
  {Schoeller}}]{piasotski_etal_prr_21}%
  \BibitemOpen
  \bibfield  {author} {\bibinfo {author} {\bibfnamefont {K.}~\bibnamefont
  {Piasotski}}, \bibinfo {author} {\bibfnamefont {M.}~\bibnamefont
  {Pletyukhov}}, \bibinfo {author} {\bibfnamefont {C.~S.}\ \bibnamefont
  {Weber}}, \bibinfo {author} {\bibfnamefont {J.}~\bibnamefont {Klinovaja}},
  \bibinfo {author} {\bibfnamefont {D.~M.}\ \bibnamefont {Kennes}},\ and\
  \bibinfo {author} {\bibfnamefont {H.}~\bibnamefont {Schoeller}},\ }\bibfield
  {title} {\bibinfo {title} {Universality of abelian and non-abelian wannier
  functions in generalized one-dimensional aubry-andr\'e-harper models},\
  }\href {https://doi.org/10.1103/PhysRevResearch.3.033167} {\bibfield
  {journal} {\bibinfo  {journal} {Phys. Rev. Research}\ }\textbf {\bibinfo
  {volume} {3}},\ \bibinfo {pages} {033167} (\bibinfo {year}
  {2021})}\BibitemShut {NoStop}%
\bibitem [{\citenamefont {Hayward}\ \emph {et~al.}(2021)\citenamefont
  {Hayward}, \citenamefont {Bertok}, \citenamefont {Schneider},\ and\
  \citenamefont {Heidrich-Meisner}}]{hayward_etal_pra_21}%
  \BibitemOpen
  \bibfield  {author} {\bibinfo {author} {\bibfnamefont {A.~L.~C.}\
  \bibnamefont {Hayward}}, \bibinfo {author} {\bibfnamefont {E.}~\bibnamefont
  {Bertok}}, \bibinfo {author} {\bibfnamefont {U.}~\bibnamefont {Schneider}},\
  and\ \bibinfo {author} {\bibfnamefont {F.}~\bibnamefont {Heidrich-Meisner}},\
  }\bibfield  {title} {\bibinfo {title} {Effect of disorder on topological
  charge pumping in the rice-mele model},\ }\href
  {https://doi.org/10.1103/PhysRevA.103.043310} {\bibfield  {journal} {\bibinfo
   {journal} {Phys. Rev. A}\ }\textbf {\bibinfo {volume} {103}},\ \bibinfo
  {pages} {043310} (\bibinfo {year} {2021})}\BibitemShut {NoStop}%
\bibitem [{SM()}]{SM}%
  \BibitemOpen
  \href@noop {} {\ }\bibinfo {note} {\hspace{-0.125cm}In the Supplemental
  Material we discuss the dimensional reduction of the 3D model; check the
  agreement of the low-energy state values in the effective 2D model between
  the tb and the continuum model calculations; provide the low-energy
  description of the hinge states; numerically verify the absent contribution
  to the linear slope from the bands $\varepsilon_{\lambda \leq -1}$, and
  provide theoretical details of the numerical studies in 3D tb model in the
  presence of the disorder potential. We also investigate the role of the
  orbital $B$-field effect and deformations of the torus surface, showing the
  robustness of the discussed 3D QHE.}\BibitemShut {Stop}%
\bibitem [{\citenamefont {Datta}(1995)}]{datta1997electronic}%
  \BibitemOpen
  \bibfield  {author} {\bibinfo {author} {\bibfnamefont {S.}~\bibnamefont
  {Datta}},\ }\bibinfo {title} {Electronic transport in mesoscopic systems}\
  (\bibinfo  {publisher} {Cambridge university press},\ \bibinfo {address}
  {Cambridge},\ \bibinfo {year} {1995})\ pp.\ \bibinfo {pages}
  {141--145}\BibitemShut {NoStop}%
\bibitem [{\citenamefont {Shen}(2017)}]{shen2017topological}%
  \BibitemOpen
  \bibfield  {author} {\bibinfo {author} {\bibfnamefont {S.-Q.}\ \bibnamefont
  {Shen}},\ }\bibinfo {title} {Topological insulators: Dirac equation in
  condensed matters, second edition}\ (\bibinfo  {publisher} {Springer Nature
  Singapore Pte Ltd},\ \bibinfo {address} {Singapore},\ \bibinfo {year}
  {2017})\BibitemShut {NoStop}%
\bibitem [{\citenamefont {Hosur}\ \emph {et~al.}(2011)\citenamefont {Hosur},
  \citenamefont {Ghaemi}, \citenamefont {Mong},\ and\ \citenamefont
  {Vishwanath}}]{hosur2011majorana}%
  \BibitemOpen
  \bibfield  {author} {\bibinfo {author} {\bibfnamefont {P.}~\bibnamefont
  {Hosur}}, \bibinfo {author} {\bibfnamefont {P.}~\bibnamefont {Ghaemi}},
  \bibinfo {author} {\bibfnamefont {R.~S.}\ \bibnamefont {Mong}},\ and\
  \bibinfo {author} {\bibfnamefont {A.}~\bibnamefont {Vishwanath}},\ }\bibfield
   {title} {\bibinfo {title} {Majorana modes at the ends of superconductor
  vortices in doped topological insulators},\ }\href@noop {} {\bibfield
  {journal} {\bibinfo  {journal} {Phys. Rev. Lett.}\ }\textbf {\bibinfo
  {volume} {107}},\ \bibinfo {pages} {097001} (\bibinfo {year}
  {2011})}\BibitemShut {NoStop}%
\bibitem [{\citenamefont {Do}(2014)}]{do_anp_14}%
  \BibitemOpen
  \bibfield  {author} {\bibinfo {author} {\bibfnamefont {V.-N.}\ \bibnamefont
  {Do}},\ }\bibfield  {title} {\bibinfo {title} {Non-equilibrium green function
  method: theory and application in simulation of nanometer electronic
  devices},\ }\href {https://doi.org/10.1088/2043-6262/5/3/033001} {\bibfield
  {journal} {\bibinfo  {journal} {Advances in Natural Sciences: Nanoscience and
  Nanotechnology}\ }\textbf {\bibinfo {volume} {5}},\ \bibinfo {pages} {033001}
  (\bibinfo {year} {2014})}\BibitemShut {NoStop}%
\end{thebibliography}%

\end{document}